%% file: WiFi_tracking.tex
\documentclass[10pt,journal,compsoc]{IEEEtran}

\ifCLASSOPTIONcompsoc
  \usepackage[nocompress]{cite}
\else
  \usepackage{cite}
\fi

\usepackage{amssymb,amsfonts}
\usepackage{algorithmic}
\usepackage{graphicx}
\usepackage{textcomp}
\usepackage{xcolor}
\usepackage{amsmath}
\usepackage{bm}
\usepackage{empheq}
\usepackage{caption}
\usepackage{subcaption}
\usepackage{xurl}
\DeclareMathOperator{\arctantwo}{atan2}

\ifCLASSINFOpdf
\else
\fi

\begin{document}

\title{Single-Target Real-Time Passive WiFi Tracking}
\author{Zhongqin Wang,
        J. Andrew Zhang, \IEEEmembership{Senior Member, IEEE}, 
        Min Xu, \IEEEmembership{Member, IEEE}, \\
        and Y. Jay Guo, \IEEEmembership{Fellow, IEEE} 

\IEEEcompsocitemizethanks{
\IEEEcompsocthanksitem Zhongqin Wang, J. Andrew Zhang (Corresponding Author), and Min Xu are with the School of Electrical and Data Engineering, University of Technology Sydney, Sydney 2007, Australia.\protect\\
E-mail: \{zhongqin.wang, andrew.zhang, min.xu\}@uts.edu.au
\IEEEcompsocthanksitem Y. Jay Guo is with the Global Big Data Technologies Centre, University of Technology Sydney, Sydney 2007, Australia.\protect\\
E-mail: jay.guo@uts.edu.au
}}

\IEEEtitleabstractindextext{%
\begin{abstract}
Device-free human tracking is an essential ingredient for ubiquitous wireless sensing. Recent passive WiFi tracking systems face the challenges of inaccurate separation of dynamic human components and time-consuming estimation of multi-dimensional signal parameters. In this work, we present a scheme named \textbf{Wi}Fi \textbf{D}oppler \textbf{F}requency \textbf{S}hift (\textbf{WiDFS}), which can achieve single-target real-time passive tracking using channel state information (CSI) collected from commercial-off-the-shelf (COTS) WiFi devices. We consider the typical system setup including a transmitter with a single antenna and a receiver with three antennas; while our scheme can be readily extended to another setup. To remove the impact of transceiver asynchronization, we first apply CSI cross-correlation between each RX antenna pair. We then combine them to estimate a Doppler frequency shift (DFS) in a short-time window. After that, we leverage the DFS estimate to separate dynamic human components from CSI self-correlation terms of each antenna, thereby separately calculating angle-of-arrival (AoA) and human reflection distance for tracking. In addition, a hardware calibration algorithm is presented to refine the spacing between RX antennas and eliminate the hardware-related phase differences between them. A prototype demonstrates that WiDFS can achieve real-time tracking with a median position error of 72.32 cm in multipath-rich environments.
\end{abstract}

}

\maketitle
\IEEEdisplaynontitleabstractindextext

%
\IEEEpeerreviewmaketitle

\IEEEraisesectionheading{\section{Introduction}\label{sec:introduction}}
\IEEEPARstart{P}{assive} WiFi tracking is a promising technique that uses WiFi signals to locate people without needing any other sensors or wearable devices. Compared to other wireless signal based solutions like radio frequency identification (RFID) \cite{ma2017minding, ma2017drone, luo20193d, xu2019adarf, wang2019computer} and millimetre wave (mmWave) radar \cite{wei2015mtrack, zhao2019mid, wu2020mmtrack, zhao2021human}, WiFi infrastructures are almost ubiquitous at public work places and homes, thereby avoiding the need of deploying dedicated wireless tracking infrastructure and devices. Wireless sensing techniques are free from light conditions and even perform well in non-line-of-sight (NLOS) scenarios where a target is blocked \cite{zhao2018rf,zhao2018through}.

Passive WiFi tracking has gained much attention from academic and industrial communities over the past years. Recent works \cite{karanam2019tracking, tadayon2019decimeter, chen2020fido, venkatnarayan2020leveraging} exploit a stable and feature-rich signal parameter for positioning, i.e., channel state information (CSI), which can be extracted from a commercial-off-the-shelf (COTS) WiFi network interface controller (NIC) (e.g., Intel 5300 \cite{halperin2010predictable} and Atheros QCA9558 \cite{xie2018precise}). Such CSI-based localization solutions could achieve finer decimeter-level localization accuracies. To the best of our knowledge, however, most existing approaches with fine tracking accuracy are hard to directly adopt in practical applications due to the following reasons. First, huge computation in each position estimation may jeopardize the system's real-time performance \cite{xie2019md}. Second, estimation errors in many Doppler-based continuous tracking solutions \cite{li2017indotrack} may accumulate over time, resulting in trajectory drift. Third, deep learning-based solutions \cite{wang2016csi, abbas2019wideep} require a huge amount of labeled CSI data in specified scenarios for training. Since different environments have different multipath interference, the trained network may not be universal.

There are also three major challenges in passive WiFi tracking, which have not been well addressed in the literature. These challenges are detailed below.

\textit{1)} WiFi transceiver clock asynchronization \cite{kotaru2015spotfi, zhuo2017perceiving, xie2018precise, tadayon2019decimeter} results in time-varying phase shifts in CSI. Many existing works \cite{qian2017inferring, qian2018widar2, jiang2020towards} exploit cross-correlation between CSIs of pairwise antennas to address the asynchronous signal processing problem. However, this operation introduces a side product, i.e., the conjugate terms of the dynamic cross-correlation terms of interest. In this case, \textit{DFS ambiguity} is created, meaning that the DFS to be estimated can be the true value or its negative value. A common solution to suppressing the ambiguity is adding a constant to the reference signal and subtracting another one to the rest signals before the cross-correlation \cite{kotaru2015spotfi, qian2017inferring}. However, this does not always work, especially in multipath-rich scenarios.

\textit{2)} Separating the dynamic human components is challenging. An existing common method \cite{li2017indotrack} aims to directly separate the dynamic human component from cross-correlation terms by adding a factor to the CSI amplitudes of the reference antenna and subtracting another factor from the CSI amplitudes of other antennas. However, this power adjustment solution cannot completely eliminate the impact of the side product in dynamic component separation. Furthermore, these dynamic components may not always be reliable for position indication. For example, human body is not a perfect reflector and may not reflect signals to a receive antenna array at some sampling time.

\textit{3)} The difference in WiFi hardware, including WiFi NIC, RX antennas, and RX antenna cables, introduces different phase shifts on each RX antenna, which significantly impact human tracking accuracy. An exisiting method \cite{xiong2013arraytrack} conducts WiFi hardware calibration using a SMA splitter. However, an unknown $\pi$-radians phase ambiguity will be induced. Also, since the RX antenna array is customized, the measured antenna spacing inevitably deviates from the actual value. Removing the $\pi$-radians phase ambiguity and estimating the true RX antenna spacing are challenging.

In this paper, we propose a WiFi single-target passive tracking scheme, called \textbf{Wi}Fi \textbf{D}oppler \textbf{F}requency \textbf{S}hift (\textbf{WiDFS}), which enables to overcome the above three challenges and can be run in real time at a medium class mini PC. Currently, a standard COTS NIC supports up to three antennas. WiDFS tracks a moving person using a COTS WiFi transceiver with one TX antenna and three RX antennas. Using three-RX-antenna is essential as we need to use their CSIs to estimate the angle-of-arrival (AoA) of incoming signals, and to remove transceiver clock asynchronization and accurately estimate the DFS caused by human movement. Once the DFS estimate is obtained, we then leverage it to separate dynamic human components. WiDFS can then track the person by estimating the AoA in the direction of human reflection and the length of the reflection path from the transmitter to the person and then the receiver. Our main contributions are as follows.

\textit{1)} We propose a DFS estimation algorithm based on cross-correlation between each RX antenna pair. Compared to existing solutions, WiDFS can eliminate the impact of the side product that is mixed with the dynamic cross-correlation term of interest. WiDFS leverages the static cross-correlation term which is usually abandoned in previous works. This term can be easily obtained by averaging over a CSI sampling window. WiDFS then conducts a straightforward transformation to resolve the DFS ambiguity caused by the side product. Since three RX antennas are separated by less than half a wavelength, it is reasonable to assume that the DFS of each antenna is almost the same. Thus, WiDFS builds a CSI observation matrix to estimate an accurate DFS using a subspace-based MUSIC algorithm in each short-time sampling window.

\textit{2)} We design a lightweight algorithm that uses the estimated DFS to separate the dynamic human components, which is more robust and accurate than the power adjustment and reference antenna solutions, especially in actual multipath-rich scenarios. In this work, we focus on each antenna's instantaneous power of channel frequency response (CFR) based on self-correlation. The CFR power is also free from the impact of transceiver clock asynchronization. WiDFS first relies on the estimated DFS to refine the CFR power in a CSI sampling window. Then it uses a simple yet effective solution to reconstruct dynamic human components by formulating a linear least-squares problem. We further propose a windowed algorithm that can deal with the problem that RX antennas cannot capture human reflections or can only capture minor reflections. WiDFS combines CSI data from multiple sampling windows for localization parameter estimation. It relies on the estimated DFS to achieve unsupervised motion sensing, which can detect the absence and presence of a moving person.

\textit{3)} We present a WiFi hardware calibration solution to estimate the hardware-dependent phase shifts and antenna spacing between any two antennas, without any specialized devices. WiDFS collects CSIs by deploying a TX antenna on each side of an RX antenna array and then uses a standard phase-distance model for calibration.

A prototype of WiDFS is implemented using a transmitter with a single TX antenna and a receiver with three RX antennas forming a linear uniform antenna array. WiDFS is programmed using Python. The experiments demonstrate that WiDFS achieves median and 90$^{th}$ percentile position errors of 72.31 cm and 170.8 cm, respectively. Such a localization accuracy exceeds a state-of-the-art technique Widar2.0 \cite{qian2018widar2} by about 36 cm and 70 cm. More importantly, WiDFS costs a mean running time of 0.076 s on a mini PC (online) and 0.024 s on a MacBook Pro (offline) to output each position estimate when collecting about 0.1 second CSI data, leading to real-time tracking in practical applications.

\begin{figure}
       \centering
        \includegraphics[width=0.4\textwidth]{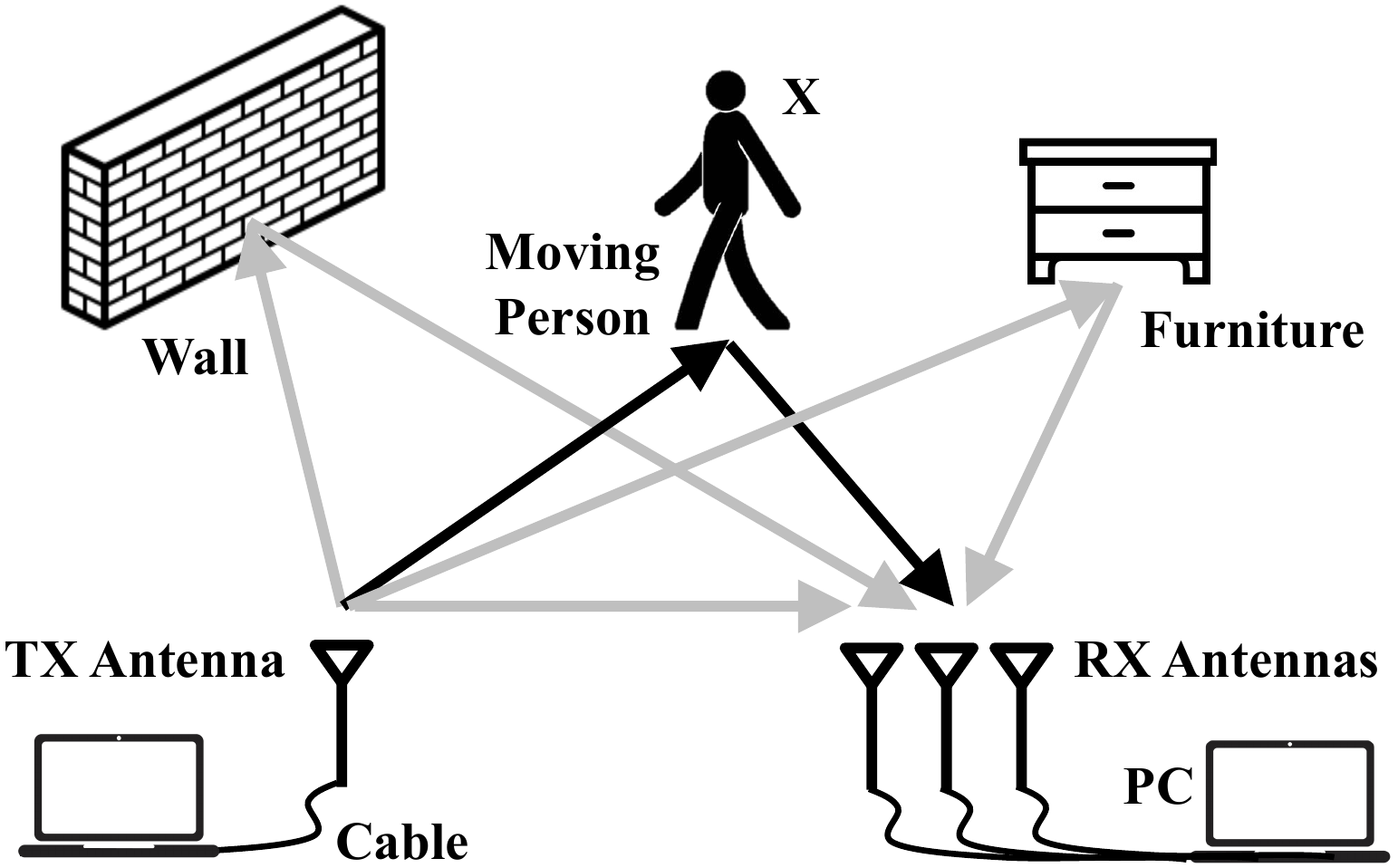}
    \caption{Signal propagation paths in dynamic environments}
    \label{Fig1}
    \vspace{-1.5em}
\end{figure}

\section{CSI Modeling in dynamic environments}
This section describes our CSI model in a dynamic environment where a person is moving in a typical room. The CSI provides information on the environment as changes of signal propagation environment cause variations of CSI over time. The CSI can be accessed on certain COTS WiFi NICs using 802.11n or Atheros CSI tools \cite{halperin2010predictable, xie2018precise}. Such a NIC typically supports up to three Tx and Rx antennas. Fig. \ref{Fig1} shows a typical WiFi-based sensing scenario. In this setup with Intel NIC 5300, we use a Tx with a single antenna and a receiver with three RX antennas forming a uniform linear antenna array (ULA). The 3-antenna ULA offers Angle-of-Arrival (AoA) estimation capability for a limited number of multipath signals.

\subsection{General CSI model in COTS WiFi systems}
The CSI characterizes the channel frequency response (CFR) of a wireless signal propagating from a transmitter to a receiver. In the IEEE 802.11n standard, a channel has a 20 MHz bandwidth with 30 subcarriers. Then the CSI matrix in our scheme contains the number of $1 \times 3 \times 30$ complex channel coefficients at each sampling time. In COTS WiFi systems, the actual CSI measurements always suffer from the additional noise caused by NIC processing imperfection and WiFi hardware diversity.

\textbf{1) WiFi NIC Processing Imperfection.} The original CSI may contain \textit{time-varying} terms associated with imperfect signal processing in NIC, such as residual time and frequency synchronization offsets due to transmitter-receiver clock asynchronism, a power control uncertainty error and an I-Q imbalance error \cite{tadayon2019decimeter}. These errors produce the same impact on all RX antennas. Let $H^e_{j,k}$ denote the CFR associated with such imperfections at the $j$-th subcarrier ($j=1,2,...,30$) and the $k$-th CSI sample (or time-slot, which typically corresponds to the $k$-th CSI packet, $k=1,2,...$),
\begin{equation}
H^e_{j,k} = {\rho _{k}^{agc}}{{e}^{-\bm{J} \left( { \varphi_{j,k}^{FO} +\varphi_{j,k}^{TO} }\right)}},
\label{equation1}
\end{equation}
where ${\rho _{k}^{agc}}$ represents the gain set at the automatic gain controller (AGC) at the receiver and it dynamically varies with different channel characteristics; $\varphi_{j,k}^{FO}$ is the time-varying phase shift caused by the frequency offset (FO) at each subcarrier frequency; $\varphi_{j,k}^{TO}$ is the time-varying phase shift caused by the time offset (TO); both $\varphi_{j,k}^{TO}$ and $\varphi_{j,k}^{FO}$ may vary across CSI samples because of the un-locked clocks between Tx and Rx. The CFO may be accurately estimated at the receiver and hence the term $\varphi_{j,k}^{FO}$ can be compensated. However, TO associated with the transmitter and receiver clocking difference is generally hard to be estimated. 

\textbf{2) WiFi Hardware Diversity.} The WiFi hardware includes external WiFi antennas, antenna cables, SMA connectors as well as NIC itself. Their impacts on receiving signals at each subcarrier keep unchanged over time. However, manufacturing imperfection may cause different phase shifts on each RX antenna. According to \cite{xiong2013arraytrack}, their variations over subcarriers can be ignored, so the WiFi hardware-related CFR $H^h_{i}$ at the $i$-th RX antenna ($i=1,2,3$) can be represented as
\begin{equation}
H^h_{i} = {\rho _{i}^h}{{e}^{-\bm{J} { \varphi _{i}^h }}},
\label{equation2}
\end{equation}
where ${\rho _{i}^h}$ and $\varphi_{i}$ are the WiFi hardware-related attenuation and phase shift at the $i$-th RX antenna.

Accounting for the two multiplicative interference terms in \eqref{equation1} and \eqref{equation2}, a general CSI model \cite{zhang2021overview, ni2021uplink} at the $i$-th RX antenna, the $j$-th subcarrier and the $k$-th CSI sample is represented as
\begin{equation}
\mathit{CSI}_{i,j,k}=H^e_{j,k}H^h_{i}\sum\limits_{l=1}^{L}\mathit{H}_{i,j,k}\lbrack l \rbrack,
\label{equation3}
\end{equation}
where
\begin{equation}
\resizebox{\linewidth}{!}{$
\mathit{H}_{i,j,k}\lbrack l \rbrack={{\rho _{i,j,k}\lbrack l \rbrack}{{e}^{-\bm{J} 2\pi \frac{f_j}{c} \left( d_k\lbrack l \rbrack+ c\frac{f_{i,k}^D\lbrack l \rbrack}{f_c}\Delta t+\left( i-1 \right)\Delta d \sin \theta_k\lbrack l \rbrack \right) }}}.
$}
\label{equation4}
\end{equation}
Some variables in Eq. \ref{equation4} are described as follows: $f_j$ is the $j$-th subcarrier frequency, $f_c$ is the center frequency, $c$ is the speed of light, $\Delta t$ is the sampling time interval, $\Delta d$ is the RX antenna spacing no more than half a wavelength. For the $l$-th multipath, $\rho_{i,j,k} \lbrack l \rbrack$ and $d_k\lbrack l \rbrack$ are the signal propagation attenuation and length, $f_{i,k}^D\lbrack l \rbrack$ is the \textbf{DFS} which is introduced to the carrier frequency at the $i$-th RX antenna due to object movement ($v_{i,k}\lbrack l \rbrack=c\frac{{f}^{D}_{i,k}\lbrack l \rbrack}{f_c}$ is called the radial speed), $\theta_k\lbrack l \rbrack$ is the \textbf{AoA} that is the direction from which a reflection is received by the ULA.

According to \cite{zhang2021overview}, we define a short-time sampling window which is typically a few milliseconds when tracked objects move at velocities up to several meters per second. In WiDFS, the length of the time window is set to about 0.1 seconds during which $N_p=100$ CSI packets are collected due to the sampling frequency $f_s$ of 1 KHz. In this window, we can assume that  the gain ${\rho _{k}^{agc}}$, attenuation $\rho _{i,j,k}^{X}$, DFS $f_{i,k}^D\lbrack l \rbrack$ in Eq. \ref{equation4} all remain almost unchanged, then we denote ${\rho _{k}^{agc}} = {\rho^{agc}}$, ${\rho _{i,j,k}} = {\rho _{i,j}}$, and $f_{i,k}^D\lbrack l \rbrack = f_i^D\lbrack l \rbrack $, respectively. In addition, we assume the AoA spacing $\Delta {{\theta}} ^X$ between two successive CSI samples, i.e., $\Delta {{\theta}} ^X = \theta_{k+1}\lbrack l \rbrack - \theta_{k}\lbrack l \rbrack$, keeps unchanged in the window. In the following, we will rely on these assumptions to revise the above general CSI model.

\subsection{Our CSI model for single-target passive tracking}
Here we extend the general CSI model and introduce ours for single-target passive tracking in a dynamic environment in a short-time window. 

When a transmitter emits WiFi signals to space, the receiver receives two categories of signals. One is the \textit{direct} signal that travels along the direct path from the transmitter to the receiver. Another is the \textit{reflected} signals that bounce off different objects like the floor, wall, furniture, and the tracked person. Furthermore, we divide these signals into \textit{static} and \textit{dynamic} signals: \textit{(1)} the former includes a direct signal that propagates from the transmitter to the receiver, and the signals that reflect off surrounding static objects to the receiver; \textit{(2)} the latter contains the reflected signals that directly reflect off the human body to the receiver, and those that firstly bounce off the human body to other surrounding objects and then travel back to the receiver. Let $H^S_{i,j}$ be the CFR corresponding to the static signals between the TX antenna and the $i$-th RX antenna at the $j$-th subcarrier, called \textit{static component}. Specially, we assume the signal strength of the direct path between the TX and RX antennas is much higher than other static multipath signals. Let $H^X_{i,j,k}$ be the CFR corresponding to the reflected signals off the moving person $X$, called \textit{dynamic human component}. Thus, we rewrite the CSI model in Eq. \ref{equation3} to formulate the problem of single-target passive tracking as
\begin{equation}
\mathit{CSI}_{i,j,k}=H^e_{j,k}H^h_{i}\left( H^S_{i,j}+H^X_{i,j,k}\right),
\label{equation5}
\end{equation}
where
\begin{equation}
\left\{ 
\begin{aligned}
&H^S_{i,j} = {\rho _{i,j}^{S}}{{e}^{-\bm{J}{2\pi \frac{{f}_{j}}{c} {d^{S}_{i}}}}}+\mathbb{N}^S_{i,j} \\
&H^X_{i,j,k}={\rho _{i,j}^{X}}{e}^{-\bm{J}2\pi\frac{f_j}{c}d_{i,k}^X}+\mathbb{N}^X_{i,j,k}\\
\end{aligned} 
\right..
\label{equation6}
\end{equation}
In the above, $\rho _{i,j}^{S}$ and $\rho _{i,j,k}^{X}$ are the propagation attenuations of the direct and reflected path; $d_{i}^S$ is the distance of the direct path between the transmitter and the $i$-th RX antenna, which can be manually measured in advance; $d_{i,k}^X$ is the distance of the reflected path where the transmit signal is reflected from the tracked person $X$ to the $i$-th RX antenna at the $k$-th sampling point (for simplicity, we call it \textit{human reflection distance}); $\mathbb{N}_{i,j}^S$ and $\mathbb{N}_{i,j,k}^X$ are the noise terms caused by other minor multipath reflections. 
\begin{figure*}
       \centering
        \includegraphics[width=0.83\textwidth]{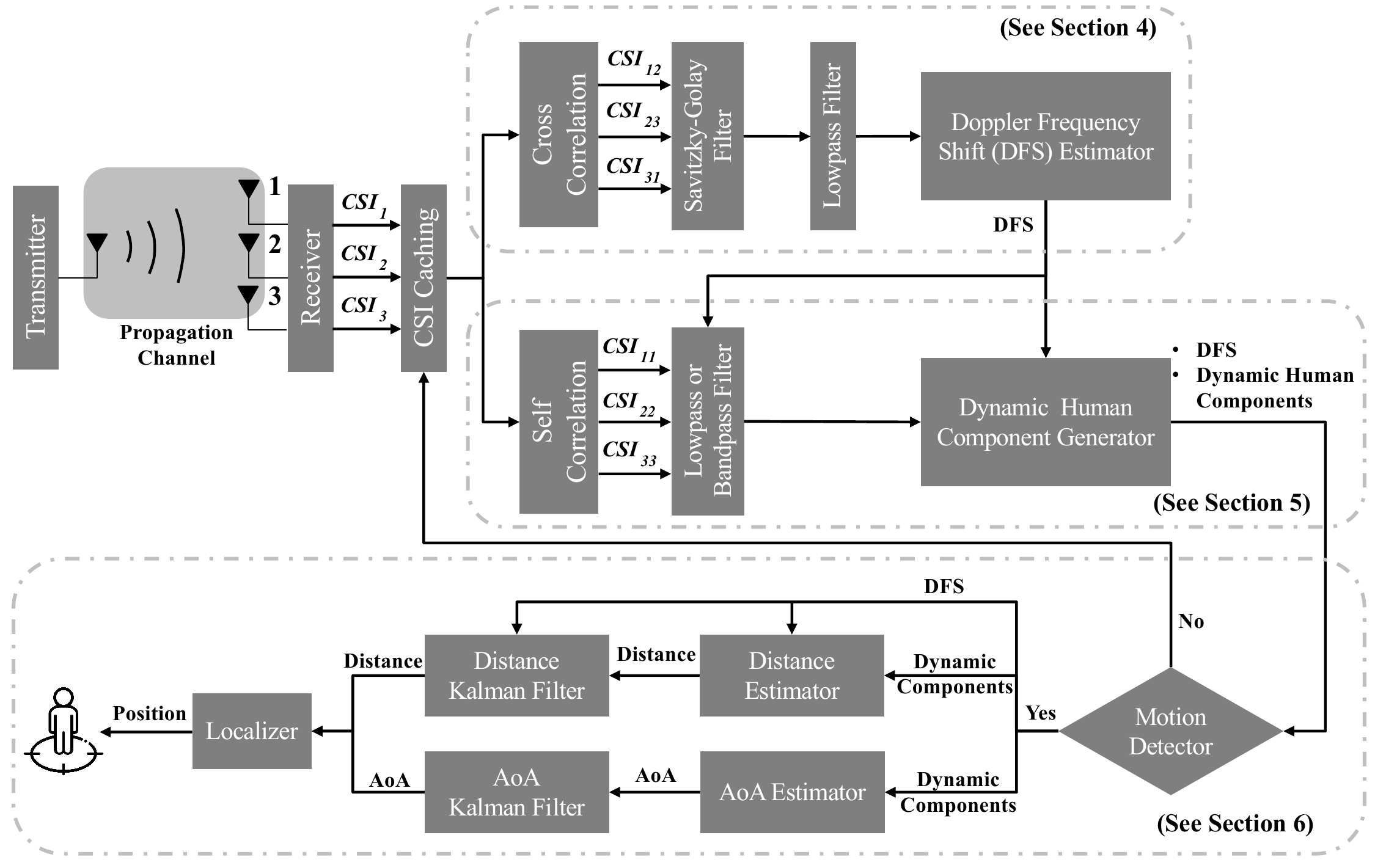}
    \caption{WiDFS workflow}
    \label{Fig2}
    \vspace{-1.5em}
\end{figure*}

In a short-time window as defined above, the human reflection distance $d_{i,k}^X$ can be represented as
\begin{equation}
d_{i,k}^X \approx d_{i,1}^X+c\frac{{f}^{D}_{i}}{f_c}\left( k-1 \right)\Delta t,
\label{equation7}
\end{equation}
where $d_{i,1}^X$ is the human reflection distance at the initial position in the window.
Further, the distance $d_{i,k}^X$, $i\in \lbrace2,3\rbrace$, can be rewritten using an initial AoA $\theta_{1}^X$,
\begin{equation}
\begin{aligned}
d_{i,k}^X \approx&  d_{1,k}^X+\left(i-1\right)\Delta d\sin\theta_{k}^X \\ 
\approx&  d_{1,1}^X+c\frac{{f}^{D}_{1}}{f_c}\left( k-1 \right)\Delta t+ \\
& \qquad \quad\left(i-1\right)\Delta d\sin\left[ \theta_{1}^X + \left( k-1 \right) \Delta {{\theta}}^X  \right].
\end{aligned} 
\label{equation8}
\end{equation}

In indoor environments, since the person does not moves too fast, $\Delta {{\theta}} ^X$ would be very small in the time interval of about 0.001 seconds. Then the derivative of $\sin \theta_{k}^X$ is 
\begin{equation}
\lim_{\Delta \theta^X \to 0} \frac{\sin\left[ \theta_{1}^X + \left( k-1 \right)\Delta {{\theta}}^X  \right]-\sin \theta_{1}^X}{ \left( k-1 \right)\Delta {{\theta}}^X}=\cos\theta_{1}^X. \\ 
\label{equation9}
\end{equation}

Then Eq. \ref{equation8} can be approximately rewritten as 
\begin{equation}
\resizebox{\linewidth}{!}{$
\begin{aligned}
d_{i,k}^X \approx & d_{1,1}^X+\left( i-1 \right)\Delta d \sin \theta_{1}^X+ \\
& \left[c\frac{{f}^{D}_{1}}{f_c}+ \left( i-1 \right)f^s \Delta d \Delta {{\theta}}^X\cos\theta_{1}^X  \right]\left( k-1 \right)\Delta t \\ 
= &  d_{1,1}^X+\left( i-1 \right)\Delta d \sin \theta_{1}^X+ \frac{c}{f_c}\left({f}^{D}_{1}+ f^{AoA}_{i,j} \right)\left( k-1 \right)\Delta t, \\ 
\end{aligned}
$}
\label{equation10}
\end{equation}
where $f^{AoA}_{i}=\left( i-1 \right)\frac{f_c}{c}f^s \Delta d\Delta {{\theta}}^X\cos\theta_{1}^X $ is called the \textit{AoA frequency shift (AFS)}. Here we note that the DFS at the $i$-th RX antenna ($i=2,3$) is ${f}^{D}_{i}={f}^{D}_{1}+f^{AoA}_{i}$. We can see that the DFS has a slight difference among different RX antennas, which is generally ignored in previous works. When a person moves at a low speed, the impact of $f^{AoA}_{i}$ is small and thereby we can ignore its impact. In Section 9, we will verify the impact of human motion speed on tracking accuracy. In the rest of this work, we denote ${f}^{D}={f}^{D}_{1} \approx {f}^{D}_{2} \approx {f}^{D}_{3}$. And we use $\theta^X$ and $d^X$ to denote the AoA and human reflection distance for simplicity.

In this case, the dynamic human component $H^X_{i,j,k}$ can be revised based on the three key signal parameters (DFS $f^D$, AoA $\theta^X$ and human reflection distance $d^X$), and we have
\begin{equation}
\resizebox{\linewidth}{!}{$
\begin{aligned}
H^X_{i,j,k}={\rho _{i,j}^{X}}{e}^{-\bm{J}2\pi\frac{f_j}{c}\left[ d^X+ c\frac{f^D}{f_c}\left( k-1 \right)\Delta t+\left( i-1 \right)\Delta d \sin \theta^X \right]}+\mathbb{N}^X_{i,j,k}.
\end{aligned}
$}
\label{equation11}
\end{equation}

\subsection{Challenges}
Our CSI model reveals that, to be able to track a single moving object, we need to estimate $\{f^D, \theta^X, d^X\}$ in the presence of multiple static multipath signals, the time-varying phase shifts $\varphi_{j,k}^{TO}$ and $\psi_{j,k}^{FO}$ in $H^e_{j,k}$ and the WiFi hardware-dependent phase shift $\varphi_{i}$. Specifically, the time-varying phase shifts are major hurdles for jointly exploiting CSIs across time in tracking, and they will also cause ambiguity in parameter estimation if not being removed. The large number of multipath will cause inefficiency in parameter estimation, particularly in AoA estimation given the limited number of Rx antennas. These issues cause that conventional localization and tracking algorithms cannot be directly applied. To achieve fine-grained passive tracking, we firstly need to minimize the impact of transceiver asynchrony, which is a critical and challenging problem in WiFi tracking; we can then separate the dynamic human component $H^X_{i,j,k}$ from the CSI measurements and then estimate, ideally, only the parameters associated with the single dynamic path reflected from the object. In the following, we aim to achieve this goal by proposing a novel scheme.

\section{System Overview}
This work presents a scheme WiDFS that enables single-target real-time passive tracking using COTS WiFi devices. In WiDFS, a COTS Intel 5300 WiFi NIC that supports up to three antennas is used in a transmitter and receiver for CSI collection. The transmitter has one TX antenna while the receiver connects to a linear array consisting of three RX antennas. Such a three-antenna deployment is necessary due to the design of our algorithm in addressing the impact of transceiver asynchronization, as well as for estimating AoA. WiDFS firstly applies CSI cross-correlation between each pair of RX antennas to remove the time-varying phase shifts in CSIs. Then WiDFS obtains an unambiguous DFS estimate from the calculated cross-correlation terms in a short-time window. After that, WiDFS adopts CSI self-correlation of each RX antenna to acquire each antenna's CFR power and then separates the dynamic human component using a simple yet effective DFS-based separation algorithm. When WiDFS detects the presence of a moving person, it enables to estimate the AoA of the tracked person relative to the RX array and the human reflection distance from the person to the TX and RX antennas. Finally, WiDFS combines the estimated signal parameters to achieve real-time tracking. The workflow of WiDFS is illustrated in Fig. \ref{Fig2}. The processing in the main modules is summerized below and will be detailed later.

\textbf{1) Doppler Frequency Shift Estimation.} WiDFS firstly adopts cross-correlation between CSIs of any two antennas to remove time-varying phase shifts. The corresponding cross-correlation terms are $\mathit{CSI}_{12}$, $\mathit{CSI}_{23}$, and $\mathit{CSI}_{31}$, respectively. In a short-time window containing 100 CSI samples (about 0.1 s due to the sampling frequency of 1 kHz), WiDFS cleans high-frequency noise via Savitzky-Golay and lowpass filters. The DFS estimator outputs an unambiguous DFS estimate based on these filtered CSI cross-correlation terms. This part is described in detail in Section 4.

\textbf{2) Dynamic Human Component Separation.} In this part, WiDFS adopts self-correlation to calculate the CFR power of each antenna and obtain $\mathit{CSI}_{11}$, $\mathit{CSI}_{22}$ and $\mathit{CSI}_{33}$, respectively. After cleaning the CFR power terms via a lowpass/bandpass filter, WiDFS combines the refined dynamic CFR powers with the estimated DFS to separate the dynamic human component.The details will be provided in Section 5.

\textbf{3) Moving Person Detection and Tracking.} Since a human body does not act as a perfect reflector, the RX array may just capture a few signal reflections. In this part, WiDFS combines the estimated dynamic human components over multiple CSI sampling windows. A DFS-based motion detector is designed to determine the absence and presence of a moving person. When the human movement is present, WiDFS separately performs AoA and distance estimation and then uses a Kalman filter to refine each parameter. Finally, the localizer uses optimized  parameters to locate the person being tracked. Section 5 will present the details.

\textbf{4) WiFi Hardware Diversity Calibration.} In addition, we propose a one-time WiFi hardware calibration algorithm in Section 6 to calibrate and compensate for the difference in hardware-related phase shifts between RX antennas. At the same time, the algorithm can also calibrate the actual spacing between two adjacent RX antennas, which may be a little different from our manually measured value in our 
\section{Estimating Doppler Frequency Shift}
To achieve passive human tracking, three key signal parameters, i.e., DFS $f^D$, AoA $\theta^X$, and human reflection distance $d^X$, will be estimated based on our CSI model. This section  mainly introduces how to apply CSI cross-correlation to estimate the DFS $f^D$ in each sampling window.

\subsection{Random phase shift removal via cross-correlation}
Recall that ${\mathit{CSI}_{i,j,k}}$ is the reported CSI by a COTS WiFi system at $i$-th antenna, $j$-th subcarrier and $k$-th CSI sampling time. The time-varying phase shifts $\varphi_{j,k}^{FO}$ and $\varphi_{j,k}^{TO}$ in CSI across packets are unknown. To remove them, WiDFS adopts CSI cross-correlation between RX antennas by multiplying a CSI for a RX antenna (e.g., ${\mathit{CSI}_{1,j,k}}$) by the conjugate of a CSI for another antenna (e.g., ${\mathit{CSI}_{2,j,k}}$) at the same subcarrier. Different to previous works \cite{kotaru2015spotfi, qian2017inferring}, we do not apply the approach of adding and subtracting constants. Although this approach aims to suppress the imaging components in the cross-correlation output, it is not always effective and can introduce more interfering terms.  

Here let us take Antenna 1 and Antenna 2 for example. The CSI cross-correlation between them is
\begin{equation}
\resizebox{\linewidth}{!}{$
\begin{aligned}
& {\mathit{CSI}_{1,j,k}}{{\overline{\mathit{CSI}}}_{2,j,k}}\\
&= \left( H_{j,k}^{e}H_{1}^{h} \right)\left( H_{1,j}^{S}+H_{1,j,k}^{X} \right)\left( \overline{H}_{j,k}^{e}\overline{H}_{i=2}^{h} \right)\left( \overline{H}_{2,j}^{S}+\overline{H}_{2,j,k}^{X} \right) \\ 
&= \underbrace{{{\left\| H_{j,k}^{e} \right\|}^{2}}H_{1}^{h}\overline{H}_{2}^{h}H_{1,j}^{S}\overline{H}_{2,j}^{S}}_{{Static \ Cross-correlation \ Term}}+ \\ 
 & \quad\underbrace{{{\left\| H_{j,k}^{e} \right\|}^{2}}H_{1}^{h}\overline{H}_{2}^{h}\left( H_{1,j}^{S}\overline{H}_{2,j,k}^{X}+H_{1,j,k}^{X}\overline{H}_{2,j}^{S}+H_{1,j,k}^{X}\overline{H}_{2,j,k}^{X} \right)}_{{Dynamic \ Cross-correlation \ Term}}, \\
\end{aligned}
$}
\label{equation12}
\end{equation}
where the function ${{\left\| \cdot \right\|}}$ denotes the operator of computing the amplitude and $\left\| H_{j,k}^{e} \right\|={\rho^{agc}}$. Likewise, we also obtain the cross-correlation terms, i.e., ${\mathit{CSI}_{2,j,k}}{{\overline{\mathit{CSI}}}_{3,j,k}}$ and ${\mathit{CSI}_{3,j,k}}{{\overline{\mathit{CSI}}}_{1,j,k}}$, from other RX antenna pairs. 

After the cross-correlation operation, we can see that the phase shifts  $\varphi_{j,k}^{FO}$ and $\varphi_{j,k}^{TO}$ in $H_{j,k}^{e}$ are eliminated. Recall that a short-time sampling window of about 0.1s is defined. WiDFS collects \textit{3 RX antennas $\times$ 30 subcarriers $\times$ $N_p$ CSI samples} in this window, where $N_p=100$. Then all CSI cross-correlation terms between each pair of RX antennas free from the impact of time-varying phase shifts are
\begin{equation}
\left\{ 
\begin{aligned}
  & {CSI_{12}}=\{ {\mathit{CSI}_{1,j,1}}{{\overline{\mathit{CSI}}}_{2,j,1}},...,{\mathit{CSI}_{1,j,N_p}}{{\overline{\mathit{CSI}}}_{2,j,N_p}} \}\\ 
  & {CSI_{23}}=\{ {\mathit{CSI}_{2,j,1}}{{\overline{\mathit{CSI}}}_{3,j,1}},...,{\mathit{CSI}_{2,j,N_p}}{{\overline{\mathit{CSI}}}_{3,j,N_p}} \}\\ 
  & {CSI_{31}}=\{ {\mathit{CSI}_{3,j,1}}{{\overline{\mathit{CSI}}}_{1,j,1}},...,{\mathit{CSI}_{3,j,N_p}}{{\overline{\mathit{CSI}}}_{1,j,N_p}} \}\\ 
\end{aligned} 
\right..
\label{equation13}
\end{equation}

\subsection{High-frequency noise removal}
We then input the CSI cross-correlation terms ${CSI_{12}}$, ${CSI_{23}}$ and ${CSI_{31}}$ into a Savitzky-Golay (SG) smoothing filter followed by a lowpass filter, which aims to remove high-frequency noise from these terms. 

According to \cite{qian2017inferring}, the high-frequency noise comes from WiFi NICs and varies much faster than the dynamic human component of interest. In the SG filter, the polynomial order and frame length are set to 3 and 5, respectively. And we assume the maximum motion speed is $3.5$ m/s in indoor scenarios, so the passband of the lowpass filter is set to $f_{pass}=\frac{3.5f_{c}}{c}\approx60$ Hz, where the  center carrier frequency is $f_{c}=5.32$ GHz in our WiFi system. Here let ${CSI^{'}_{12}}$, ${CSI^{'}_{23}}$ and ${CSI^{'}_{31}}$ be the filtered cross-correlation terms.

\subsection{Doppler frequency shift estimation} 
In the following, we use the filtered cross-correlation terms to estimate a DFS $f^D$ in a short-time sampling window.

\textbf{Static and Dynamic Cross-correlation Term Separation.} The static component ${H}_{i,j}^{S}$ is a constant in the window and its power is much higher than that of the dynamic human component ${H}_{i,j,k}^{X}$, so the static cross-correlation term (denoted as ${\bm{U}_{12,j}}$) can be separated by calculating the mean value of the cross-correlation terms. The remaining dynamic cross-correlation term (denoted as ${\bm{V}_{12,j,k}}$) at each CSI sample is obtained by subtracting the mean value,
\begin{equation}
\resizebox{\linewidth}{!}{$
\left\{ 
\begin{aligned}
  & {{\left\| H_{j,k}^{e} \right\|}^{2}}H_{1}^{h}\overline{H}_{2}^{h}H_{1,j}^{S}\overline{H}_{2,j}^{S}={\bm{U}_{12,j}} \\ 
 & {{\left\| H_{j,k}^{e} \right\|}^{2}}H_{1}^{h}\overline{H}_{2}^{h}\left( H_{1,j}^{S}\overline{H}_{2,j,k}^{X}+H_{1,j,k}^{X}\overline{H}_{2,j}^{S}+H_{1,j,k}^{X}\overline{H}_{2,j,k}^{X} \right)={\bm{V}_{12,j,k}}\\ 
\end{aligned} 
\right.,
$}
\label{equation14}
\end{equation}
where 
\begin{equation}
\left\{ 
\begin{aligned}
  & {\bm{U}_{12,j}} =  \frac{1}{N_p}\sum\limits_{k=1}^{N_p}{\mathit{CSI}^{'}_{1,j,k}}{{\overline{\mathit{CSI}}}^{'}_{2,j,k}}\\ 
 & {\bm{V}_{12,j,k}}={CSI}^{'}_{1,j,k}{{\overline{CSI}}^{'}_{2,j,k}}-{\bm{U}_{12,j}} \\ 
\end{aligned} 
\right..
\label{equation15}
\end{equation}
Since the amplitude of ${H}_{i,j}^{S}$ may be far larger than that of ${H}_{i,j,k}^{X}$, we ignore the product term $H_{1,j,k}^{X}\overline{H}_{2,j,k}^{X}$ in Eq. \ref{equation14}.
\begin{figure}
       \centering
        \includegraphics[width=0.5\textwidth]{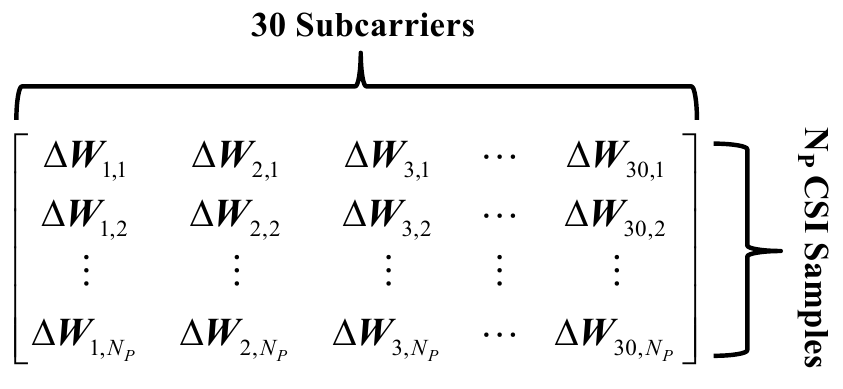}
    \caption{Observation matrix for unambiguous DFS estimation}
    \label{Fig3}
    \vspace{-1.5em}
\end{figure}

\textbf{DFS Estimation using Subspace-based Method.} From Eq. \ref{equation14}, we can see that the dynamic cross-correlation term of interest (e.g., $H_{1,j,k}^{X}\overline{H}_{2,j}^{S} $) is mixed with a side product (e.g., $\overline{H}_{2,j,k}^{X}H_{1,j}^{S}$), so the estimated DFS $f^D$ would be an actual one or its negative. This phenomenon is called the \textit{DFS ambiguity}. To tackle this issue, many exisiting works \cite{li2017indotrack, qian2018widar2, jiang2020towards} adjust the CSI amplitude of each antenna by adding or subtracting a real value. However, this intuitive method seems to be signal-dependent and is not always effective in multipath-rich scenarios. In this work, WiDFS achieves unambiguous DFS estimation as follows:

First, we divide the equations in Eq. \ref{equation14} to remove the unknown term ${{\left\| H_{j,k}^{e} \right\|}^{2}}H_{1,j}^{h}\overline{H}_{2,j}^{h}$,
\begin{equation}
\frac{H_{1,j,k}^{X}}{H_{1,j}^{S}}+\frac{\overline{H}_{2,j,k}^{X}}{\overline{H}_{2,j}^{S}}=\frac{{\bm{V}_{12,j,k}}}{{\bm{U}_{12,j}}}.
\label{equation16}
\end{equation}
Likewise, we can obtain
\begin{subequations}
\begin{empheq}[left=\empheqlbrace]{align}
& \frac{H_{2,j,k}^{X}}{H_{2,j}^{S}}+\frac{\overline{H}_{3,j,k}^{X}}{\overline{H}_{3,j}^{S}}=\frac{{\bm{V}_{23,j,k}}}{{\bm{U}_{23,j}}},  \label{equation17a}\\
& \frac{H_{3,j,k}^{X}}{H_{3,j}^{S}}+\frac{\overline{H}_{1,j,k}^{X}}{\overline{H}_{1,j}^{S}}=\frac{{\bm{V}_{31,j,k}}}{{\bm{U}_{31,j}}}.  \label{equation17b}
\end{empheq}
\end{subequations}

By subtracting Eq. \ref{equation17a} from the conjugate of Eq. \ref{equation17b} and subtracting Eq. \ref{equation17b} from the conjugate of Eq. \ref{equation16}, we have
\begin{equation}
\left\{ 
\begin{aligned}
& \frac{H_{1,j,k}^{X}}{H_{1,j}^{S}}-\frac{H_{2,j,k}^{X}}{H_{2,j}^{S}}=\frac{{\bm{\overline{V}}_{31,j,k}}}{{\bm{\overline{U}}_{31,j}}}-\frac{{\bm{V}_{23,j,k}}}{{\bm{U}_{23,j}}}={\Delta \bm{W}_{12,j,k}} \\ 
& \frac{H_{2,j,k}^{X}}{H_{2,j}^{S}}-\frac{H_{3,j,k}^{X}}{H_{3,j}^{S}}=\frac{{\bm{\overline{V}}_{12,j,k}}}{{\bm{\overline{U}}_{12,j}}}-\frac{{\bm{V}_{31,j,k}}}{{\bm{U}_{31,j}}}={\Delta \bm{W}_{23,j,k}} \\ 
\end{aligned} 
\right..
\label{equation18}
\end{equation}

Then we compute the difference of the above equations,
\begin{equation}
\begin{aligned}
\Delta{\bm{W}_{j,k}}&={\Delta \bm{W}_{12,j,k}}-{\Delta \bm{W}_{23,j,k}}\\ & =\frac{H_{1,j,k}^{X}}{H_{1,j}^{S}}-2\frac{H_{2,j,k}^{X}}{H_{2,j}^{S}}+\frac{H_{3,j,k}^{X}}{H_{3,j}^{S}} \\ 
 & =\frac{H_{2,j,k}^{X}}{H_{2,j}^{S}}\left( \frac{H_{1,j,k}^{X}}{H_{2,j,k}^{X}}\frac{H_{2,j}^{S}}{H_{1,j}^{S}}+\frac{H_{3,j,k}^{X}}{H_{2,j,k}^{X}}\frac{H_{2,j}^{S}}{H_{3,j}^{S}}-2 \right) \\ 
\end{aligned}.
\label{equation19}
\end{equation}
 
In free space without multipath interference, we can apply far-field and narrowband assumptions to obtain
\begin{equation}
\left\{
\begin{aligned}
& \left\| H_{1,j,k}^{X} \right\| \approx \left\| H_{2,j,k}^{X} \right\| \approx \left\| H_{3,j,k}^{X} \right\| \\
& \left\| H_{1,j}^{S} \right\| \approx \left\| H_{2,j}^{S} \right\|\approx \left\| H_{3,j}^{S} \right\| \\
& \angle \frac{H_{1,j,k}^{X}}{H_{2,j,k}^{X}}\approx \angle \frac{H_{2,j,k}^{X}}{H_{3,j,k}^{X}}\\
& \angle \frac{H_{1,j}^{S}}{H_{2,j}^{S}} \approx \angle \frac{H_{2,j}^{S}}{H_{3,j}^{S}}
\end{aligned} 
\right.,
\label{equation20}
\end{equation}
where the function $\angle \cdot$ denotes the operator of computing the phase shift of a complex number. Thus, we have
\begin{equation}
\bm {\Gamma}_{j,k}=\frac{H_{1,j,k}^{X}}{H_{2,j,k}^{X}}\frac{H_{2,j}^{S}}{H_{1,j}^{S}}+\frac{H_{3,j,k}^{X}}{H_{2,j,k}^{X}}\frac{H_{2,j}^{S}}{H_{3,j}^{S}} \approx 0.
\label{equation21}
\end{equation}
Unfortunately, there exist multiple signal propagation paths in an actual environment. However, for the following two reasons, we can assume that $\bm {\Gamma}_{j,k}$ keeps unchanged, i.e., $\bm {\Gamma}_{j,k}=\bm {\Gamma}_{j} $. \textit{(1)} $H_{i,j}^{S}$ is static environment-related and remain unchanged, then $ {H_{1,j}^{S}}/{H_{2,j}^{S}}$ and $ {H_{2,j}^{S}}/{H_{3,j}^{S}}$ are approximately constant. \textit{(2)} Recall that we ignore the change in the amplitude of $H_{i,j,k}^{X}$ in the window. The values of ${H_{1,j,k}^{X}}/{H_{2,j,k}^{X}}$ and ${H_{2,j,k}^{X}}/{H_{3,j,k}^{X}}$ also do not vary over time. Thus, $\Delta{\bm{W}_{j,k}}$ can be rewritten as
\begin{equation}
\resizebox{\linewidth}{!}{$
{\Delta \bm{W}_{j,k}}\approx\frac{\bm {\Gamma}_{j} -2}{H_{2,j}^{S}} [{\rho _{2,j}^{X}{{e}^{-\bm{J}\lbrack 2\pi\frac{f_j}{c}d_{2}^X+2\pi f^{D}\left( k-1 \right)\Delta t \rbrack}}} +\mathbb{N}^X_{2,j,k}].
$}
\label{equation22}
\end{equation}

\begin{figure}
     \centering
         \includegraphics[width=0.45\textwidth]{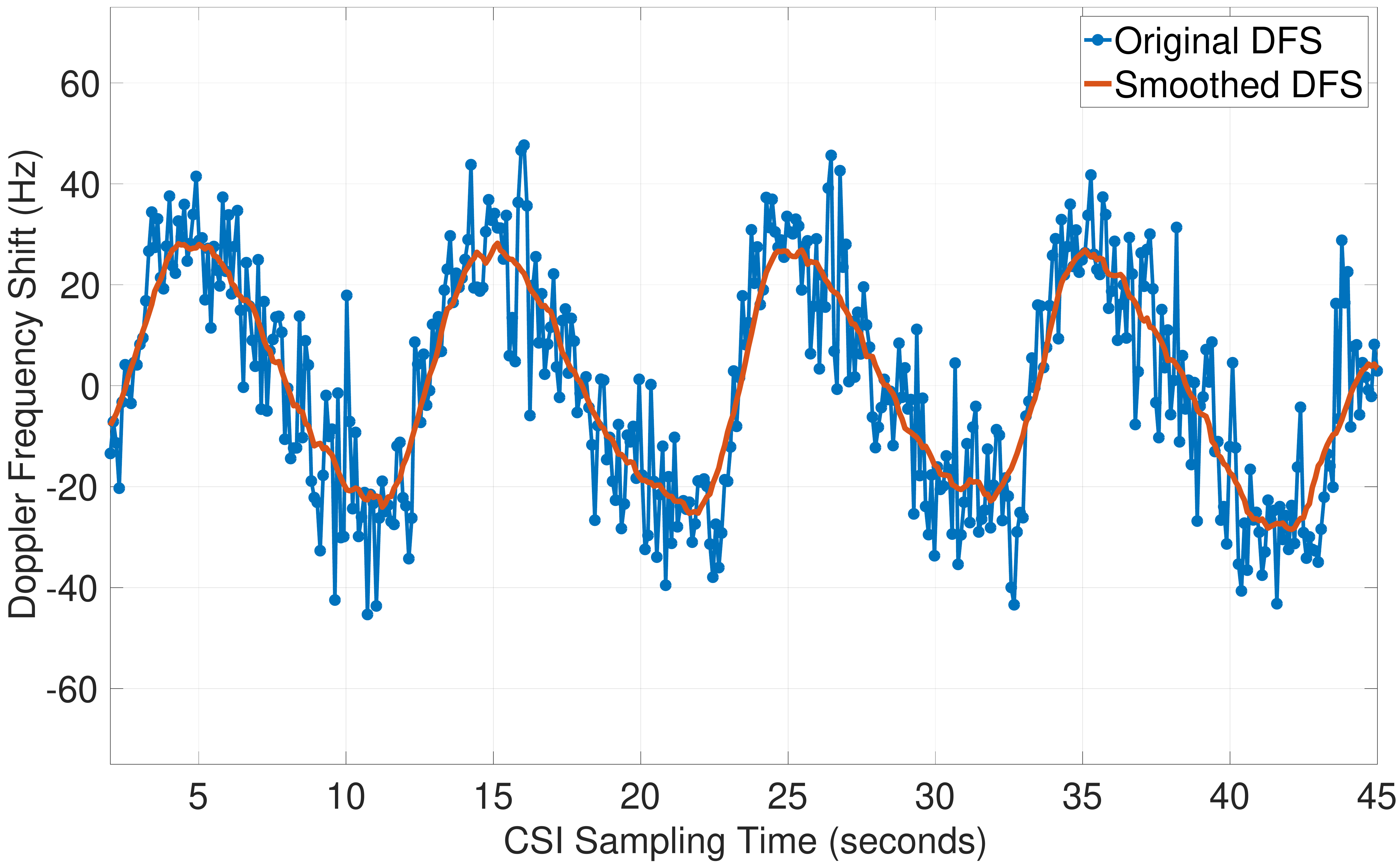}
         \caption{Measured DFS when a person walks four times along an approximate elliptical trajectory}
         \label{Fig4}
     \vspace{-1.5em}
\end{figure}

The Fast Fourier Transform (FFT) is commonly used to identify signal frequencies. However, given the small sample size (i.e., 100 CSI samples), FFT may not provide enough high estimation accuracy. Instead, WiDFS uses root multiple signal classification (Root-MUSIC) algorithm \cite{rao1989performance} to estimate $f^{D}$. It outputs frequency estimates along with the corresponding signal power estimates. As shown in Fig. \ref{Fig3}, we build an observation matrix using ${\Delta \bm{W}_{j,k}}$ in Eq. \ref{equation22} at all subcarriers. Each column represents a separate observation. The number of snapshots is 30. To determine the subspace dimension, we calculate the maximum eigenvalue $eig_{max}$ of the correlation matrix and then find the number of eigenvalues above an empirical threshold $0.6eig_{max}$. Finally, we select a frequency estimate associate with the maximum signal power as an optimal one. Fig.\ref{Fig4} shows the measured DFSs when a person walks four times along an elliptical trajectory. The results are consistent with human movement, and the smoothed results clearly reveal the DFS variation.

\section{Separating Dynamic Human Component}
This section uses the estimated DFS and CSI self-correlation to build the complex-valued dynamic human components.

\subsection{Random phase shift removal via self-correlation}
First of all, the time-varying phase shifts $\varphi_{j,k}^{FO}$ and $\varphi_{j,k}^{TO}$ in $H_{j,k}^{e}$ are removed by applying a self-correlation operation in WiDFS, that is, we multiply ${\mathit{CSI}_{i,j,k}}$ by its complex conjugate ${{\overline{\mathit{CSI}}}_{i,j,k}}$ and obtain $\left\| {\mathit{CSI}_{i,j,k}}\right\|^2$. The product is also called the CFR power, which is
\begin{equation}
\resizebox{\linewidth}{!}{$
\begin{aligned}
\left\| {\mathit{CSI}_{i,j,k}}\right\|^2& =\left( H_{j,k}^{e}H_{i}^{h} \right)\left( H_{i,j}^{S}+H_{i,j,k}^{X} \right)\left( \overline{H}_{j,k}^{e}\overline{H}_{i}^{h} \right)\left( \overline{H}_{i,j}^{S}+\overline{H}_{i,j,k}^{X} \right) \\ 
& =\underbrace{{{\left\| H_{j,k}^{e}H_{i}^{h} H_{i,j}^{S} \right\|}^{2}}}_{\bm{Staic \ Power\ Term}}+\\
 & \quad\underbrace{{{\left\| H_{j,k}^{e}H_{i}^{h} \right\|}^{2}}\left[ 2\left\| H_{i,j}^{S}{H}_{i,j,k}^{X} \right\|\cos \angle \left( \overline H_{i,j}^{S} {H}_{i,j,k}^{X} \right)+{{\left\| H_{i,j,k}^{X} \right\|}^{2}} \right]}_{\bm{Dynamic \ Power\ Term}}, \\
\end{aligned}
$}
\label{equation23}
\end{equation}

In the pre-defined sampling window with 100 CSI samples, the calculated self-correlation terms at each RX antenna ($i=1,2,3$) and subcarrier ($j=1,2,..,30$) are denoted as ${CSI_{ii}}=\lbrace \left\| {\mathit{CSI}_{i,j,1}}\right\|^2,...,\left\| {\mathit{CSI}_{i,j,N_p}}\right\|^2 \rbrace$.

\subsection{Dynamic power separation and refinement}
Next, WiDFS uses the calculated ${CSI_{11}}$, ${CSI_{22}}$ and ${CSI_{33}}$ to separate dynamic power terms in the window and then refines them using the estimated DFS $f^D$.

\textbf{Separation of Static and Dynamic Power Terms.} We follow the same operation in Eq. \ref{equation15} to calculate the static and dynamic power terms (denoted as ${\bm{u}_{i,j}}$ and ${\bm{v}_{i,j,k}}$, respectively)
\begin{equation}
\resizebox{\linewidth}{!}{$
\left\{
\begin{aligned}
& {{\left\| H_{j,k}^{e}H_{i}^{h}H_{i,j}^{S} \right\|}^{2}}={\bm{u}_{i,j}} \\ 
& {{{\left\| H_{j,k}^{e}H_{i}^{h} \right\|}^{2}}\left[ 2\left\| H_{i,j}^{S}{H}_{i,j,k}^{X} \right\|\cos \angle \left( \overline H_{i,j}^{S} {H}_{i,j,k}^{X} \right)+{{\left\| H_{i,j,k}^{X} \right\|}^{2}} \right]}\approx{\bm{v}_{i,j,k}} \\ 
\end{aligned}
\right.,
$}
\label{equation24}
\end{equation}
where 
\begin{equation}
\left\{
\begin{aligned}
  & {\bm{u}_{i,j}} =  \frac{1}{N_p}\sum\limits_{k=1}^{N_p}{{\left\| {{{\mathit{CSI}}}_{i,j,k}} \right\|}^{2}}\\ 
 & {\bm{v}_{i,j,k}}={{\left\| {{{\mathit{CSI}}}_{i,j,k}} \right\|}^{2}}-{\bm{u}_{i,j}} \\ 
\end{aligned} .
\right.
\label{equation25}
\end{equation}
The term ${{\left\| H_{i,j,k}^{X} \right\|}^{2}}$ can be ignored since it is much smaller than other terms. 

\textbf{Filtering of Dynamic Power Terms of Interest.} Since the dynamic power terms ${\bm{v}_{i,j,k}}$ of interest may be around the estimated DFS $f^D$, we refine ${{\bm{v}_{i,j,k}}}$ using a lowpass filter with the passband frequency of $\left| f^D\right|+\Delta f$, where the empirical frequency $\Delta f$ is set to 10 Hz in our scheme. And if $\left| f^D \right|>15$ Hz, we further apply a highpass filter with the passband frequency of $\left|f^D \right|-\Delta f$. The refined dynamic power terms at each RX antenna and subcarrier in the window are denoted as ${{{v}^{'}_{ii}}}=\{ {{\bm{v}^{'}_{i,j,1}}},...,{{\bm{v}^{'}_{i,j,N_p}}} \}$. \textit{For single-object tracking, it is reasonable to assume that there is only a dominant human reflection in each ${{\bm{v}^{'}_{i,j,k}}}$.}

\begin{figure*}
     \centering
         \includegraphics[width=0.85\textwidth]{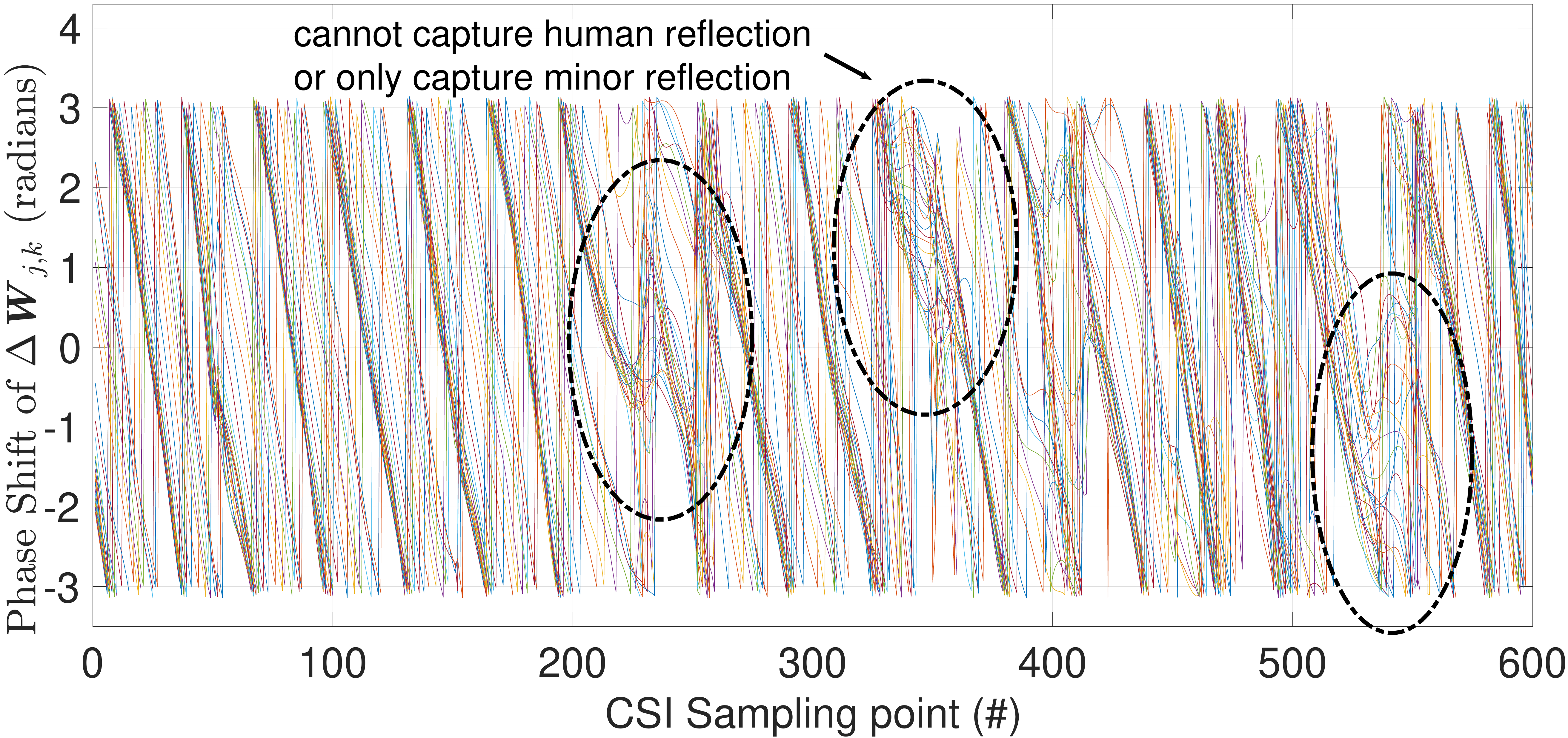}
         \caption{Signal reflections off human body. Recall from Eq. \ref{equation19}, $\Delta \bm{W}_{j,k}$ represents the dynamic component caused by human motion. Since human body has a complex surface, the transmit signals may not always be reflected to RX antennas.}
         \label{Fig5}
     \vspace{-1.5em}
\end{figure*}

\subsection{Dynamic human component reconstruction}
Here WiDFS depends on the filtered ${v}^{'}_{ii}$ and the estimated DFS $f^D$ to reconstruct the complex-valued dynamic human component at each RX antenna and subcarrier.

\textbf{Static-Dynamic Associated Component Estimation.} At first, we remove the unknown term ${{\left\|H_{j,k}^{e}H_{i}^{h} \right\|}}$ by
\begin{equation}
\frac{{\bm{v}^{'}_{i,j,k}}}{{\bm{u}_{i,j}}}={2{\left\|  \frac{{H}_{i,j,k}^{X}}{H_{i,j}^{S}} \right\|}} \cos \angle \left( \overline{H}_{i,j}^{S}{H}_{i,j,k}^{X} \right).
\label{equation26}
\end{equation}

According to our CSI model described in Section 2.3, let $\varphi_{i,j}^{X}=2 \pi \frac{f_j}{c}d_{i}^X $ be the phase shift of the $i$-th antenna at the $j$-th subcarrier, then the above equation can be rewritten as
\begin{equation}
\resizebox{\linewidth}{!}{$
\begin{aligned}
\frac{{\bm{v}^{'}_{i,j,k}}}{{\bm{u}_{i,j}}}
&=2{{\left\| \frac{{H}_{i,j,k}^{X}}{H_{i,j}^{S}} \right\|}}\cos \left[ \angle \overline H_{i,j}^{S}-\varphi_{i,j}^{X} -2\pi {{f}^{D}}\left( k-1 \right)\Delta t \right] \\ 
 & ={{x}_{i,j}}\cos \left[ 2\pi {{f}^{D}}\left( k-1 \right) \Delta t\right]+{{y}_{i,j}}\sin \left[ 2\pi {{f}^{D}}\left( k-1 \right) \Delta t\right], \\ 
\end{aligned} 
$}
\label{equation27}
\end{equation}
where
\begin{equation}
\left\{ 
\begin{aligned}
& {{x}_{i,j}}=2{{\left\| \frac{{H}_{i,j,k}^{X}}{H_{i,j}^{S}} \right\|}} \cos \left(\angle \overline H_{i,j}^{S}-\varphi_{i,j}^{X} \right) \\ 
& {{y}_{i,j}}=2{{\left\| \frac{{H}_{i,j,k}^{X}}{H_{i,j}^{S}} \right\|}}\sin\left(\angle \overline H_{i,j}^{S}-\varphi_{i,j}^{X} \right) \\ 
\end{aligned} 
\right..
\label{equation28}
\end{equation}

Given $N_p=100$ CSI samples at a RX antenna and subcarrier in the window, we write Eq. \ref{equation27} in a matrix form,
\begin{equation}
\resizebox{\linewidth}{!}{$
\left[ 
\begin{matrix}
1 & 0  \\
\cos \left( 2\pi {{f}^{D}}\Delta t \right) & \sin \left( 2\pi {{f}^{D}}\Delta t \right)  \\
\vdots  & \vdots   \\
\cos \left[ 2\pi {{f}^{D}}\left( N_p-1 \right)\Delta t \right] & \sin \left[ 2\pi {{f}^{D}}\left( N_p-1 \right)\Delta t \right]  \\
\end{matrix} \right]\left[ 
\begin{matrix}
   {{x}_{i,j}} \\ {{y}_{i,j}}  \\
\end{matrix} \right]=\left[ 
\begin{matrix}
   {\frac{{\bm{v}^{'}_{i,j,1}}}{{\bm{u}_{i,j}}}}  \\
   {\frac{{\bm{v}^{'}_{i,j,2}}}{{\bm{u}_{i,j}}}}  \\
   \vdots   \\
   {\frac{{\bm{v}^{'}_{i,j,N_p}}}{{\bm{u}_{i,j}}}}  \\
\end{matrix} \right].
$}
\label{equation29}
\end{equation}
A least-squares method is applied to easily solve ${{x}_{i,j}}$ and ${{y}_{i,j}}$, so the static-dynamic associated component $\bm Z_{i,j}^{S,X}$ at the $i$-th antenna and the $j$-th subcarrier in the window is 
\begin{equation}
\begin{aligned}
\bm Z_{i,j}^{S,X} =w_{i,j} {e}^{\bm{J} \arctantwo\left({y}_{i,j},{x}_{i,j} \right)}=w_{i,j} {e}^{\bm{J} \left(\angle \overline H_{i,j}^{S} -2 \pi \frac{f_j}{c}d_{i}^X \right)},
\end{aligned} 
\label{equation30}
\end{equation}
where $w_{i,j}$ denotes the weight derived from the sum of squared residuals. When the residual is minimum, we normalize the weight to be maximum (or vice versa).

\textbf{Dynamic Human Component Separation.} Further, let $\Delta\varphi^{h}_{12}$ and $\Delta\varphi^{h}_{31}$ be the phase differences caused by WiFi hardware diversity, which can be pre-estimated using our proposed approach to be detailed in Section 7, 
\begin{equation}
\left\{ 
\begin{aligned}
 & \Delta\varphi^{h}_{12}={\angle \left(H_{1}^{h}\overline{H}_{2}^{h} \right)}\\
 & \Delta\varphi^{h}_{23}={\angle \left(H_{2}^{h}\overline{H}_{3}^{h} \right)}\\
 & \Delta\varphi^{h}_{31}={\angle \left(H_{3}^{h}\overline{H}_{1}^{h} \right)}\\
\end{aligned}. 
\right.
\label{equation31}
\end{equation}
We combine them with the calculated static cross-correlation terms $\bm{U}_{12,j}$ and $\bm{U}_{31,j}$ in Section 4 to conduct a transformation on ${{\bm{Z}_{2,j}^{S,X}}}$ and ${{\bm{Z}^{S,X}_{3,j}}}$,
\begin{equation}
\left\{
\begin{aligned}
 &{w_{2,j} {{e}^{\bm{J}
\left( \angle \overline H_{1,j}^{S}-\frac{2\pi {{f}_{j}}}{c}d^{X}_{2} \right) }} }={{{\bm{Z}^{S,X}_{2,j}}{{e}^{-\bm{J} \left( \angle{\bm{U}_{12,j}}-\Delta\varphi^{h}_{12} \right)   }}  }}\\
 & {w_{3,j}{{e}^{\bm{J}\left( \angle \overline H_{1,j}^{S}-\frac{2\pi {{f}_{j}}}{c}d^{X}_{3}\right)  }}}={{{\bm{Z}^{S,X}_{3,j}}{{e}^{\bm{J} \left( \angle{\bm{U}_{31,j}}-\Delta\varphi^{h}_{31}\right)  }}  }} \\
\end{aligned} 
\right.,
\label{equation32}
\end{equation}
where
\begin{equation}
\left\{ 
\begin{aligned}
 &\angle{\bm{U}_{12,j}}-\Delta\varphi^{h}_{12} =\angle H_{1,j}^{S}+\angle \overline H_{2,j}^{S} \\
 &\angle{\bm{U}_{31,j}}-\Delta\varphi^{h}_{31} = \angle H_{3,j}^{S}+\angle \overline H_{1,j}^{S} \\
\end{aligned} 
\right..
\label{equation33}
\end{equation}

Recall from Section 2 that $\Delta d $ is the spacing between two adjacent RX antennas and $\theta^X$ is the AoA of the tracked human
relative to the RX antenna array. Then we have $\Delta d \sin \theta^X \approx d^{X}_{2}-d^{X}_{1} \approx  d^{X}_{3}-d^{X}_{2}$. Thus, the dynamic human components ${\bm{Z}^X_{1,j}}$, ${\bm{Z}^X_{2,j}}$, and ${\bm{Z}^X_{3,j}}$ at the $j$-th subcarrier in the window are reconstructed as
\begin{equation}
\resizebox{\linewidth}{!}{$
\left\{ 
\begin{aligned}
 & {\bm{Z}_{1,j}^X}=\frac{{\bm{Z}_{1,j}^{S,X}}}{\eta_j}=w_{1,j} {{{e}^{-\bm{J}\frac{2\pi {{f}_{j}}}{c}d^{X} }}} \\
 & {\bm{Z}_{2,j}^X}=\frac{{{{\bm{Z}_{2,j}^{S,X}}  }}}{\eta_j}{{e}^{-\bm{J} \left( \angle{\bm{U}_{12,j}}-\Delta\varphi^{h}_{12} \right)   }}={ w_{2,j} {{e}^{-\bm{J}\frac{2\pi {{f}_{j}}}{c}\left(d^{X}+\Delta d \sin \theta^X \right) }} }\\
 & {\bm{Z}_{3,j}^X}=\frac{{{\bm{Z}_{3,j}^{S,X}}  }}{\eta_j}{{e}^{\bm{J} \left( \angle{\bm{U}_{31,j}}-\Delta\varphi^{h}_{31}\right)  }}=w_{3,j} {{{e}^{-\bm{J}\frac{2\pi {{f}_{j}}}{c}\left(d^{X}+2\Delta d \sin \theta^X \right) }}}\\
\end{aligned} 
\right.,
$}
\label{equation34}
\end{equation}
where ${\eta_j}={e}^{\bm{J} \angle \overline H_{1,j}^{S} }$, and $d_1^{X}$ is expressed as $d^{X}$ for simplicity. Since the distance $d^{S}_{1}$ between the transmitter and Antenna 1 can be measured in advance, we can get the approximation $\angle \overline H_{1,j}^{S} \approx\ mod\left( {2\pi \frac{f_j}{c}d^{S}_{1}},2\pi \right)$.
\begin{figure}
     \centering
         \includegraphics[width=0.5\textwidth]{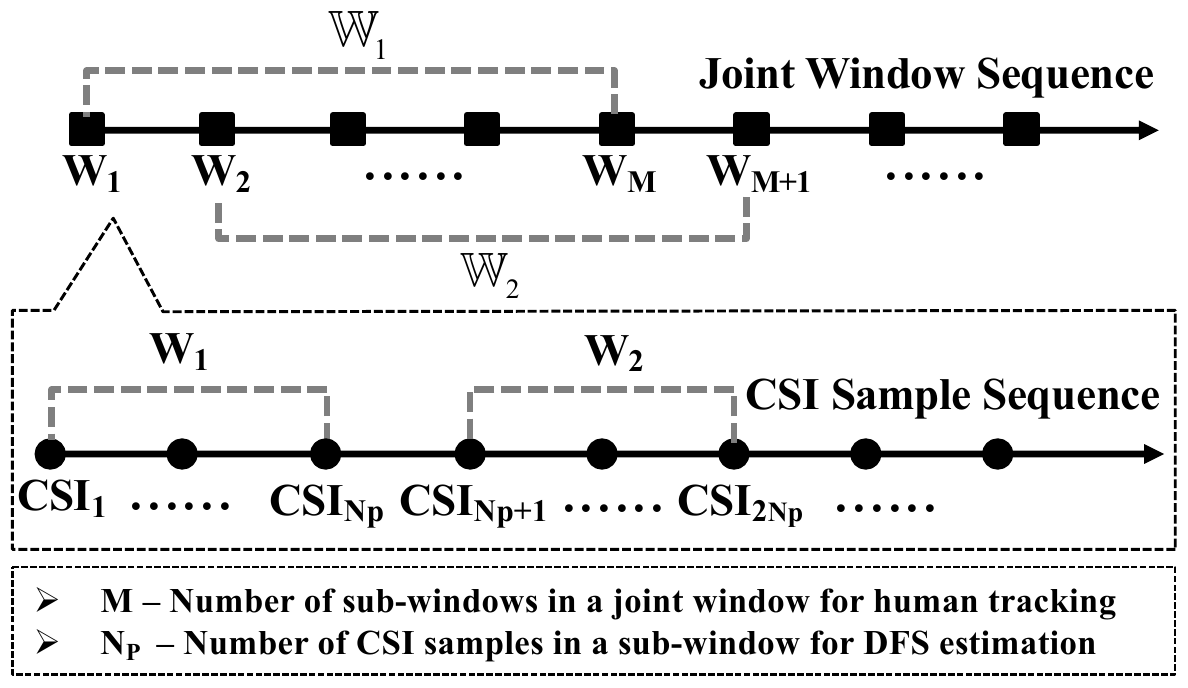}
         \caption{Combining multiple CSI sampling sub-windows for reliable motion detection and human tracking}
         \label{Fig6}
     \vspace{-1.5em}
\end{figure}

\section{Detecting and Tracking Moving Person}
This section describes how to leverage the estimated DFS $f^D$ to detect the presence of a moving person and introduces how to track a moving person using the calculated dynamic human components ${\bm{Z}^X_{i,j}}$ over multiple sampling windows.

\subsection{Capture of Human body reflection}
When the transmitter sends WiFi signals to space, the RX antenna array may not capture all reflection signals off human since signal propagation is complicated by reflections from the human body surface. To intuitively illustrate this phenomenon, we plot the phase change of $\Delta{\bm{W}_{j,k}}$ in Eq. \ref{equation19} at 30 subcarriers in Fig. \ref{Fig5}. We collect 6 seconds of CSI data in a dynamic scenario that a person moves at a speed of about 1 m/s away from the receiver. It shows the phase variation cannot match our expectation in some sampling time due to the fact that some reflections may be invisible to the receiver. 

To achieve robust passive human tracking, we combine the human reflections over multiple CSI sampling sub-windows $\bm{W}$ (as shown in Fig. \ref{Fig6}), where each sub-window has $N_P=100$ CSI samples as described before. And the window $\mathbb{W}$ consisting of $M$ sub-windows is called the \textit{joint window}. The two adjacent joint windows overlap with $\left(M-1\right)$ sub-windows. The average computation time should be less than 0.1 s to enable real-time tracking.
\begin{figure}
     \centering
         \includegraphics[width=0.5\textwidth]{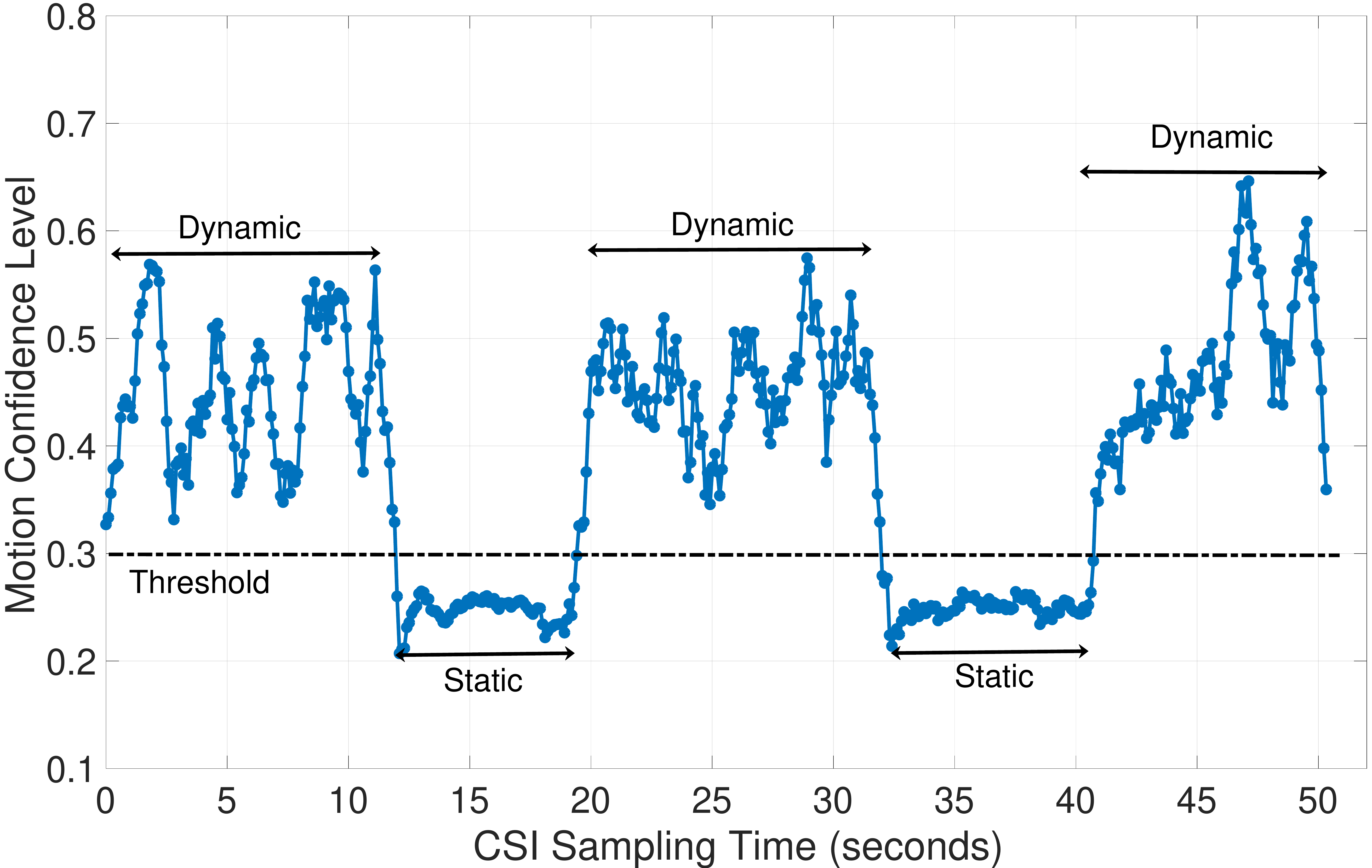}
         \caption{Detecting the absence or presence of human motion}
         \label{Fig7}
     \vspace{-1.5em}
\end{figure}
\begin{figure*}
\centering
\begin{minipage}[t]{0.329\linewidth}
\centering
	\includegraphics[width=\textwidth]{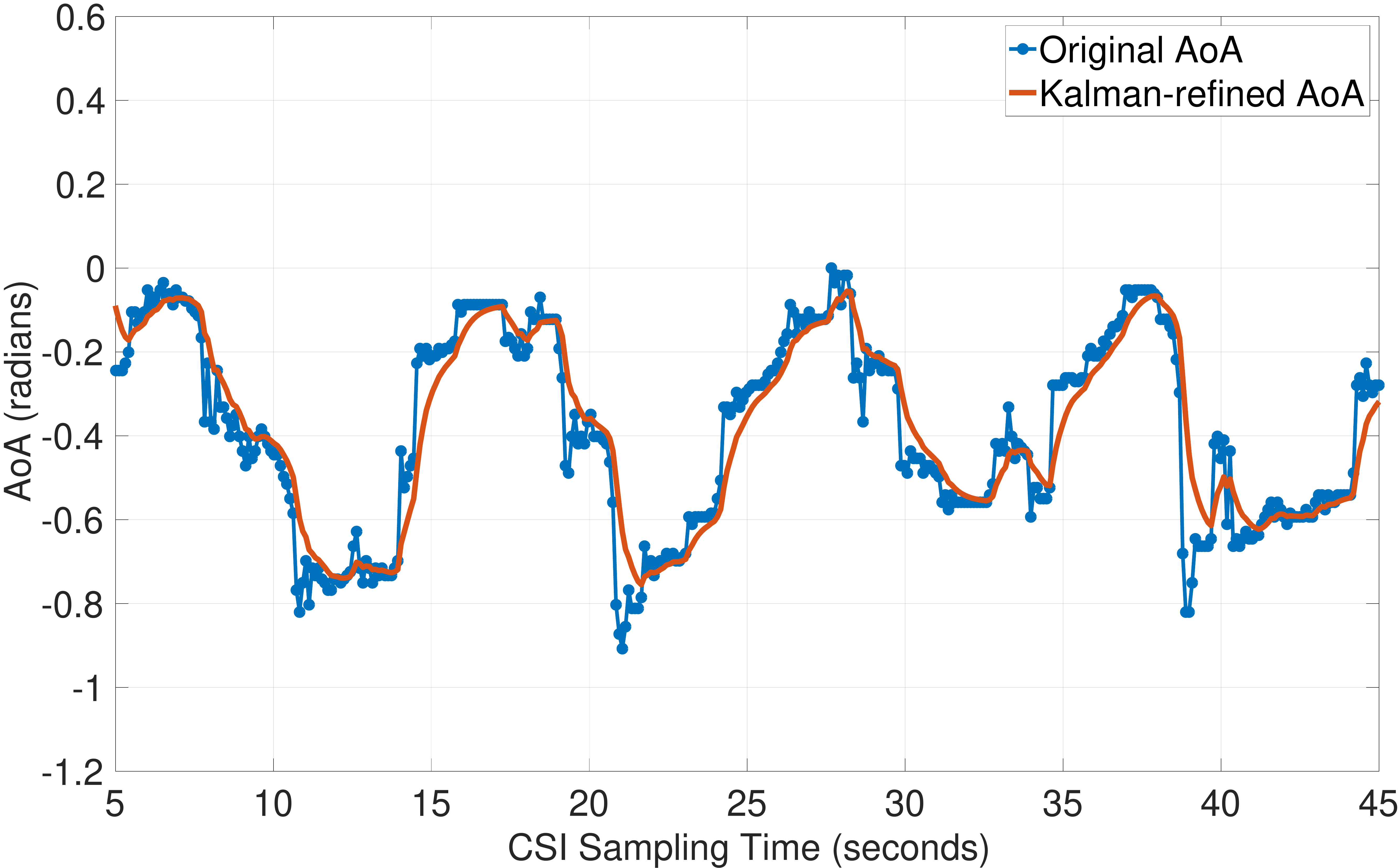}
	\caption{Human AoA relative to receiver}
	\label{Fig8}
\end{minipage}
\begin{minipage}[t]{0.329\linewidth}
\centering
	\includegraphics[width=\textwidth]{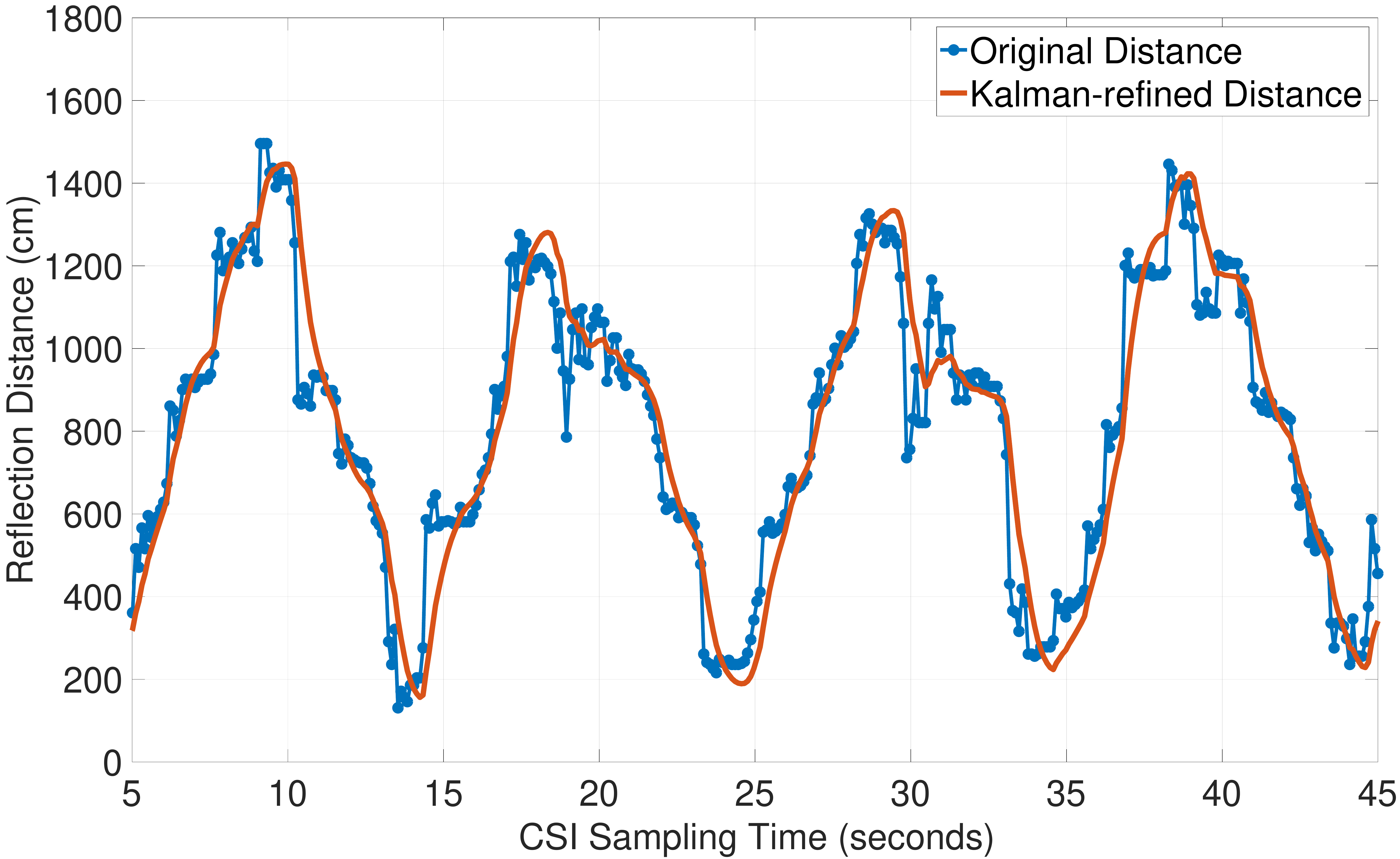}
	\caption{Human reflection distance}
	\label{Fig9}
\end{minipage}
\begin{minipage}[t]{0.329\linewidth}
\centering
	\includegraphics[width=\textwidth]{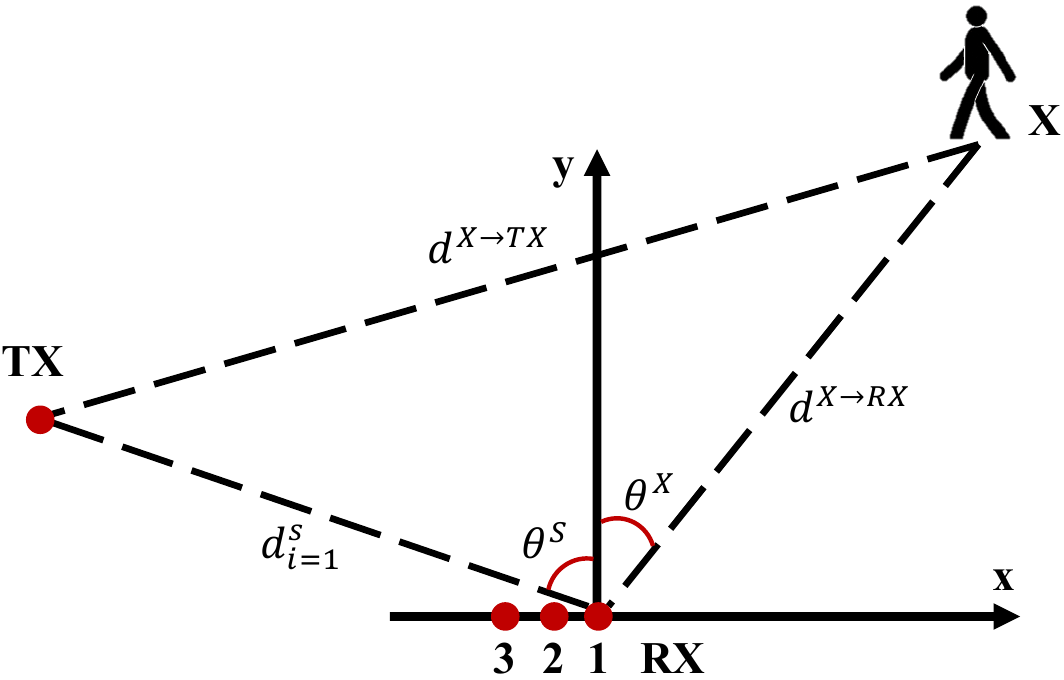}
	\caption{Human localization}
	\label{Fig10}
\end{minipage}
	\vspace{-1.5em}
\end{figure*}
\subsection{Dynamic environment sensing}
WiDFS performs passive human tracking only when a moving person is present. It uses the estimated ${f}^{D}$ and ${\Delta \bm{W}_{j,k}}$ in Eq. \ref{equation19} in Section 4 to determine the absence or presence of a moving person in the region of interest by
\begin{equation}
\resizebox{\linewidth}{!}{$
\mathbb{P}=\frac{1}{30\times MN_p}\sum\limits_{l=1}^{M}\sum\limits_{j=1}^{30}{\left\| \sum\limits_{k=1}^{N_p}{{{e}^{\bm{J} \angle{\Delta \bm{W}_{j,k}}\lbrack l \rbrack }}{{e}^{\bm{J}2\pi {{f}^{D}\lbrack l \rbrack}\left( k-1 \right)\Delta t}}} \right\|},
$}
\label{equation35}
\end{equation}
where ${\Delta \bm{W}_{j,k}} \lbrack l \rbrack$ and ${{f}^{D}\lbrack l \rbrack}$ represent the corresponding estimates in the $l$-th sub-window. The range of $\mathbb{P}$ is within $\left(0,1\right)$. In a static scenario, the extracted $\Delta \bm{W}_{j,k}$ is a noise term, so $\mathbb{P}$ maintains in a low level. However, when a person moves in the region, $\mathbb{P}$ will be close to 1. Our judging criteria is that if $\mathbb{P}$ is lower (or higher) than a pre-defined threshold, it indicates the absence (or presence) of a moving person.

In Fig. \ref{Fig7}, we use an example to show the result of dynamic environment sensing. A person moves randomly and then sits on a chair for a while. The volunteer repeats the two motions. We find that the sensing function could effectively indicate the human motion status via a threshold. In Section 9.4, a comprehensive experiment will be conducted to determine this threshold in different scenarios.

\subsection{Localization parameter estimation}
WiDFS performs tracking using two fundamental parameters. One is the AoA $\theta^X$, the direction that the human reflections are received from at the RX antenna array. Another is the reflection distance $d^X$ of a transmitted signal being reflected by the tracked person to an RX antenna. To guarantee our system's real-time performance, we separately calculate the two parameters.
\subsubsection{AoA} 
Based on the calculated dynamic human component $Z^X_{i,j}$, the AoA ${{\theta}^{X}}$ in a joint window is estimated by
\begin{equation}
\underset{{{\theta}^{X}}\in \lbrack -{{90}^{\circ }},{{90}^{\circ }} \rbrack}{\mathop{\text{argmax}}}\,{\sum\limits_{l=1}^{M}{{\sum\limits_{j=1}^{30}{\left\|\sum\limits_{i=1}^{3} {{{\bm{Z}^X_{i,j}}\lbrack l \rbrack} {{e}^{\bm{J}2\pi \frac{{{f}_{j}}}{c}(i-1)  \Delta d \sin \theta^X   }} } \right\|}}}}.
\label{equation38}
\end{equation}
Due to the $2\pi$ phase ambiguity, the searching range of AoA should be within $\lbrack -{{90}^{{}^\circ }},{{90}^{{}^\circ }} \rbrack$. In this way, the tracked person is required to move on one side of the receive antenna array. The angular searching spacing is set as $1^\circ$ in WiDFS.
\subsubsection{Human reflection distance} 
The human reflection distance $d^{X}$ in a joint window is calculated by
\begin{equation}
\underset{{{d}^{X}}\in \left( d_{min}^{X}, d_{max}^{X} \right]}{\mathop{\text{argmax}}}\, {\sum\limits_{l=1}^{M}{ \sum\limits_{i=1}^{3}{\left\|\sum\limits_{j=1}^{30}    {{{\bm{Z}^X_{i,j}}\lbrack l \rbrack} {{\bm Z}^{D} \lbrack l \rbrack} {{{e}^{\bm{J}2\pi \frac{{f}_{j}}{c} d^{X} }}} } \right\|}}},
\label{equation36}
\end{equation}
where 
\begin{equation}
\begin{cases}
{{\bm Z}^{D}}\lbrack l \rbrack={{e}^{-\bm J 2\pi \sum\limits_{\ell=l}^{\epsilon-1}{{f}^{D}\lbrack \ell \rbrack}N\Delta t}}, & l <\epsilon  \\
{{\bm Z}^{D}}\lbrack l \rbrack=1, & l = \epsilon \\
{{\bm Z}^{D}}\lbrack l \rbrack={{e}^{\bm J 2\pi \sum\limits_{\ell=\epsilon}^{l-1}{{f}^{D}[\ell ]}N\Delta t}}, & l >\epsilon  \\
\end{cases}.
\label{equation37}
\end{equation}
Here $\epsilon=\frac{M+1}{2}$ and $M$ is an odd number. The minimum distance $d_{min}^{X}$ is equal to the length $d^{S}$ of the direct path between the TX and RX antennas since the human reflection distance is larger than $d^{S}$. The maximum distance $d_{max}^{X}$ is $20$ m, which is sufficiently large for indoor scenarios. The distance searching spacing is set as 5 cm in WiDFS.

\subsection{Localization parameter refinement}
Since the estimated $\theta^{X}$ and $d^{X}$ may be relatively noisy, we refine them using Kalman smoothers. 

Let $ \theta^{X}\left[ L \right]$ and $ d^{X}\left[ L \right]$ be the current AoA and human reflection distance in the $L$-th joint window after outlier rejection. The corresponding Kalman model is built as follows,
\begin{equation}
\left\{ 
\begin{aligned}
 & \sin\left( \theta^{X}\left[ L+1 \right]\right)\approx \sin\left(\theta^{X}\left[L \right]\right)\\
&  d^{X}\lbrack L+1 \rbrack= d^{X}\left[ L \right]+c\frac{{\bm f }^{D}\left[ L \right]}{{{f}_{c}}}\frac{N_p}{f^s}\\
\end{aligned} 
\right.,
\label{equation39}
\end{equation}
where ${\bm f}^{D}\left[ L \right]$ is the median of $\lbrace  {f}^{D}\lbrack 1 \rbrack,{f}^{D}\lbrack 2 \rbrack, ..., {f}^{D}\lbrack M\rbrack  \rbrace $ in the $L$-th joint window. In the AoA Kalman smoother, the noise covariance of the sine of AoA is set to 0.2. In the distance Kalman smoother, the noise covariance of the reflection distance and DFS are set as 20 cm and $5$ Hz, respectively. Fig. \ref{Fig8} and Fig. \ref{Fig9} show the estimated AoA and human reflection distance without and with Kalman smoothing.

\subsection{Human localization}
So far, the two core localization parameters, i.e., human reflection distance $d^X$ and AoA $\theta^X$, have been calculated. The person position in a 2D coordinate system can then be estimated using these estimated parameters.

As shown in Fig. \ref{Fig10}, let $d^{X \rightarrow TX}$ and $d^{X \rightarrow RX}$ be the distance of the tracked person to the transmitter and receiver, respectively, where $d^{X}=d^{X \rightarrow TX}+d^{X \rightarrow RX}$. And let $\theta^S$ be the AoA of the transmitter relative to the RX antenna array, i.e., $\Delta d \sin \theta^S \approx d^{S}_{2}-d^{S}_{1} \approx  d^{S}_{3}-d^{S}_{2}$. Since the phase differences caused by WiFi hardware diversity are pre-estimated, the AoA $\theta^S$ is obtained by 
\begin{equation}
\underset{{{\theta}^{S}}\in \lbrack -{{90}^{{}^\circ }},{{90}^{{}^\circ }} \rbrack}{\mathop{\text{argmax}}}\,{\sum\limits_{l=1}^{M}{{\sum\limits_{j=1}^{30}{\left\| \left( {\bm Z^S_{12}\lbrack l \rbrack+\bm Z^S_{23}\lbrack l \rbrack} \right) {{e}^{-\bm{J}\Delta d \sin \theta^S }}\right\|}}}},
\label{equation40}
\end{equation}
where
\begin{equation}
\left\{ 
\begin{aligned}
 & \bm Z^S_{12}\lbrack l \rbrack={{e}^{\bm{J} \left(\angle \bm{U}_{12,j}\lbrack l \rbrack-\Delta\varphi^{h}_{12}\right)  }}\\
 & \bm Z^S_{23}\lbrack l \rbrack={{e}^{\bm{J} \left(\angle \bm{U}_{23,j}\lbrack l \rbrack-\Delta\varphi^{h}_{23}\right)  }}\\
\end{aligned} 
\right..
\label{equation41}
\end{equation}
\begin{figure}
	\centering
    \begin{subfigure}{0.235\textwidth}
        \centering
        \includegraphics[width=0.88\textwidth]{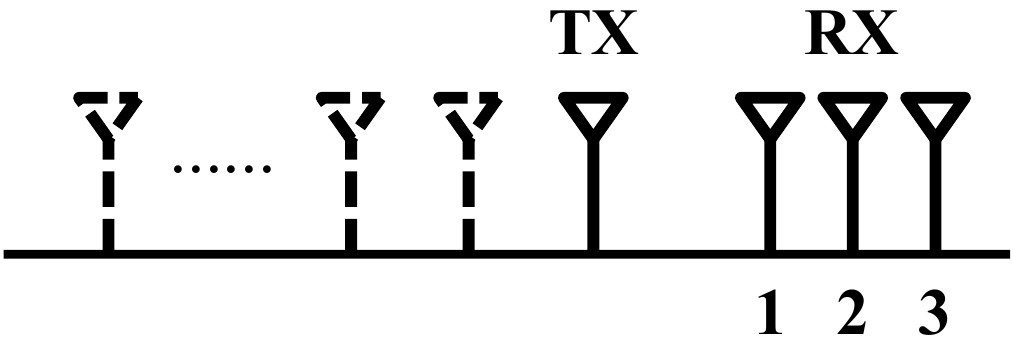}
        \subcaption{Transmitter placed on the left of the RX antenna array}
        \label{Fig11a}
    \end{subfigure}
~
	\begin{subfigure}{0.235\textwidth}
        \centering
        \includegraphics[width=\textwidth]{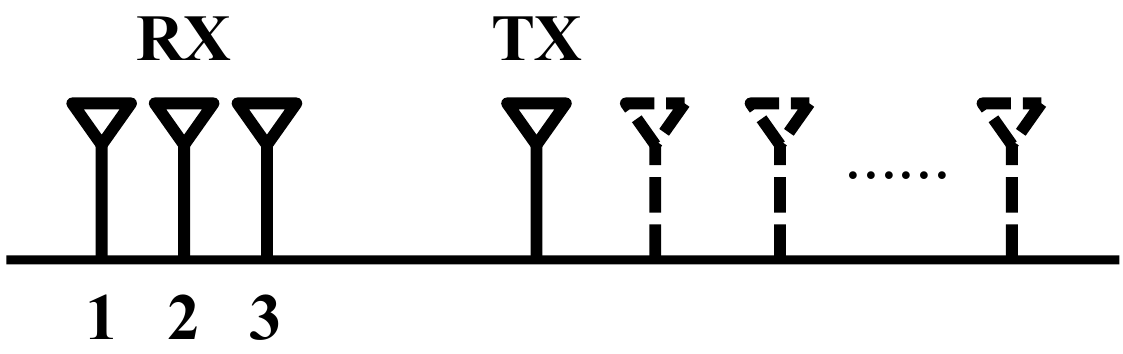}
        \subcaption{Transmitter placed on the right of the RX antenna array}
        \label{Fig11b}
	\end{subfigure}
\caption{Static component measurement on the both sides of receive antennas for hardware diversity calibration}
    \label{Fig11}
    \vspace{-2.5em}
\end{figure}

\begin{figure*}
			\centering
    		\begin{subfigure}{0.329\textwidth}
        		\centering
        		\includegraphics[width=1.01\textwidth]{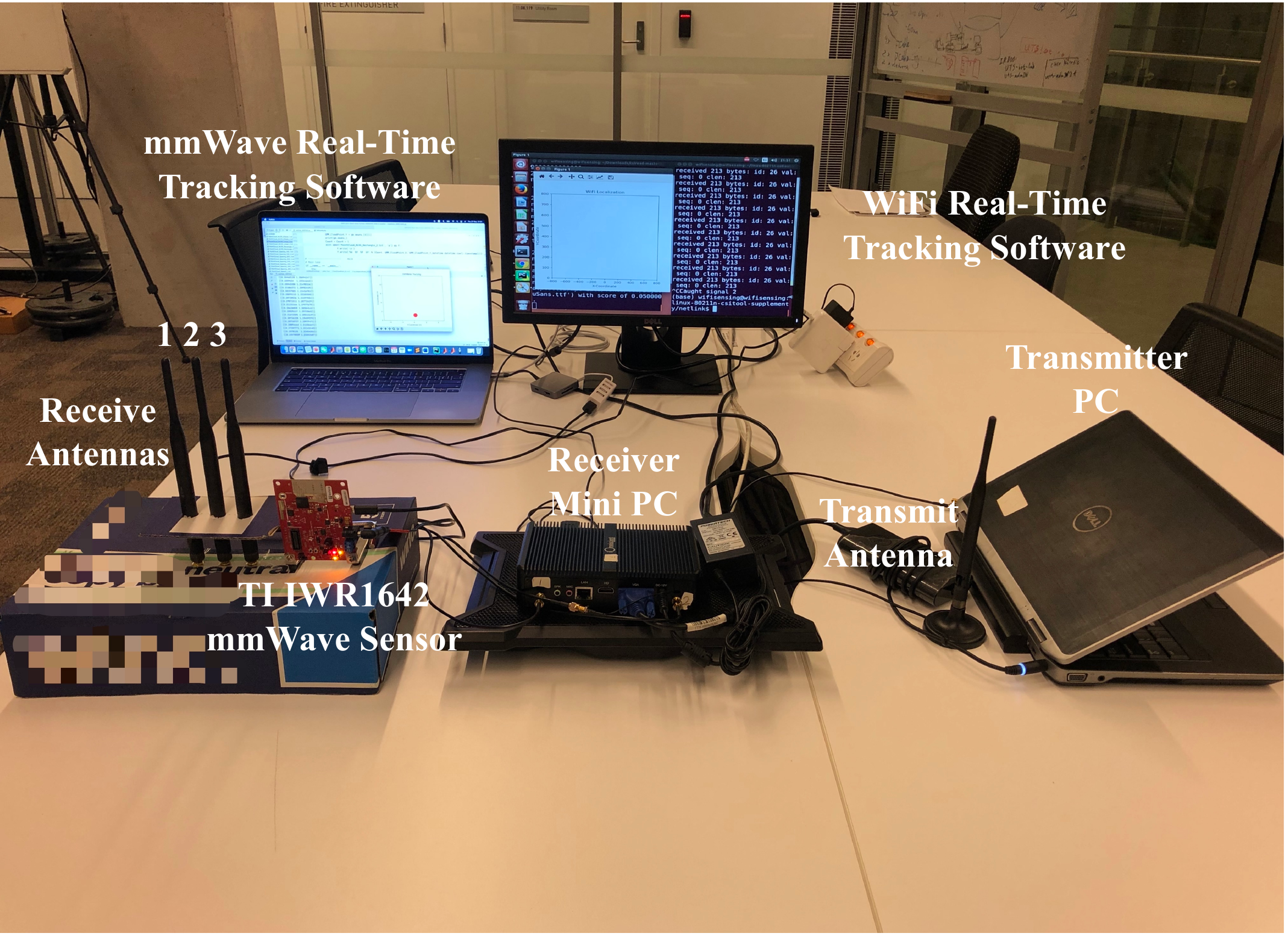}
        		\subcaption{Hardware and Software}
        		\label{Fig12a}
    		\end{subfigure}
			\begin{subfigure}{0.329\textwidth}
        		\centering
        		\includegraphics[width=0.98\textwidth]{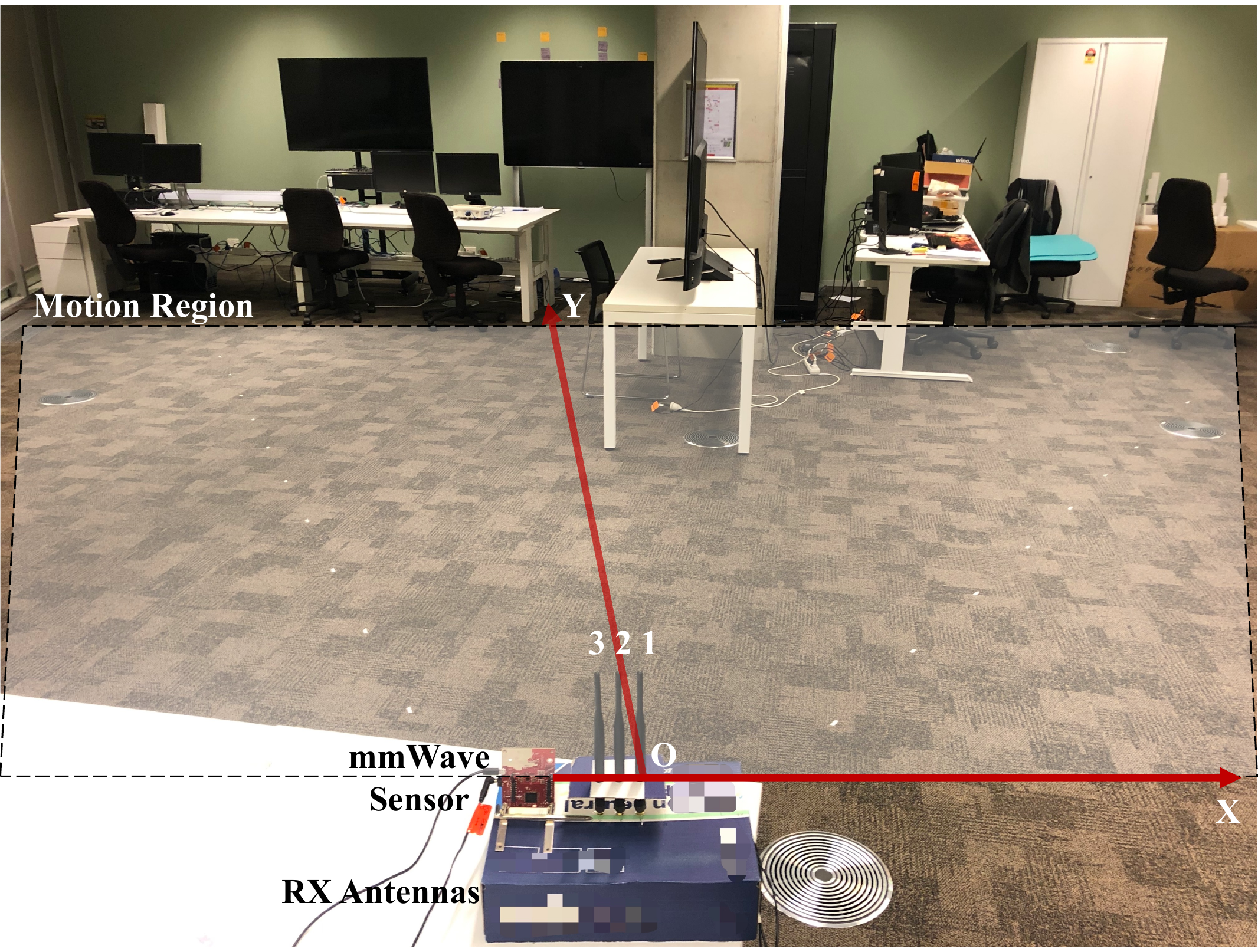}
        		\subcaption{Tracking coordinate system}
        		\label{Fig12b}
			\end{subfigure}
			\begin{subfigure}{0.329\textwidth}
        		\centering
        		\includegraphics[width=0.99\textwidth]{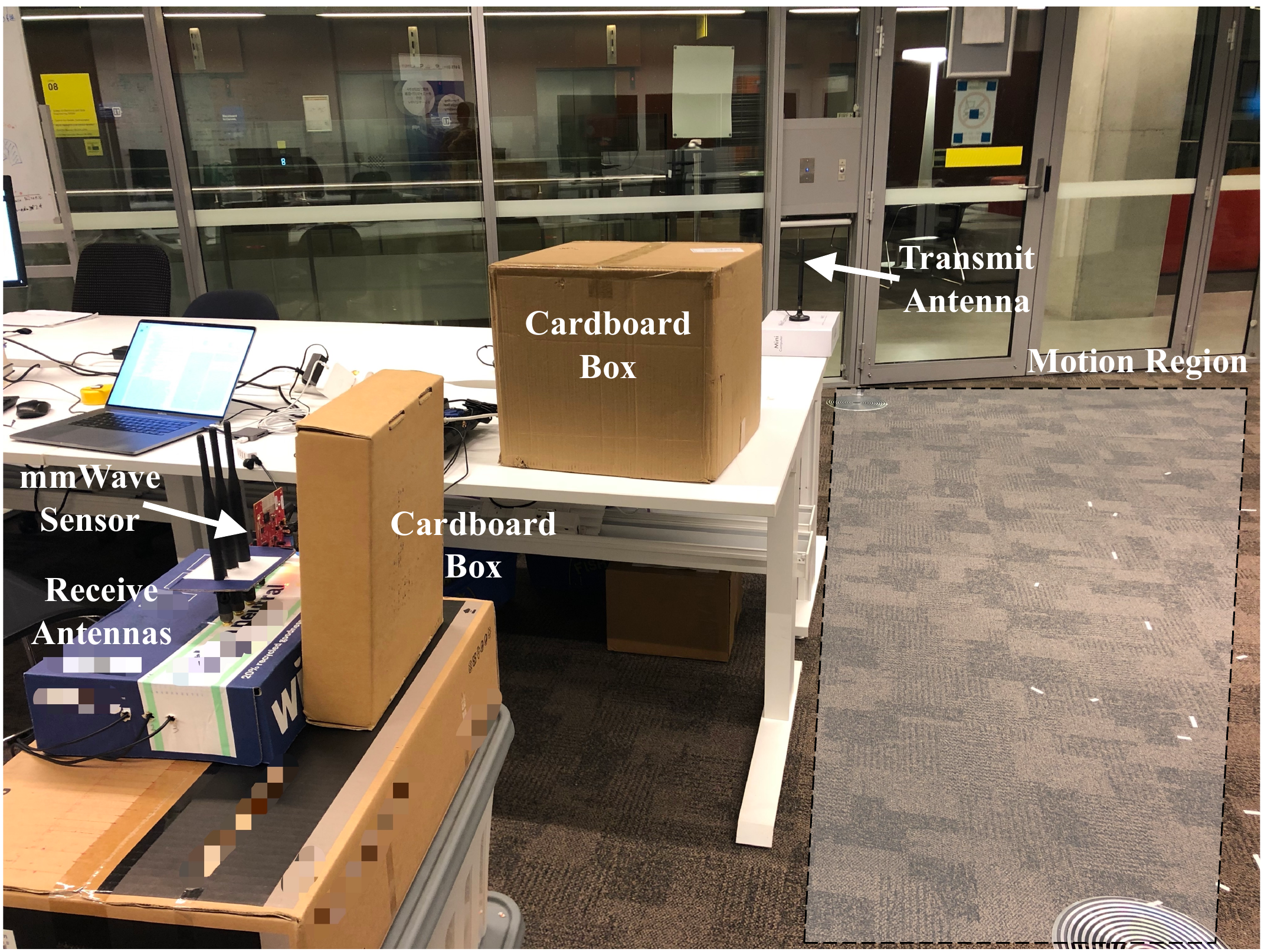}
        		\subcaption{NLOS scenario}
        		\label{Fig12c}
    		\end{subfigure}
		\caption{WiDFS Implementation}
		\label{Fig12}
		\vspace{-1.5em}
\end{figure*}

According to the cosine formula, the distance $d^{X \rightarrow RX}$ can be calculated by
\begin{equation}
d^{X \rightarrow RX}=\frac{\left({d^{X}}\right)^2-\left({d^{S}_{1}}\right)^2}{2\lbrack d^{X}-{d^{S}_{1}}\cos\left( \theta^X-\theta^S\right)\rbrack}.
\label{equation42}
\end{equation}


Thus, the person's position in a x-y coordinate system is
\begin{equation}
\left\{ 
\begin{aligned}
 & x=d^{X \rightarrow RX}\sin \theta^X\\
 & y=d^{X \rightarrow RX}\cos \theta^X\\
\end{aligned} 
\right..
\label{equation43}
\end{equation}

\section{WiFi hardware diversity calibration} 
In \cite{xiong2013arraytrack}, a method was proposed to calibrate the phase difference between antennas caused by manufacturing imperfection. However, the calibrated results are always wrapped by an unknown $\pi$-radians ambiguity, meaning that the estimated phase difference is the actual value or the actual value plus $\pi$ radians. Also, this method cannot calibrate the antenna spacing. For example, a spacing error of 0.5 cm will cause a phase deviation of about 0.557 radians, which will significantly impact the AoA estimation accuracy. Here we develop a novel WiFi hardware calibration algorithm. It is a one-time procedure after the system is set up. 

Here we use RX Antenna 1 and Antenna 2 as an example. As shown in Fig. \ref{Fig11}, we firstly place the TX antenna at the right side of the RX antenna array. They are deployed in a straight line. We measure multiple static components denoted as $\lbrace \bm{U}_{12,j}^{R}\lbrack 1 \rbrack,  \bm{U}_{12,j}^{R}\lbrack 2 \rbrack,... \rbrace$. We then place the TX antenna at the left side of the array and obtain $\lbrace \bm{U}_{12,j}^{L}\lbrack 1 \rbrack,  \bm{U}_{12,j}^{L}\lbrack 2 \rbrack,... \rbrace$. The data collection is conducted in a very low-multipath environment. Let $\Delta d_{12}$ be the spacing between Antenna 1 and Antenna 2, then 
\begin{equation}
\left\{ 
\begin{aligned}
 & \angle \bm{U}_{12,j}^{L}=\Delta\varphi^{h}_{12}+\frac{2\pi f_c}{c}\Delta d_{12}+2\pi K_{12}^{L}\\
 & \angle \bm{U}_{12,j}^{R}=\Delta\varphi^{h}_{12}-\frac{2\pi f_c}{c}\Delta d_{12}+2\pi K_{12}^{R}\\
\end{aligned} 
\right.,
\label{equation44}
\end{equation}
where $K_{12}^{L}$ and $K_{12}^{R}$ are unknown integers.

Then the antenna spacing $\Delta d_{12}$ is calculated by
\begin{equation}
\resizebox{\linewidth}{!}{$
\Delta d_{12}=\frac{c}{2\pi f_c}\left[ \frac{1}{2}\angle\left( {\sum\limits_{l} {e}^{\bm{J}{\left(\angle \bm{U}_{12,j}^{L}\lbrack l \rbrack-\angle \bm{U}_{12,j}^{R}\lbrack l \rbrack \right) }}} \right)+\pi K_{12} \right],
$}
\label{equation45}
\end{equation}
where $K_{12}$ is an integer making $\Delta d_{12}$ closest to our pre-defined spacing (half a wavelength of about 2.8 cm). 

Once $\Delta d_{12}$ is determined, we substitute it into Eq. \ref{equation44} to estimate hardware-related phase difference $\Delta\varphi^{h}_{12}$,
\begin{equation}
\resizebox{\linewidth}{!}{$
\Delta\varphi^{h}_{12}=\angle\sum\limits_{l}\left[{e}^{\bm{J}\left(\angle \bm{U}_{12,j}^{L}\lbrack l \rbrack-\frac{2\pi f_c}{c}\Delta d_{12}\right)}+{e}^{\bm{J}\left(\angle \bm{U}_{12,j}^{R}\lbrack l \rbrack+\frac{2\pi f_c}{c}\Delta d_{12}\right)}\right].
$}
\label{equation46}
\end{equation}

Similarly, the distance $\Delta d_{23}$ and phase difference $\Delta\varphi^{h}_{23}$ can also be measured based on the above process.

\section{Implementation \& Evaluation}
\textbf{Implementation:} \textit{(1) Hardware.} We implement WiDFS using two computers separately equipped with an Intel 5300 NIC, shown in Fig. \ref{Fig12}. One is served as a transmitter and has one WiFi antenna, while another is a receiver that has three external antennas to form a linear uniform antenna array. These antennas are all omnidirectional and have 2 dBi gain. \textit{(2) Software.} The operating system of each laptop is Ubuntu 14.04 LTS with 3.16.0-30-generic Linux kernel version. They are configured in the monitor mode via \textit{Linux 802.11n CSI Tool} \cite{halperin2010predictable, dhalperi2014, dhalperi2014supplementary}. The CSI sampling rate is 1 kHz. The center frequency is 5.32 GHz. WiDFS is programmed by \textit{Python 3.8} and implemented on a Mini PC with Intel(R) Core CPU i5-7300U 2.6 GHz$\times$4 and 3.8G memory. We adopt \textit{csiread package}\footnote{\begin{scriptsize} \url{https://github.com/citysu/csiread} \end{scriptsize}} to parse CSI data in real time.

\textbf{WiFi Hardware Diversity Calibration.} The manually measured antenna spacing in the RX antenna array is 2.8 cm. However, our hardware calibration algorithm outputs that the spacing between Antenna 1 and Antenna 2 is 2.618 cm while the spacing between Antenna 2 and Antenna 3 is 2.391 cm. It also shows the hardware-related phase difference between Antenna 1 and Antenna 2 is 5.956 radians while that between Antenna 2 and Antenna 3 is 1.418 radians. 

\textbf{Default Configuration.} The transmitter and receiver are placed at the same height, and their separation distance is 235 cm. The AoA of the transmitter relative to the receiver is about $-70^\circ$. A CSI sampling window contains $100\times30\times3$ samples for 3 RX antennas and 30 subcarriers, and a joint window contains 9 CSI sampling sub-windows.

\textbf{Evaluation.} We evaluate WiDFS in a multipath-rich office environment with various types of strong reflectors such as tables, chairs, metal lockers, computers, large-size displays, concrete ceiling, and tempered glass/hollow walls. In each experiment, we ask a person to walk along with three types of trajectories (shown in Fig. \ref{Fig14a}, Fig. \ref{Fig15a} and Fig. \ref{Fig16a}) in different office regions under different multipath interference. For a NLOS experiment, we use cardboard boxes to block RX antennas, shown in Fig. \ref{Fig12c}, so there is no LOS path from the person and transmitter to the receiver. 
\begin{figure}
     \centering
         \includegraphics[width=0.435\textwidth]{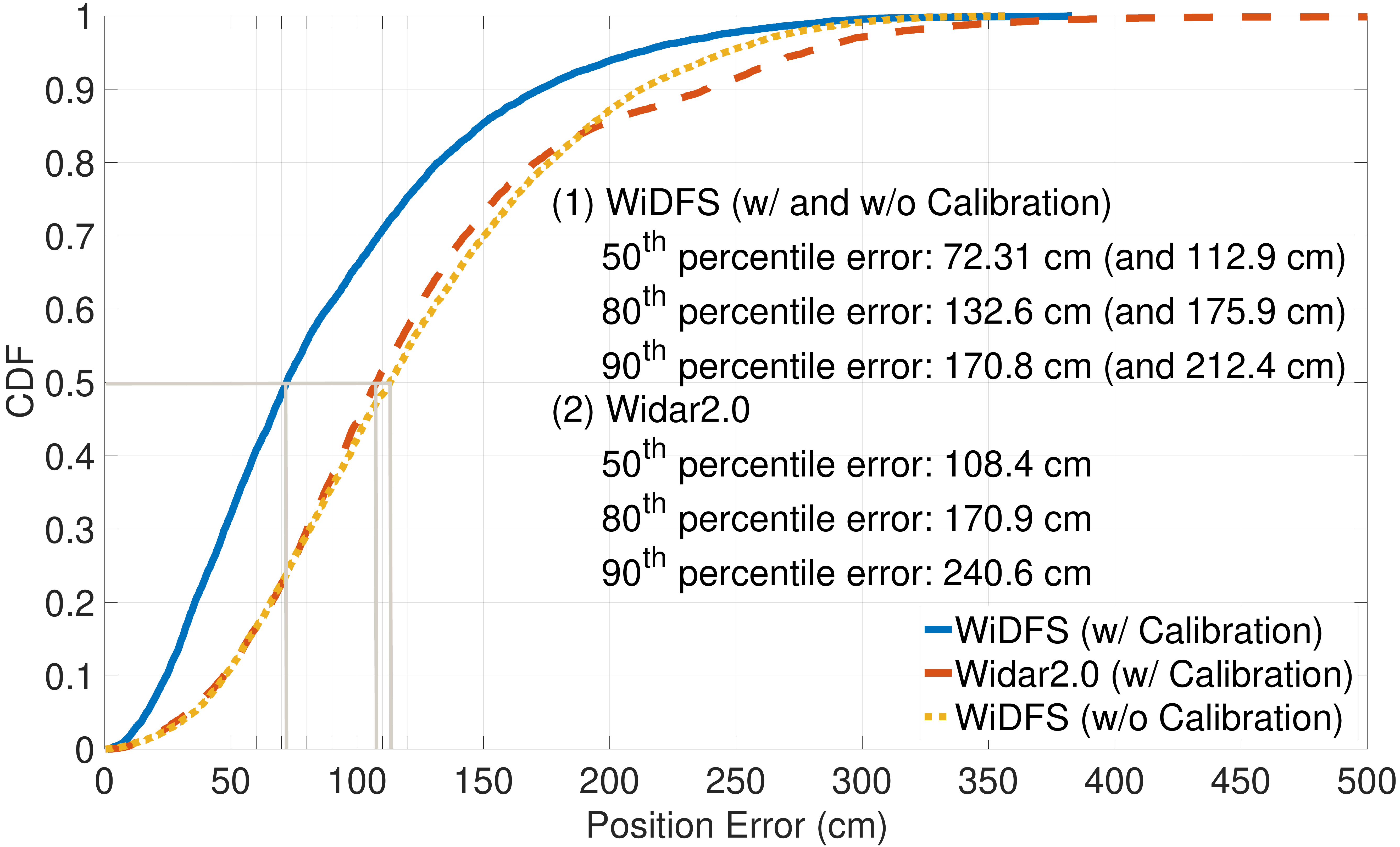}
         \caption{CDF of tracking accuracy of WiDFS and Widar2.0}
         \label{Fig13}
     \vspace{-1.5em}
\end{figure}
\begin{figure*}
     \centering
		\begin{minipage}[t]{0.329\linewidth}
			\centering
    		\begin{subfigure}{\textwidth}
        		\centering
        		\includegraphics[width=0.85\textwidth]{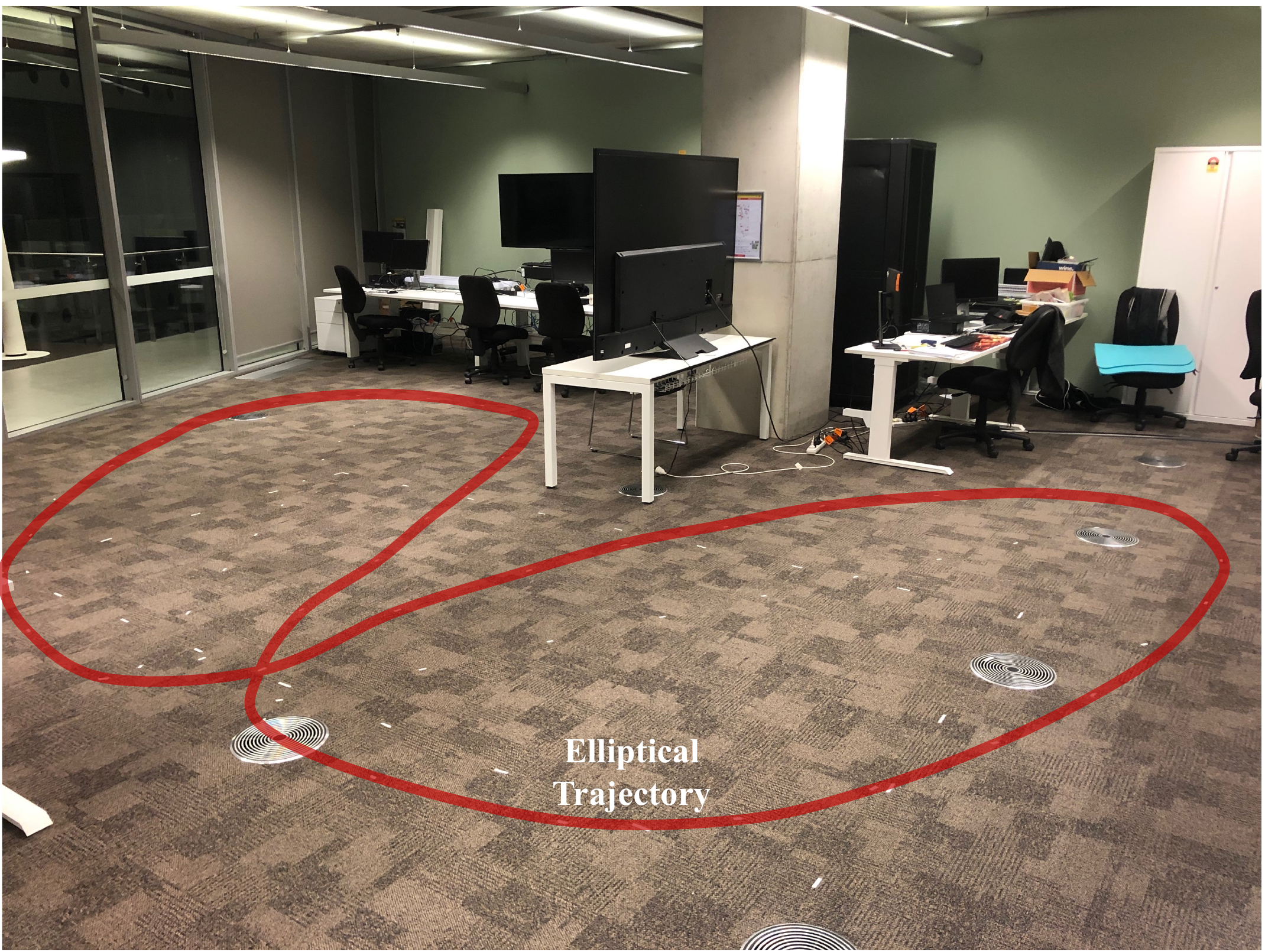}
        		\subcaption{Elliptical trajectory}
        		\label{Fig14a}
    		\end{subfigure}\\
			\begin{subfigure}{\textwidth}
        		\centering
        		\includegraphics[width=\textwidth]{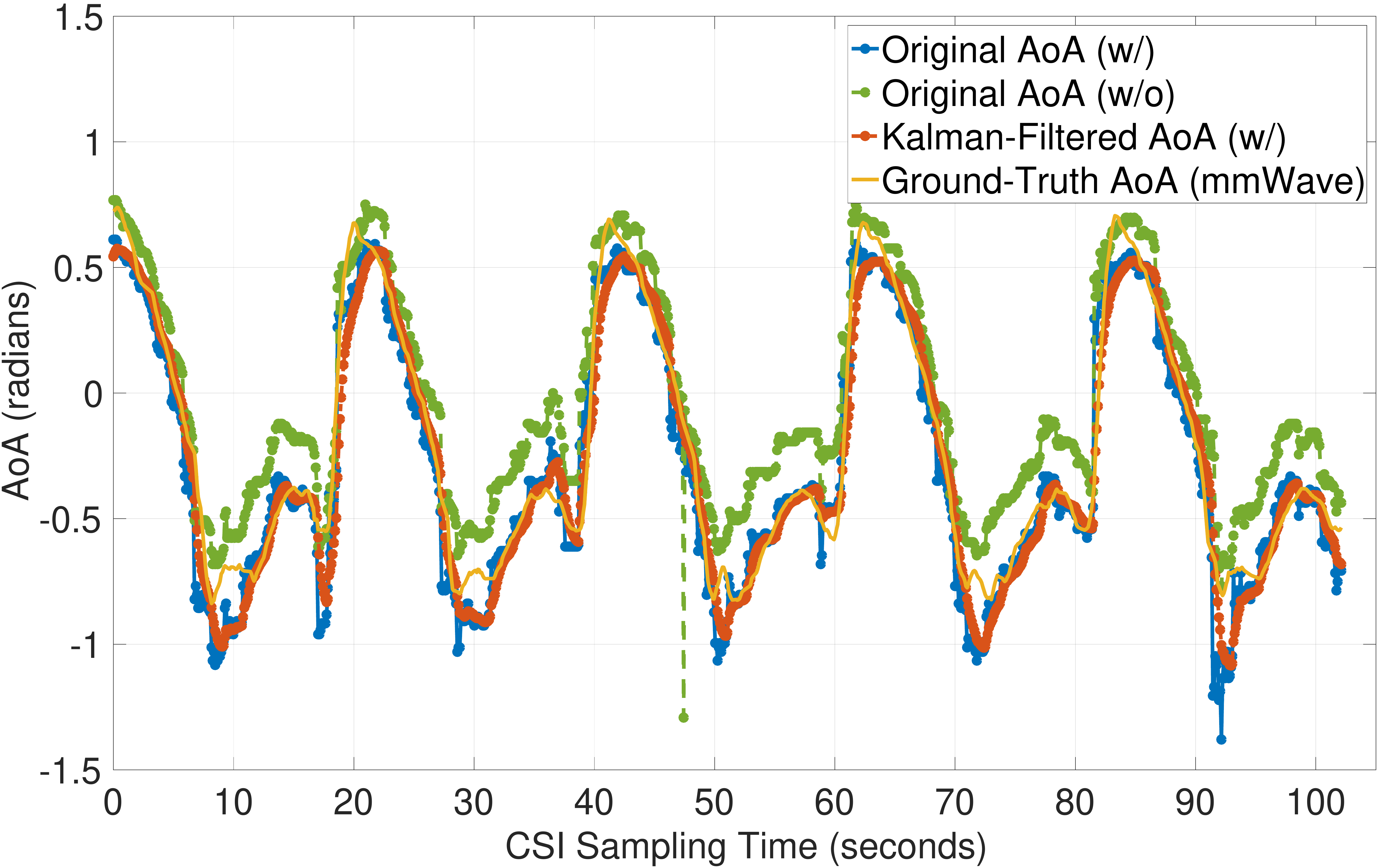}
        		\subcaption{AoA}
        		\label{Fig14b}
			\end{subfigure}\\
			\begin{subfigure}{\textwidth}
        		\centering
        		\includegraphics[width=\textwidth]{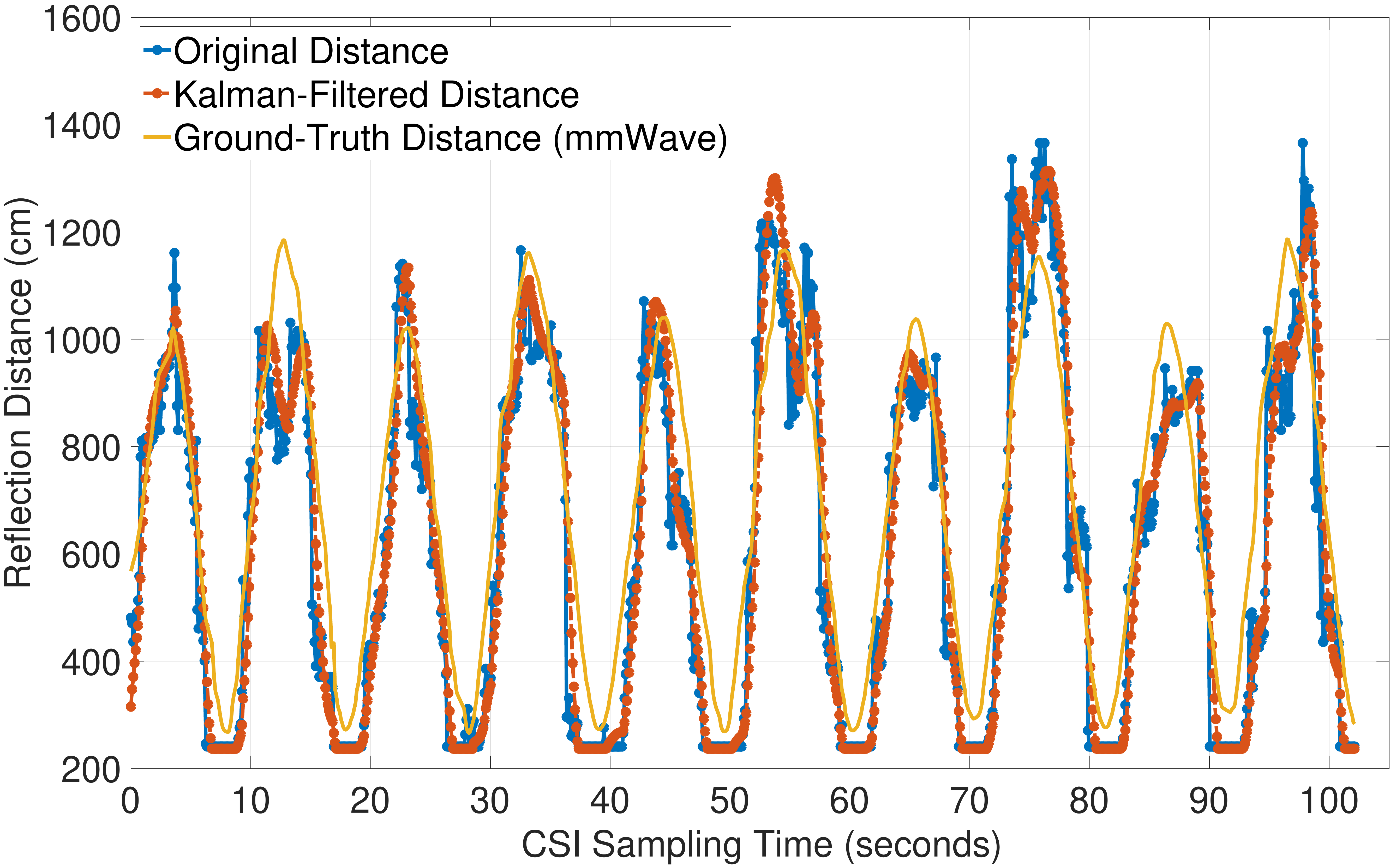}
        		\subcaption{Reflection distance}
        		\label{Fig14c}
    		\end{subfigure}\\
			\begin{subfigure}{\textwidth}
        		\centering
        		\includegraphics[width=\textwidth]{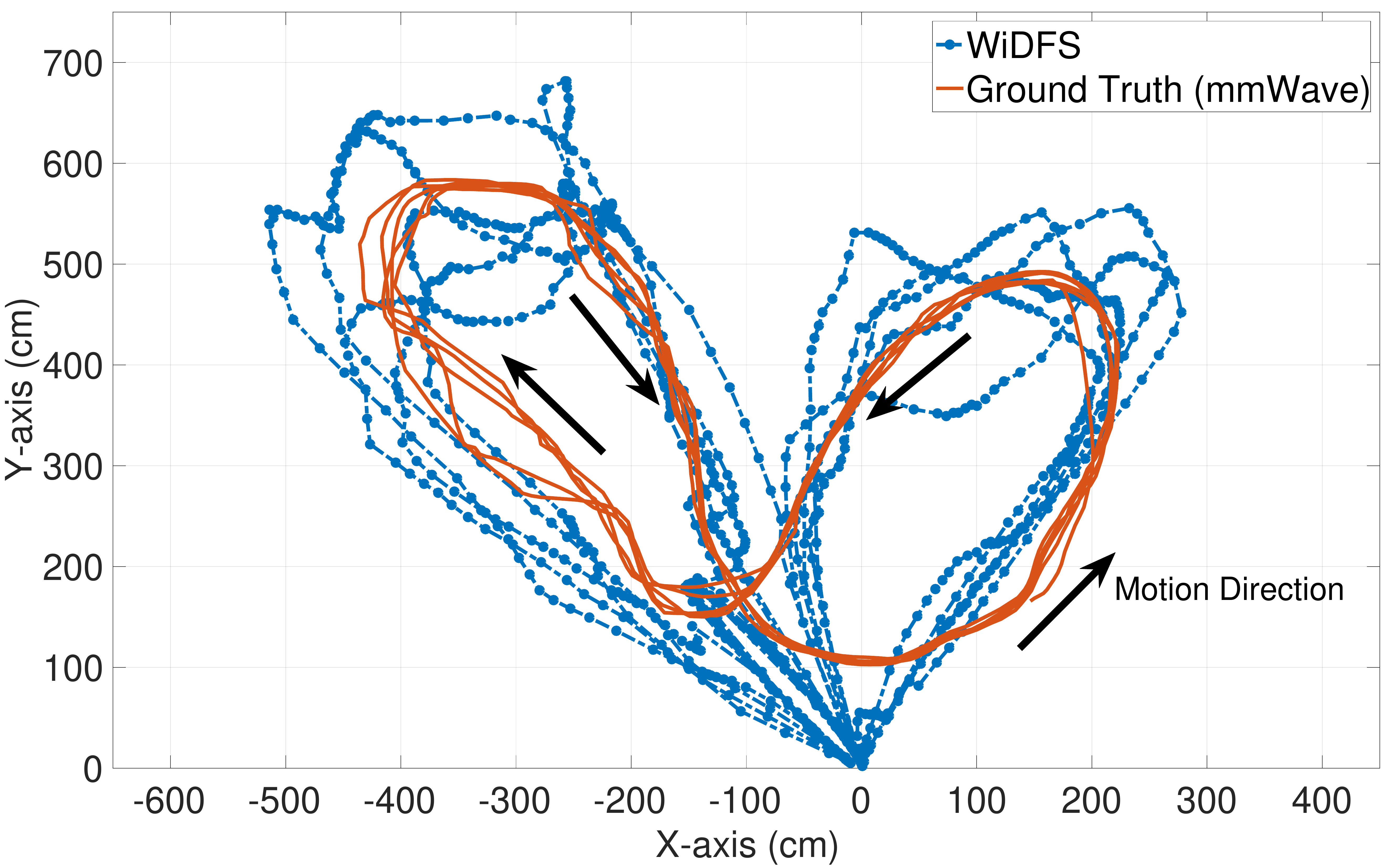}
        		\subcaption{Estimated ellipse trajectory}
        		\label{Fig14d}
			\end{subfigure}
		\caption{A case of ellipse path tracking}
		\label{Fig14}
		\end{minipage}
		\begin{minipage}[t]{0.329\linewidth}
			\centering
    		\begin{subfigure}{\textwidth}
        		\centering
        		\includegraphics[width=0.84\textwidth]{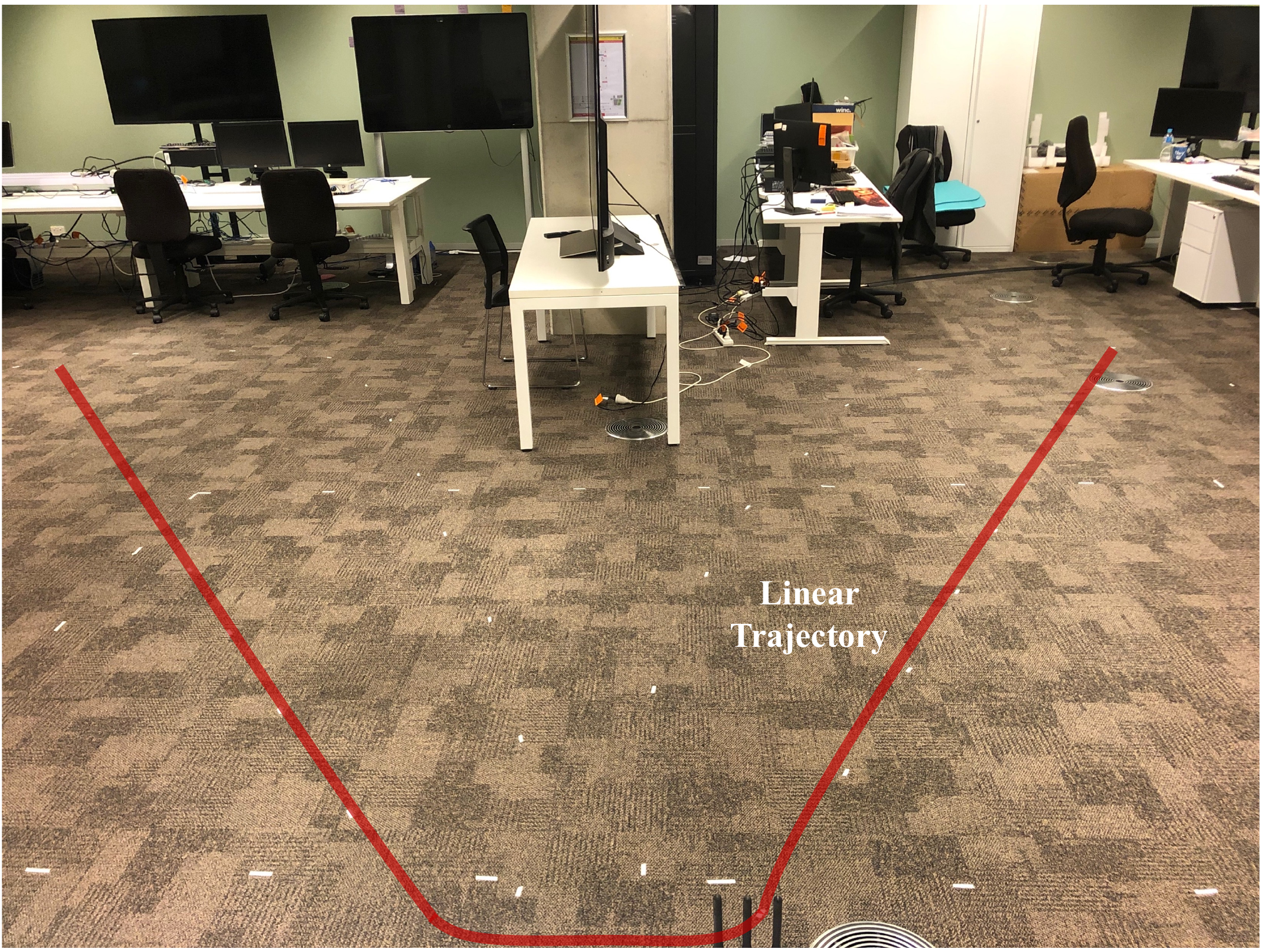}
        		\subcaption{Linear trajectory}
        		\label{Fig15a}
    		\end{subfigure}\\
			\begin{subfigure}{\textwidth}
        		\centering
        		\includegraphics[width=\textwidth]{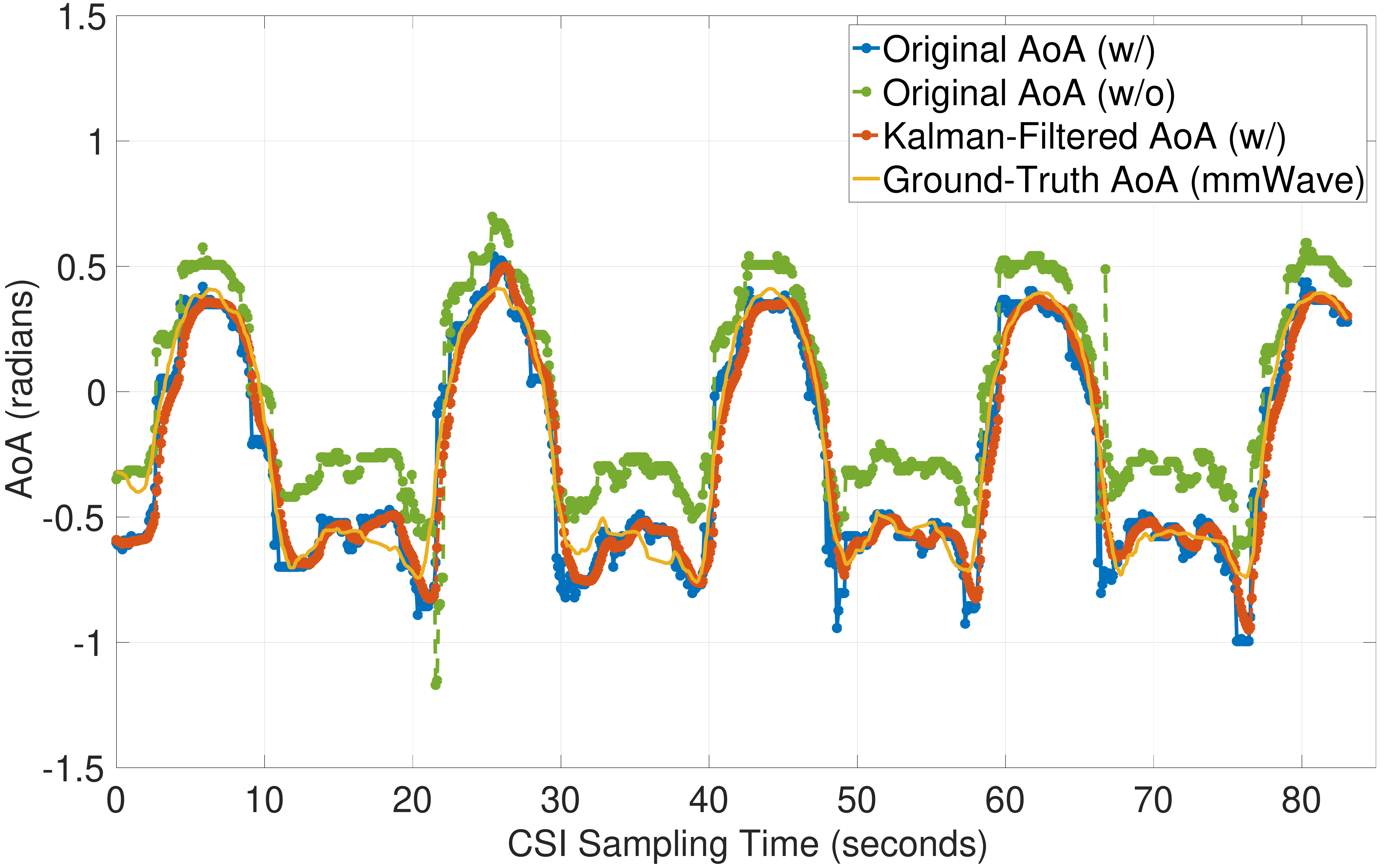}
        		\subcaption{AoA}
        		\label{Fig15b}
			\end{subfigure}\\
			\begin{subfigure}{\textwidth}
        		\centering
        		\includegraphics[width=\textwidth]{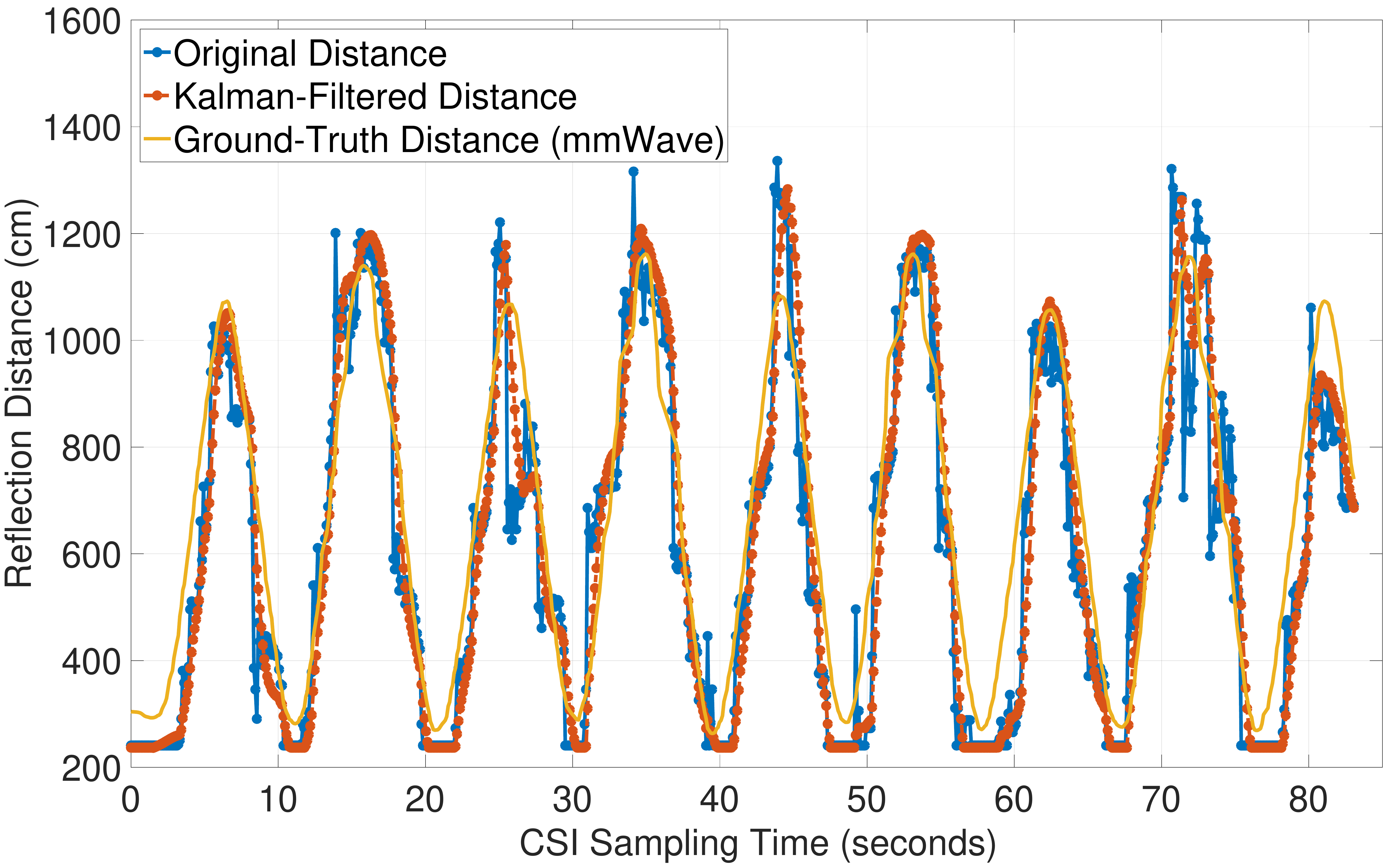}
        		\subcaption{Reflection distance}
        		\label{Fig15c}
    		\end{subfigure}\\
			\begin{subfigure}{\textwidth}
        		\centering
        		\includegraphics[width=0.99\textwidth]{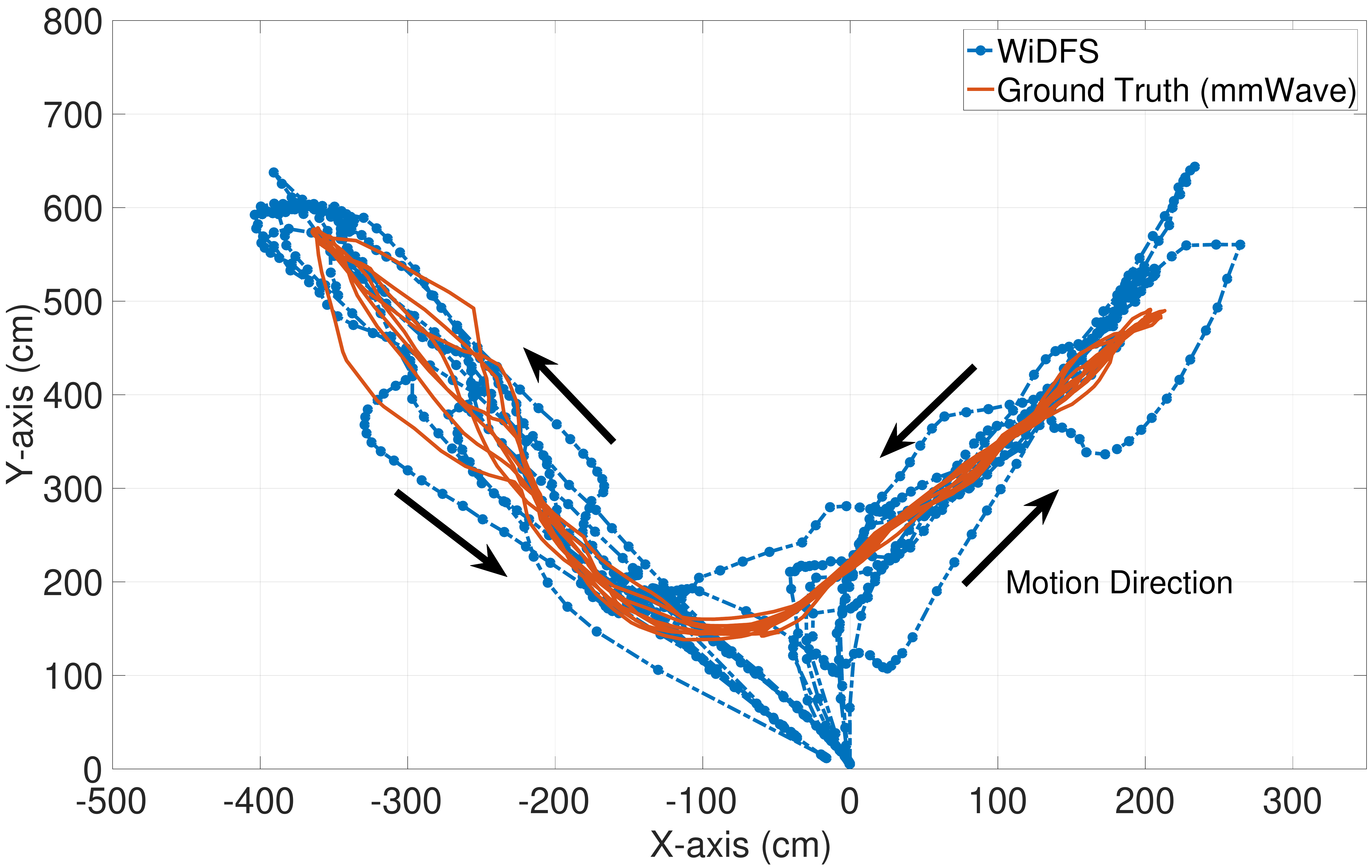}
        		\subcaption{Estimated line trajectory}
        		\label{Fig15d}
			\end{subfigure}
		\caption{A case of line path tracking}
		\label{Fig15}
		\end{minipage}
\begin{minipage}[t]{0.329\linewidth}
			\centering
    		\begin{subfigure}{\textwidth}
        		\centering
        		\includegraphics[width=1.05\textwidth]{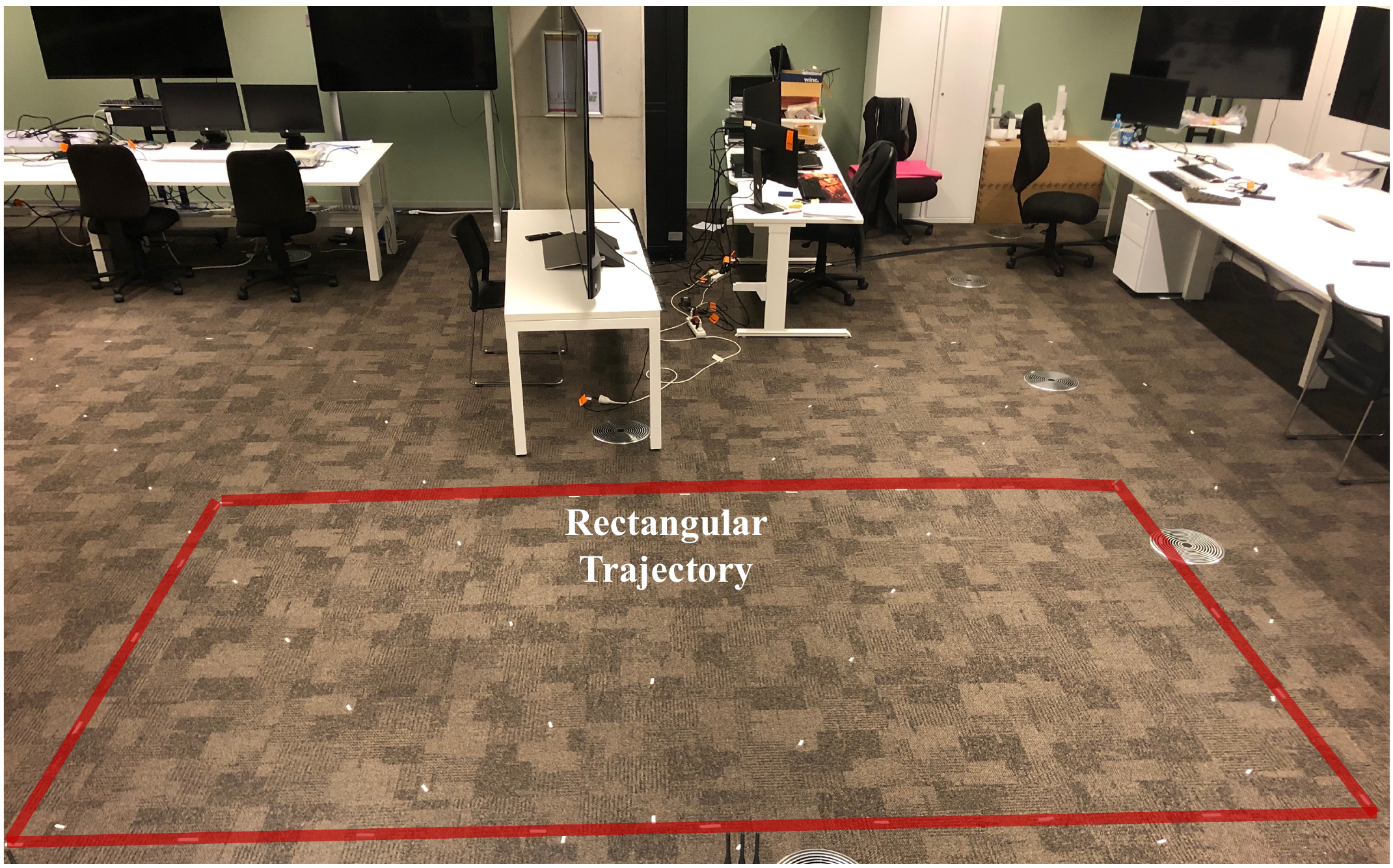}
        		\subcaption{Rectangular trajectory}
        		\label{Fig16a}
    		\end{subfigure}\\
			\begin{subfigure}{\textwidth}
        		\centering
        		\includegraphics[width=\textwidth]{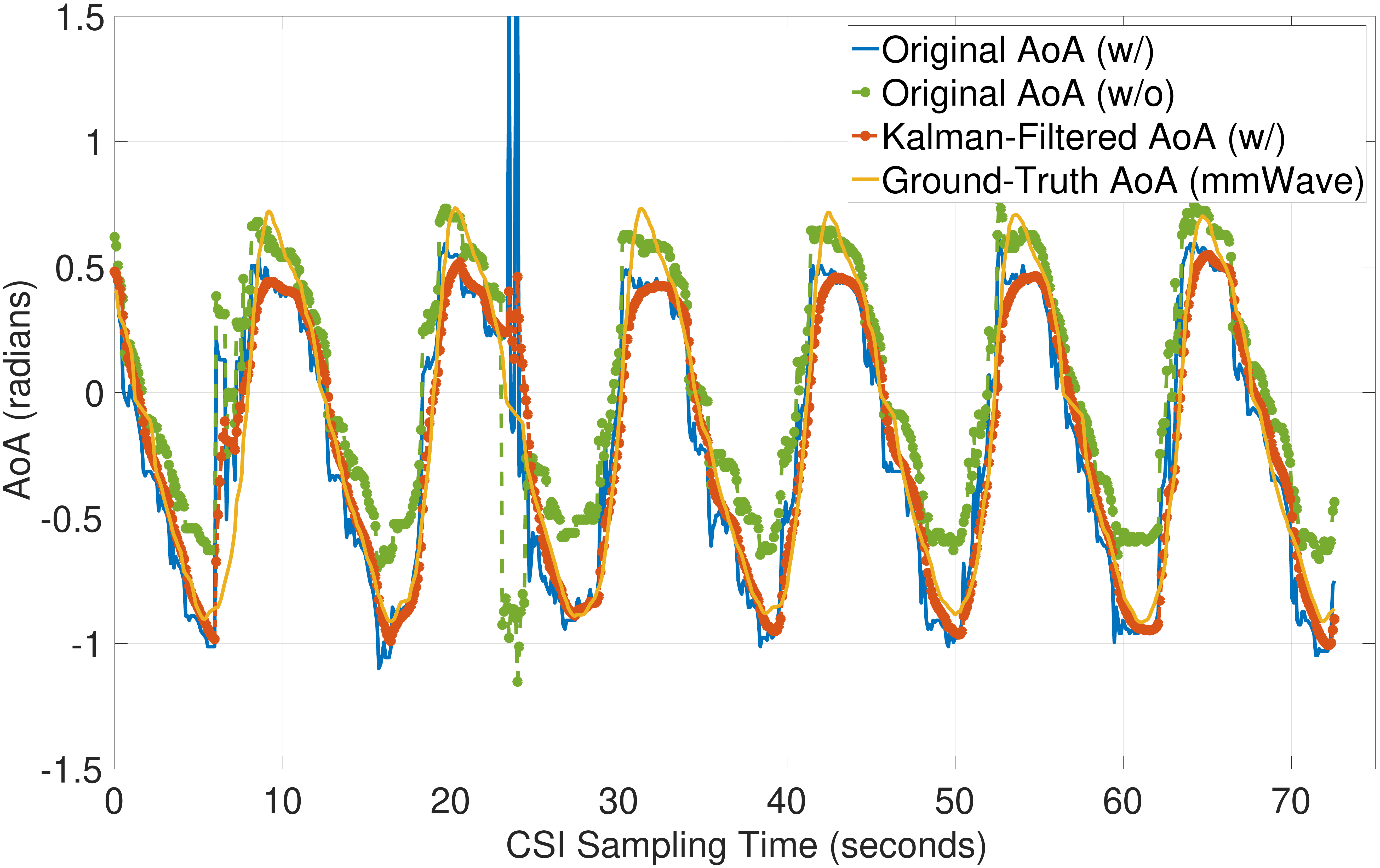}
        		\subcaption{AoA}
        		\label{Fig16b}
			\end{subfigure}\\
			\begin{subfigure}{\textwidth}
        		\centering
        		\includegraphics[width=\textwidth]{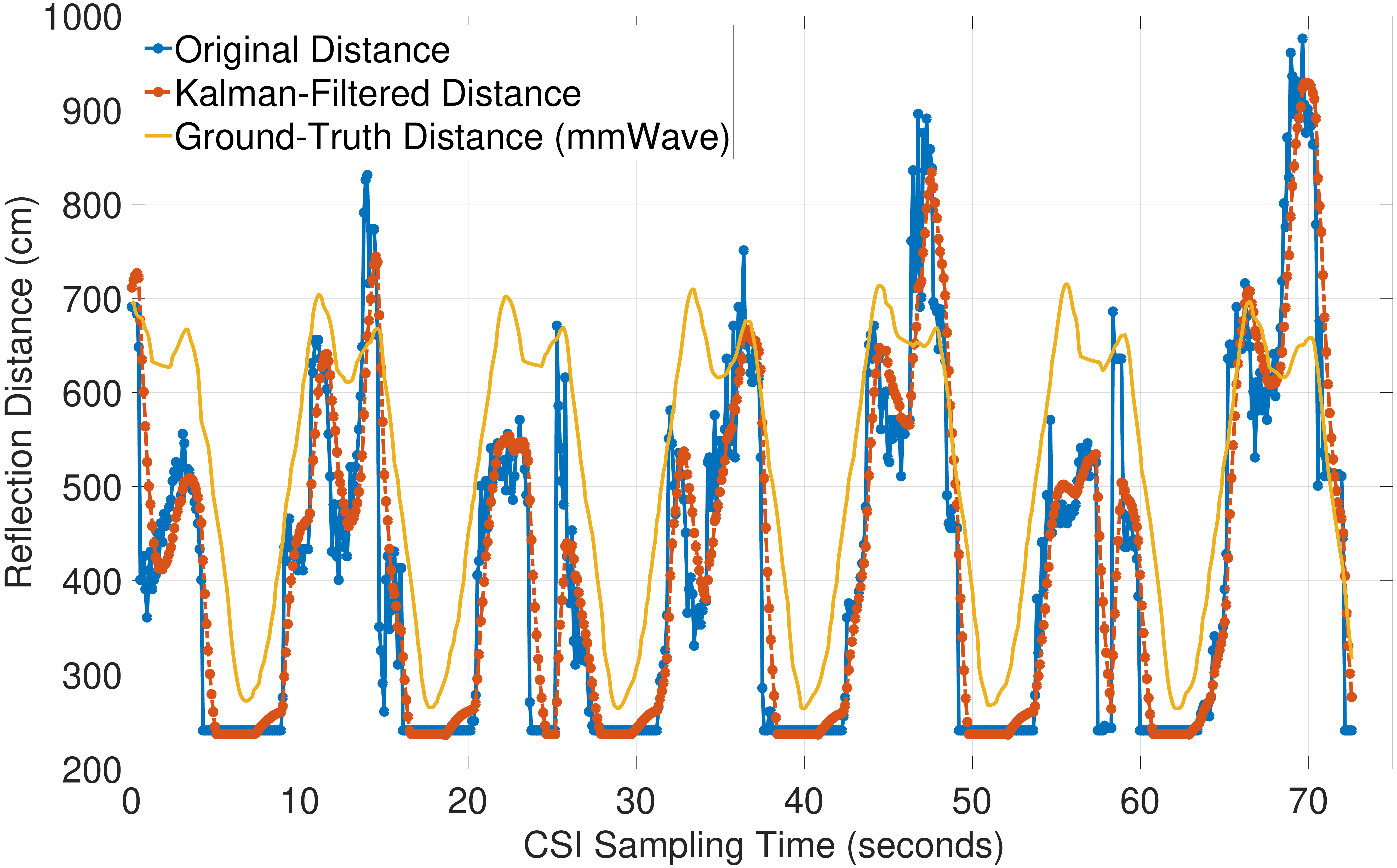}
        		\subcaption{Reflection distance}
        		\label{Fig16c}
    		\end{subfigure}\\
			\begin{subfigure}{\textwidth}
        		\centering
        		\includegraphics[width=\textwidth]{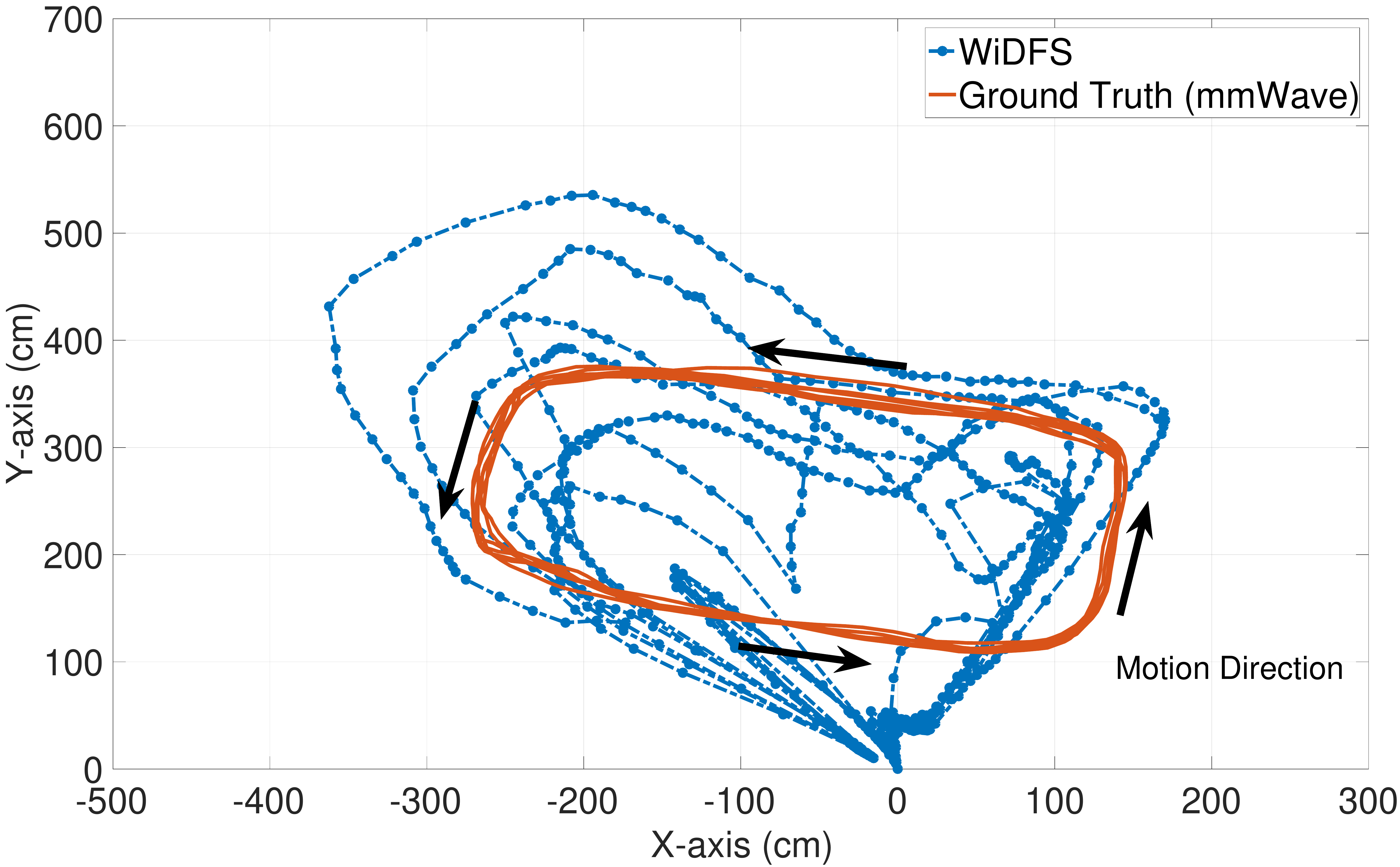}
        		\subcaption{Estimated rectangle trajectory}
        		\label{Fig17d}
			\end{subfigure}
		\caption{A case of rectangle path tracking}
		\label{Fig16}
		\end{minipage}
     \vspace{-1.5em}
\end{figure*}

\textbf{Baseline.} We compare WiDFS to a state-of-the-art unsupervised method Widar2.0 \cite{qian2018widar2} with its implementation software downloaded from the website\footnote{\begin{scriptsize}{http://tns.thss.tsinghua.edu.cn/wif\mbox{}iradar/Widar2.0Project.zip} \end{scriptsize}}. The baseline has the same system setup as ours. It also works with three parameters, i.e., DFS, AoA, and human reflection distance. The Widar2.0 software does not support real-time implementation, and hence its results are obtained offline. For fairness, we substitute hardware calibration parameters into Widar2.0 and use its algorithm in Matlab to track a person.

\textbf{Ground Truth.} We use a TI IWR1642 mmWave radar to measure the person's trajectory as ground truth. Such a radar could achieve centimeter-level tracking accuracy. It is placed beside the WiFi RX antenna array. We modify the project\footnote{{\begin{scriptsize}\url{https://github.com/ibaiGorordo/AWR1642-Read-Data-Python-MMWAVE-SDK-2} \end{scriptsize}}} in Python to perform single-target real-time mmWave tracking as follows. We firstly obtain point cloud data and use DBSCAN clustering algorithm to eliminate some noisy points. We then adopt a Gaussian mixture to capture the density center of refined cloud data as the tracked target's position. The localization results of WiFi and mmWave are synchronized using 1-D data interpolation based on their real-world sampling timestamps.

\section{Results}
\subsection{Comparison to state-of-the-art}
We start by comparing WiDFS's performance to Widar2.0. Fig. \ref{Fig13} shows the CDF of the tracking error of WiDFS and Widar 2.0 after hardware diversity calibration. It shows that Widar2.0 achieves a median tracking error of 108.4 cm and a $90^{th}$ percentile error of 240.6 cm. Comparatively, the median and $90^{th}$ percentile error of WiDFS are reduced to 72.31 cm and 170.8 cm, respectively, which could achieve about 36 cm and 70 cm improvement over Widar2.0. It is further noted that Widar2.0 performs backward smoothing by using posterior localization estimates to inversely refine the previous, which could result in large processing delay.

To demonstrate the significance of WiFi hardware diversity calibration, we plot the results of WiDFS without conducting calibration for comparison. In this case, a median accuracy of 112.9 cm and a $90^{th}$ percentile error of 212.4 cm are achieved. The tracking error is significantly higher than that after applying calibration. Fig. \ref{Fig14}, Fig. \ref{Fig15} and Fig. \ref{Fig16} show three cases of the trajectories estimated by WiDFS. Most localization results could match the ground-truth trajectory. It can be observed when the tracking target turns around (at peak positions in these figures), some estimated reflection distances exhibits larger errors than other cases. This may mainly be attributed to that part of human reflections are not be reflected to RX antennas.

\subsection{Passive tracking accuracy in LOS and NLOS}
\begin{figure*}
\centering
	\begin{minipage}[t]{0.329\linewidth}
	\centering
		\includegraphics[width=\textwidth]{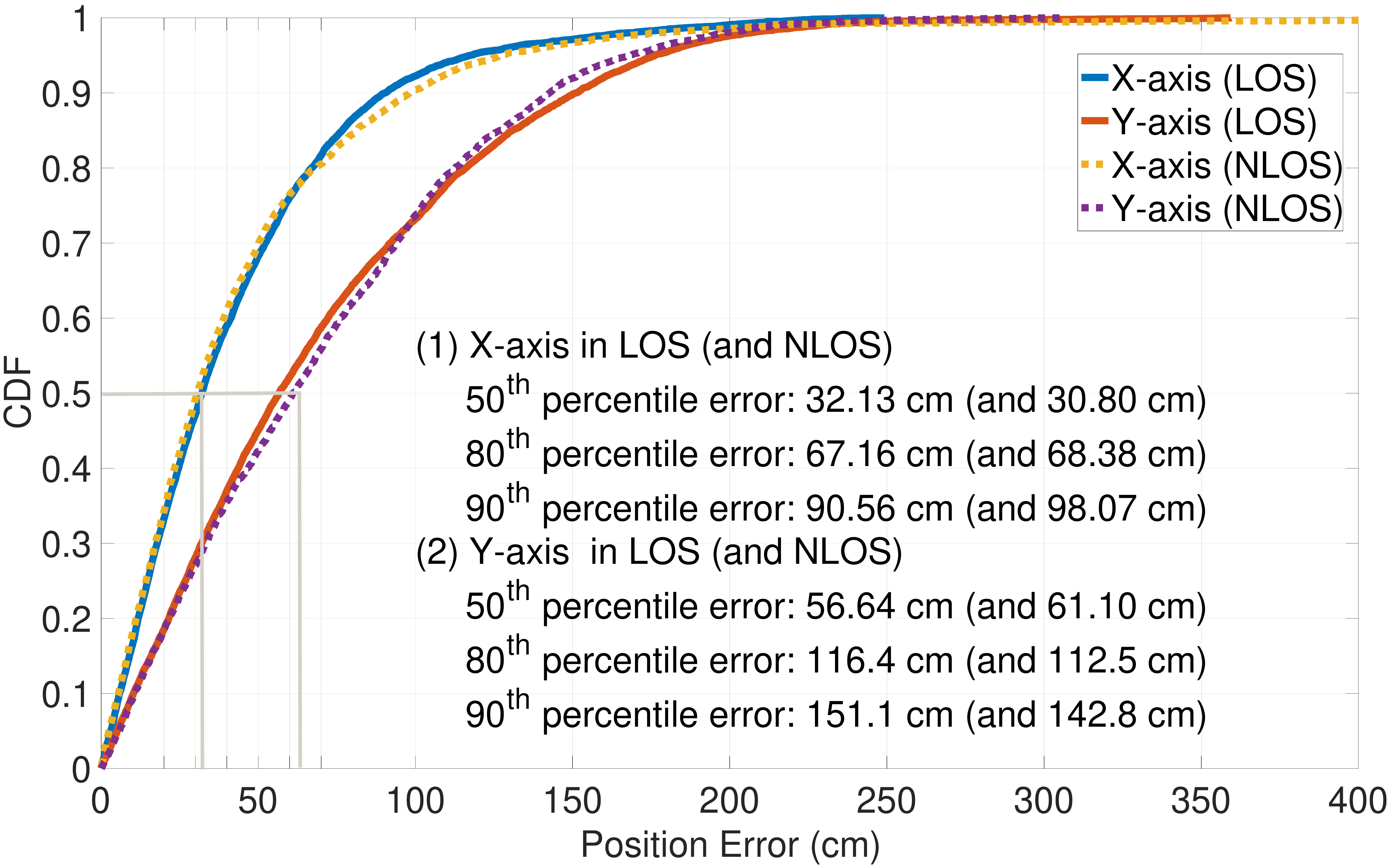}
		\caption{Tracking error in LOS and NLOS}
		\label{Fig17}
	\end{minipage}
	\begin{minipage}[t]{0.329\linewidth}
	\centering
		\includegraphics[width=\textwidth]{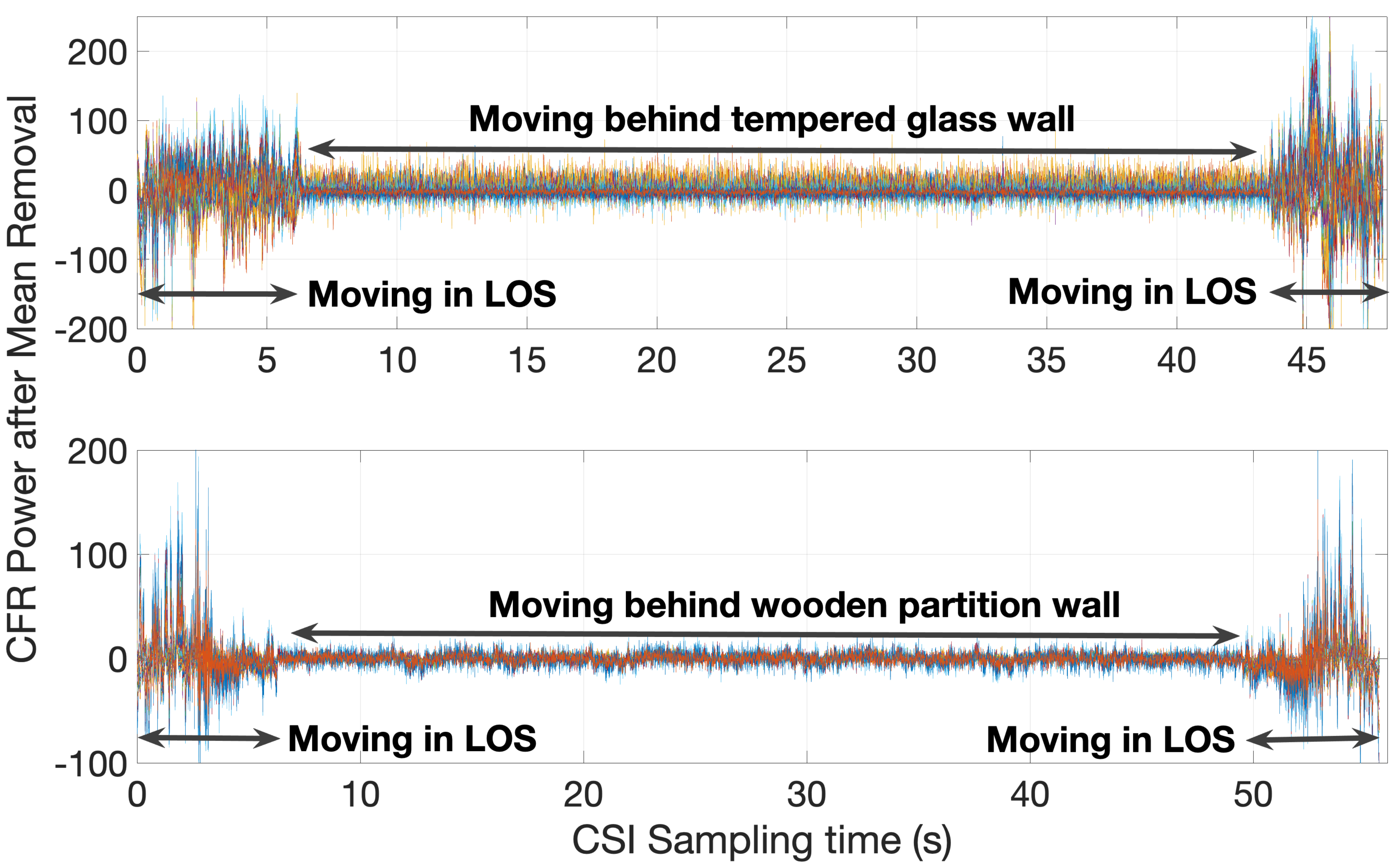}
		\caption{CFR power change through wall}
		\label{Fig18}
	\end{minipage}
	\begin{minipage}[t]{0.329\linewidth}
	\centering
		\includegraphics[width=\textwidth]{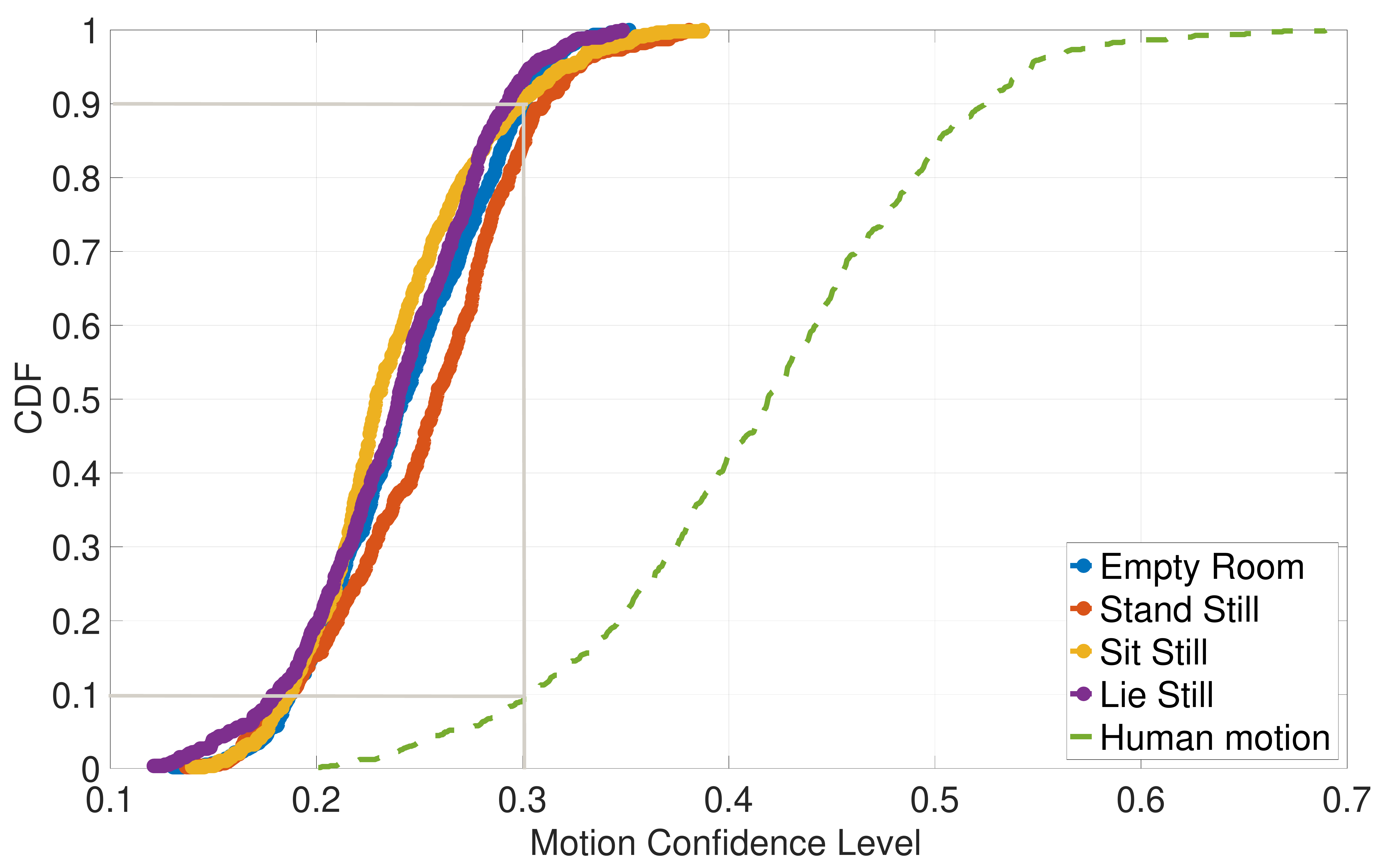}
		\caption{Human motion threshold}
		\label{Fig19}
	\end{minipage}
\vspace{-1em}
\end{figure*}

Next, Fig. \ref{Fig17} plots the CDF of the tracking error along the x- and y-axis dimensions in LOS and NLOS cases. The median errors along x- and y-axis are less than about 32 cm and 61 cm while the $90^{th}$ percentile errors along x- and y-axis are less than about 98 cm and 151 cm. Obviously, the x-axis has a lower error than the y-axis, which is likely attributed to that three RX antennas form a linear array along the x-axis. In addition, the tracking performance in NLOS settings is very close to the LOS case. In our scenario, WiFi signals may penetrate the cardboard boxes. Ideally, as long as the receiver could capture reliable human body reflections in NLOS cases, we believe that WiDFS enables performing accurate tracking since an obstacle (e.g., wall) may introduce a same phase shift on each subcarrier \cite{adib20143d, adib2015multi}.

We also conduct experiments to test WiDFS in NLOS scenarios where a person moves behind a glass and partition wall. Fig. \ref{Fig18} shows the change in CSI amplitudes at different subcarriers (after lowpass filtering and mean removal). The amplitudes change with human movement in the LOS region, while they almost keep constant in NLOS. This means the receiver may not capture human reflections due to higher signal attenuation through the wall.

\subsection{Accuracy vs. different parameters}
The following will evaluate WiDFS's tracking accuracy as a function of different system parameters: transmitter-to-receiver distance, motion speed, and joint window size.

\begin{table}
\caption{Running time of WiDFS and Widar2.0}
\vspace{-1.5em}
\label{table1}
\center
\begin{tabular}{ccccc}
\hline
\bfseries{Algorithm} & \bfseries{Language} & \bfseries{Platform}  & \bfseries {Mean} & \bfseries {Std.} \\
\hline
WiDFS    & Python &  Mini PC              &  0.076 s  &  0.018 s\\
WiDFS    & Python &  MacBook Pro 2019     &  0.024 s  &  0.002 s\\
WiDFS    & Matlab &  MacBook Pro 2019     &  0.016 s  &  0.002 s\\
Widar2.0 & Matlab &  MacBook Pro 2019     &  0.136 s  &  0.021 s\\
\hline
\end{tabular}
\vspace{-1.5em}
\end{table}

\begin{figure*}
\centering
	\begin{minipage}[t]{0.329\linewidth}
	\centering
		\includegraphics[width=\textwidth]{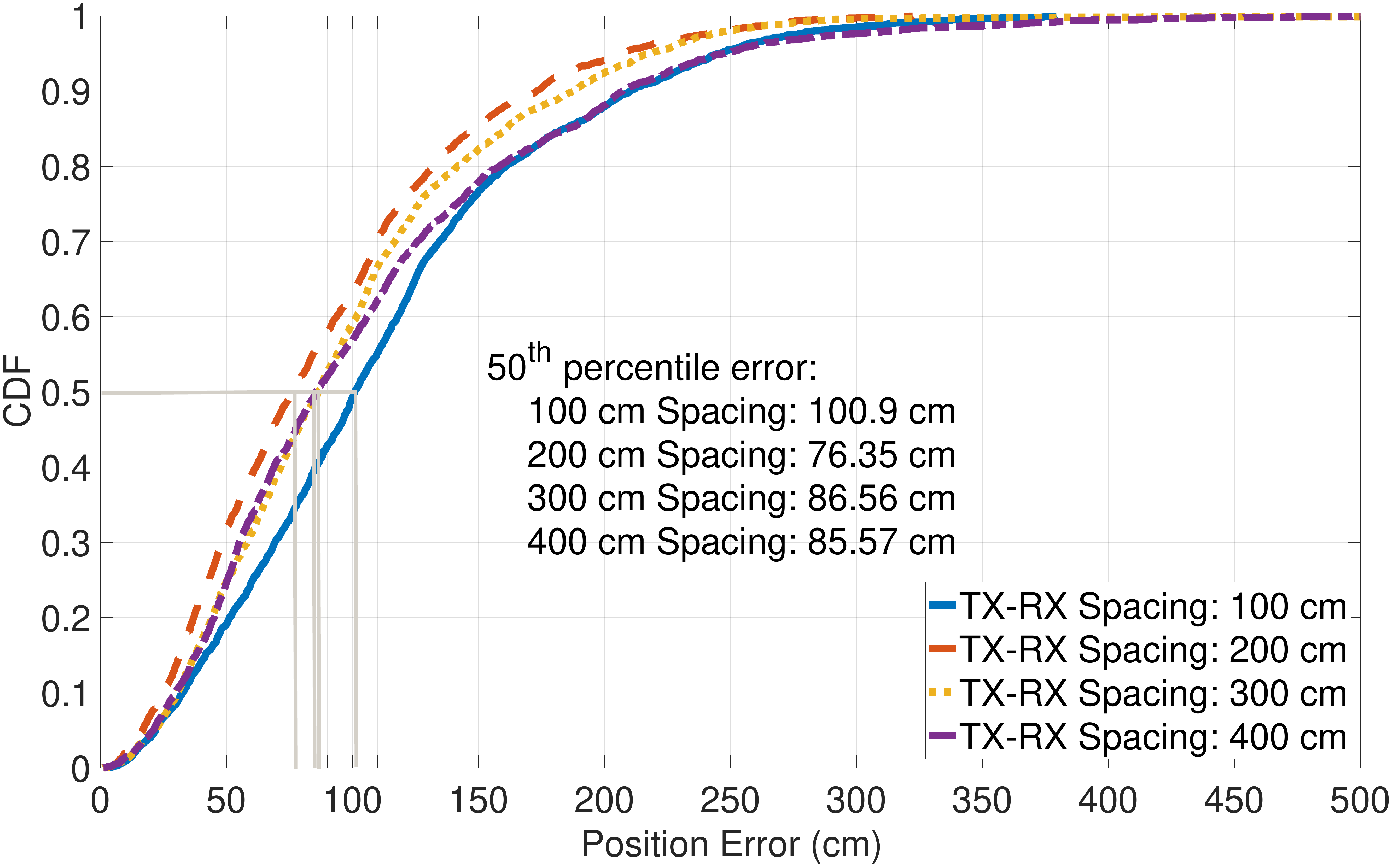}
		\caption{Impact of TX-RX spacing}
		\label{Fig20}
	\end{minipage}
	\begin{minipage}[t]{0.329\linewidth}
	\centering
		\includegraphics[width=\textwidth]{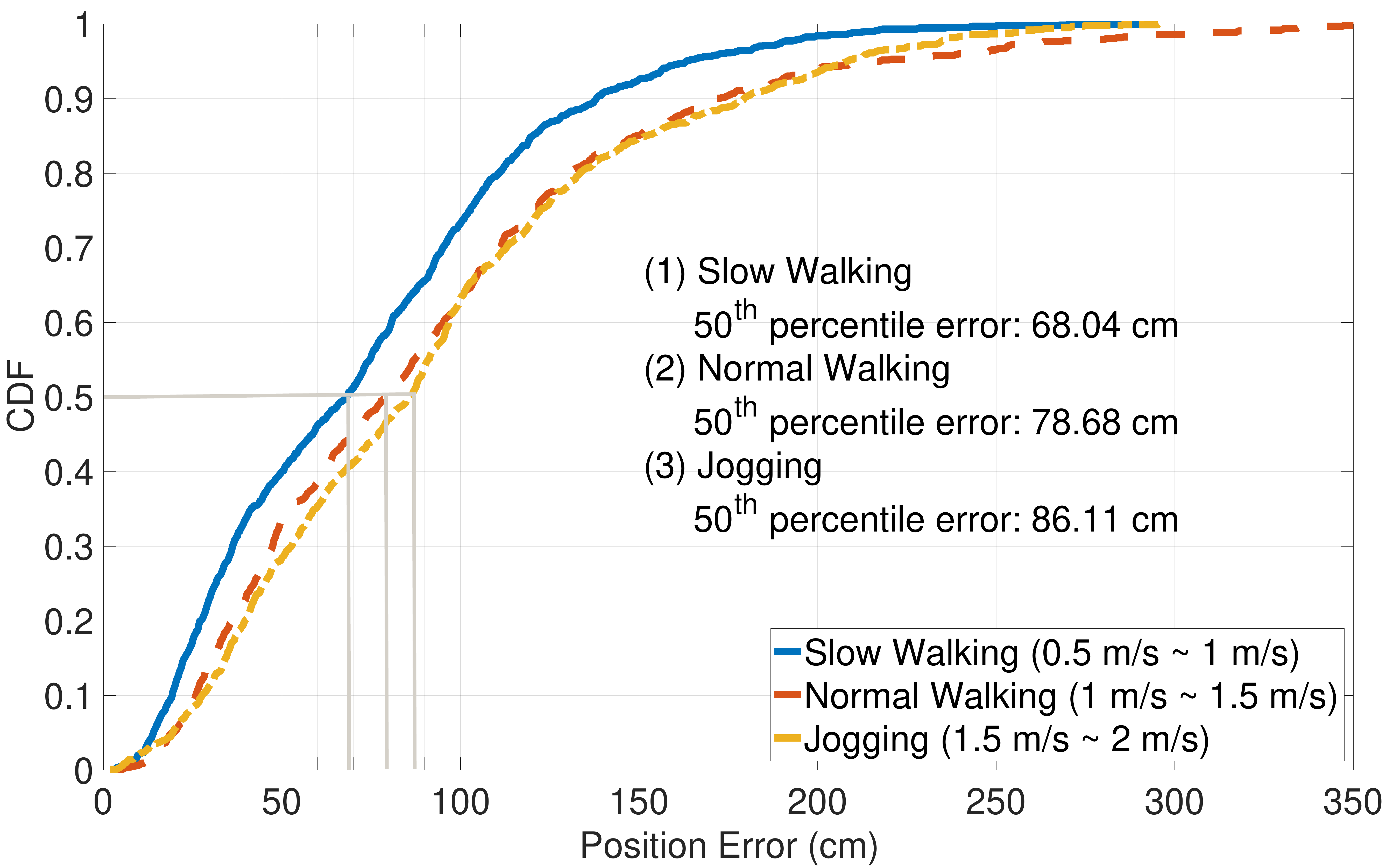}
		\caption{Impact of motion speed}
		\label{Fig21}
		\vspace{-1em}
	\end{minipage}
	\begin{minipage}[t]{0.329\linewidth}
	\centering
		\includegraphics[width=\textwidth]{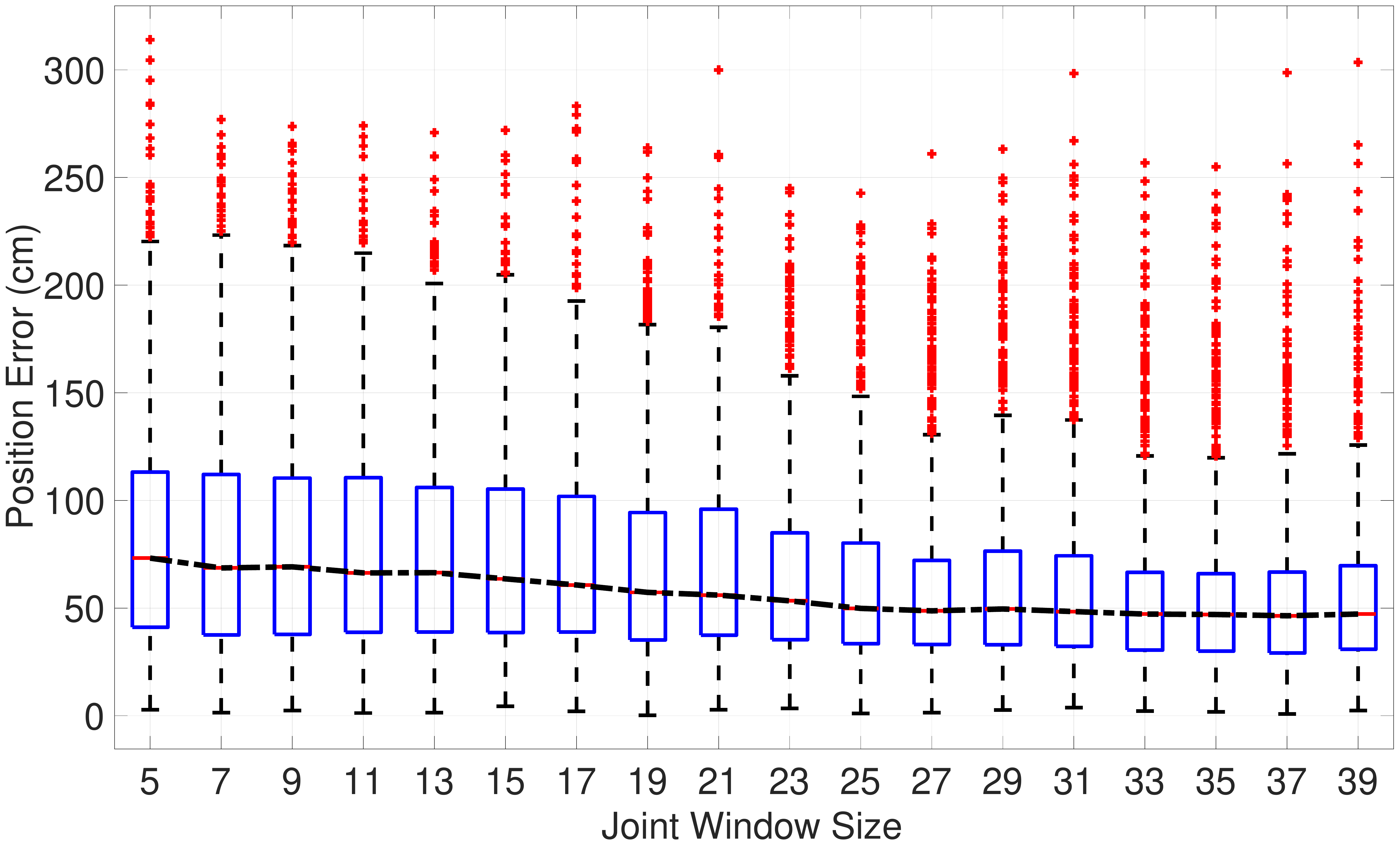}
		\caption{Impact of joint window size}
		\label{Fig22}
	\end{minipage}
\vspace{-1em}
\end{figure*}
\begin{figure*}
\centering
	\begin{subfigure}[t]{0.329\linewidth}
	\centering
		\includegraphics[width=\textwidth]{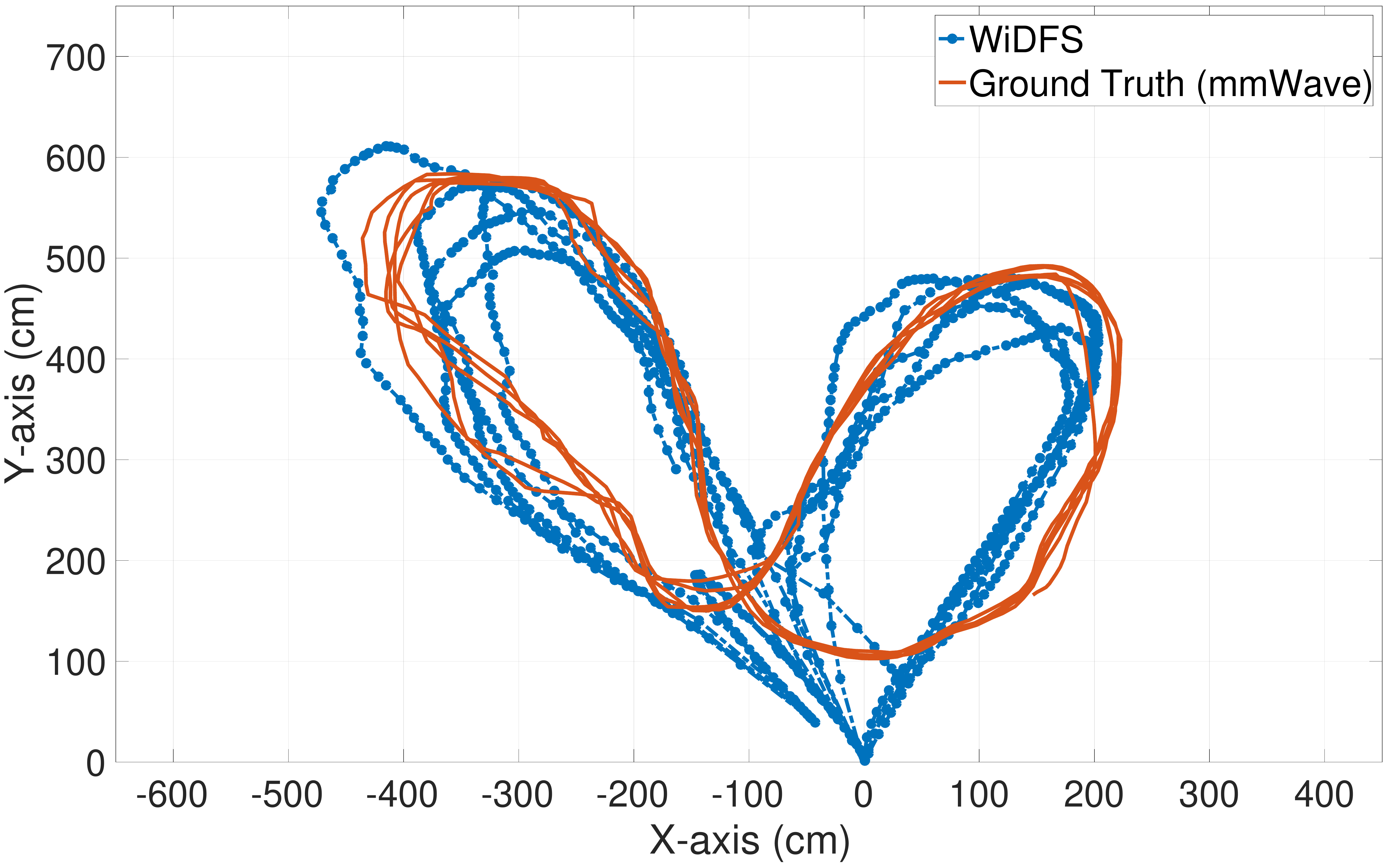}
		\caption{Elliptical Trajectory}
		\label{Fig23a}
	\end{subfigure}
	\begin{subfigure}[t]{0.329\linewidth}
	\centering
		\includegraphics[width=\textwidth]{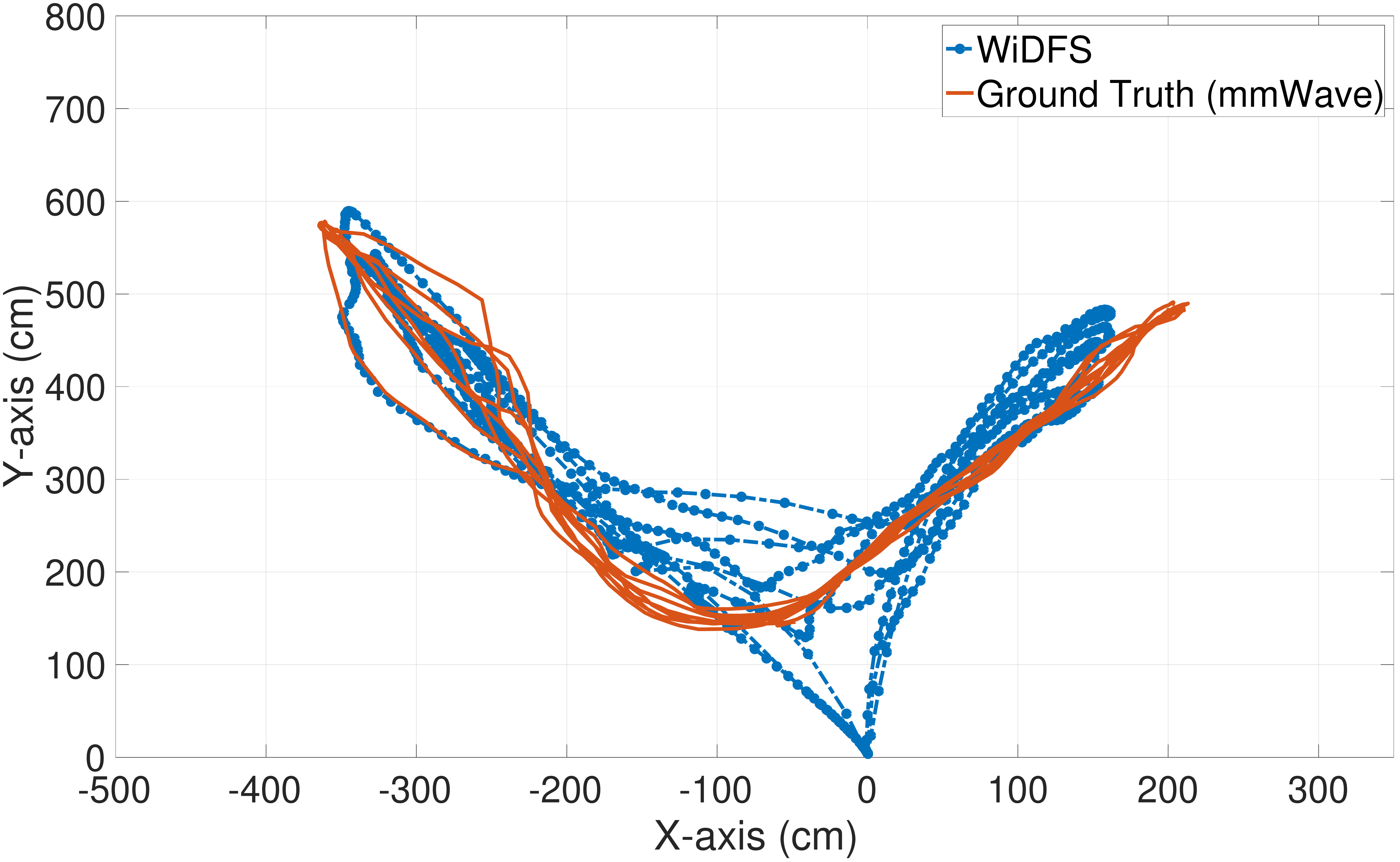}
		\caption{Linear Trajectory}
		\label{Fig23b}
	\end{subfigure}
	\begin{subfigure}[t]{0.329\linewidth}
	\centering
		\includegraphics[width=\textwidth]{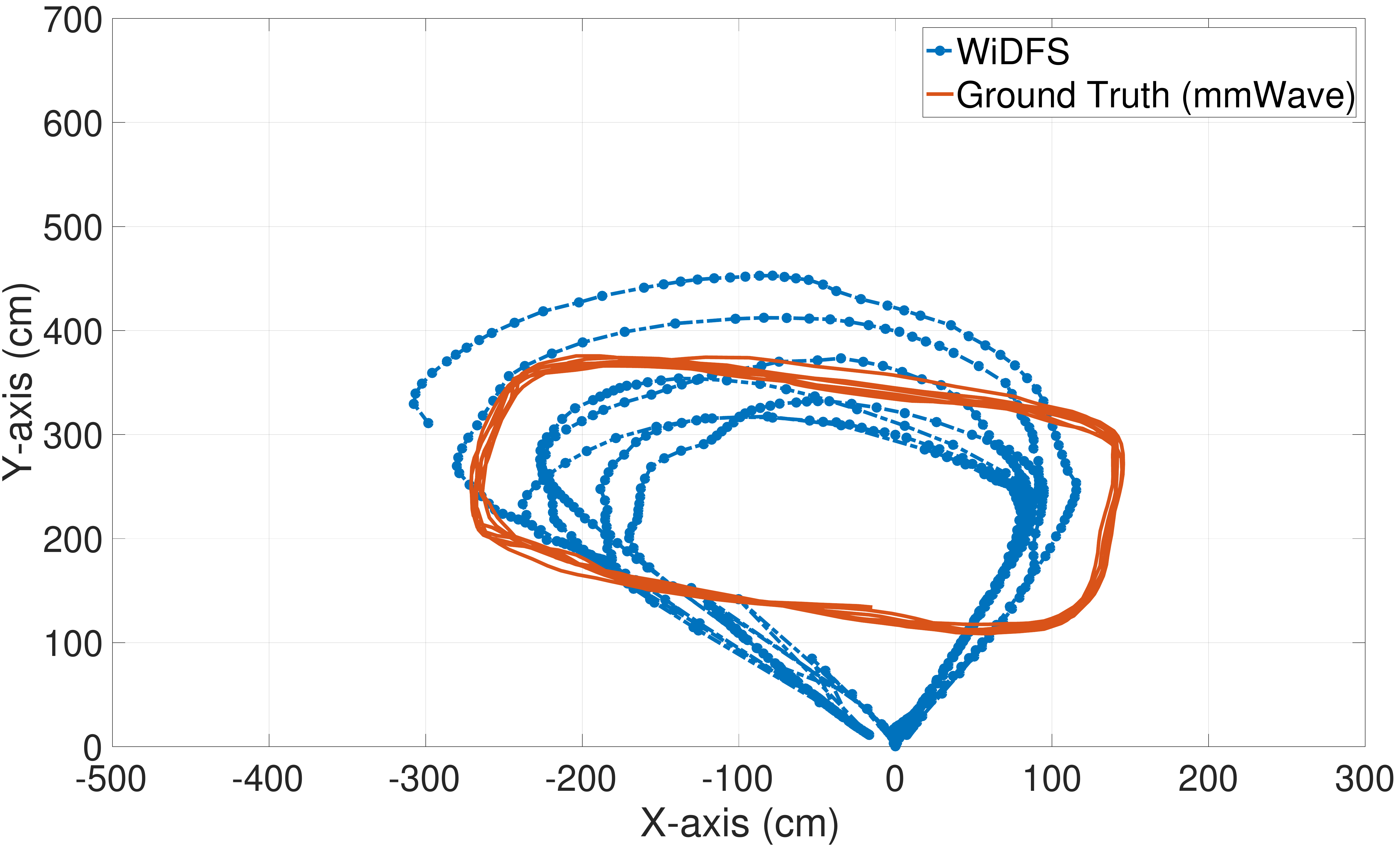}
		\caption{Rectangular Trajectory}
		\label{Fig23c}
	\end{subfigure}
	\caption{Three trajectories under a joint window containing 31 CSI sampling sub-windows. Compared to the results in Section 9.1, they could better match the ground truth. However, WiDFS needs more CSI samples for initialization to form a joint window. The delay in each estimation will be increased to about 3s, but it will not accumulate over time.}
		\label{Fig23}
		\vspace{-1.5em}
\end{figure*}

\subsection{Realtime performance}
WiDFS outputs a position estimate when receiving $100\times30\times3$ CSI samples from 3 antennas and 30 subcarriers. The sampling interval is about 0.1 s. Any computation delay larger than this upper bound may affect WiDFS's real-time performance. Table. \ref{table1} shows that the mean running time of WiDFS (in Python) is 0.076 s on a Mini PC. We also run WiDFS on a MacBook Pro with Intel Core i7 2.6 GHz $\times 6$. The csiread package is used to offline parse CSI data. The running time is reduced to 0.024 s. Also, we program WiDFS using Matlab. We parse raw CSI data to a .MAT file and then exploit it to perform WiDFS and Widar2.0. The computation time of WiDFS on MacBook Pro is 0.016 s, which is about 10 times faster than Widar2.0. This reduction benefits from the fact that WiDFS does not need to simultaneously estimate multi-dimensional localization parameters.

\subsection{Motion Threshold Selection}
This experiment aims to verify our motion detection method. At first, we ask each of the volunteers to separately stand still, sit still, and lie still. And we also keep the office empty for testing. Their $90^{th}$-percentile confidence levels are 0.3094, 0.2998, 0.2928, and 0.2983, respectively. Then each volunteer moves randomly at different velocities and motion directions. We can see that the $10^{th}$-percentile motion confidence level is less than 0.3. In this case, the motion threshold of 0.3 is selected to detect if there is a moving person. In the presence of human motion, WiDFS will automatically perform our tracking algorithm.

\subsubsection{Impact of transmitter-to-receiver distance} 
We vary the transmitter-to-receiver distance from 100 cm to 400 cm at a spacing of 100 cm. The RX antenna array is fixed. We only change the position of the TX antenna in this experiment. In each case, a person moves along the same trajectory. Fig. \ref{Fig20} shows the minimum tracking error occurs at a distance of 200 cm. This shows that there exist a tradeoff between the separation distance and tracking accuracy. When the transmitter-to-receiver separation distance is enlarged, the geometry of the tracked person relative to the transmitter and receiver is good, which means that the sensitivity to the distance measurement error will accordingly decrease. However, if the separation distance is too large, our assumption that there exists one dominant direct signal between the transmitter and receiver may become violated. Surrounding objects produce static multipath signals that we cannot neglect. In contrast, when the transmitter and receiver are placed too closely, the strength of the direct signal between them would overwhelm dynamic multipath.

\subsubsection{Impact of motion speed}
Fig. \ref{Fig21} shows the median tracking error when a person moves along the same trajectory at different speeds, i.e., Slow walking (about 0.5$\sim$1 m/s), Normal walking (1$\sim$1.5 m/s), and Jogging (1.5$\sim$2 m/s). The median error increases gradually with the higher velocity. Recall from our algorithm that we ignore the difference in DFS among different RX antennas in a CSI sampling window. However, the difference will increase at a higher movement speed. Our future solution is to extend WiDFS to detect the presence of a target moving very quickly and estimate the AFS. In this way, the DFS for each RX antenna would be separately estimated, thereby achieving more accurate tracking.

\subsubsection{Impact of joint window size} 
We next vary the size of a joint window from 5 to 39 CSI sampling sub-windows. Fig. \ref{Fig22} shows that the tracking error gradually decreases as the number of sub-windows increases. Such a result is expected since RX antennas may not capture enough reliable human reflections in some sub-windows, and it is challenging for WiDFS to dilute the impact of the inaccurate dynamic components in a smaller joint window size. This problem can be mitigated by utilizing a larger joint window size. Fig. \ref{Fig23} shows three trajectories used in Section 9.1 for evaluation when a joint window contains 31 sub-windows. It shows that these trajectories are all refined and can better match the ground truth. However, a 3 s delay is introduced for each estimate under the joint window with 31 sub-windows. \textit{Note that since the mean estimation time is less than a sampling sub-window time, the delay will not be accumulated over time.}

\section{Related Work}
This section reviews various techniques of WiFi-based tracking and sensing, including \textit{signal model-based active tracking}, \textit{signal model-based passive tracking}, and \textit{deep learning-based passive tracking}. The signal model-based solutions are designed by purely analyzing CSI models, while the deep learning-based approaches generally require the use of pre-collected training data. The difference between \textit{active} and \textit{passive} systems is that the active tracking requires a target carrying a WiFi device while the passive tracking is free from this limitation and achieved only based on human body reflections. Compared to previous works, WiDFS is the first unsupervised real-time WiFi tracking system that leverages novel DFS estimation and dynamic component separation techniques to enable passive tracking.

\subsection{Signal model-based active WiFi tracking} 
The topic of WiFi localization has attracted much attention in past years. Typically, ArrayTrack \cite{xiong2013arraytrack} constructs a specialized MIMO-based WiFi receiver platform to estimate AoA from a transmitter's incoming signal, which could pinpoint the transmitter at decimeter-level accuracy. However, its localization error is highly dependent on the number of receivers and the relative geometry of the receivers to the transmitter. SpotFi \cite{kotaru2015spotfi} is also a AoA-based solution using COTS WiFi NICs. It estimates AoA and ToF of a transmitter's signal arriving at a three-antenna receiver via smoothed MUSCI algorithm. However, this ToF is not an actual value distorted by unknown time and frequency shifts between the transmitter and receiver. Chronos \cite{vasisht2016decimeter} is a ToF-based localization solution via COTS devices, which achieves decimeter-level localization only using a receiver equipped with multiple RX antennas. It works by combining multiple frequency bands of 2.4 GHz and 5 GHz to compute the ToF between the transmitter to each antenna of the receiver. In practice, however, it is challenging to span multiple frequency bands for COTS NICs. WiCapture \cite{kotaru2017position} is designed based on COTS WiFi devices. It estimates AoA and complex attenuation of each signal propagation path and then combines them to calculate the relative displacement of a transmitter. It just estimates a transmitter's relative trajectory rather than absolute positions. Navid et al. \cite{tadayon2019decimeter} proposes a series of pre-processing methods to eliminate random phase shifts due to transceiver clock asynchronization. Then a MUSIC algorithm is applied to obtain ToF estimates and thereby achieve decimeter-level localization. Our work focuses on passive WiFi tracking and leverages `bad' multipath interference to track a moving target's trajectory.

\subsection{Signal model-based passive WiFi tracking} 
The device-free sensing is a promising technique since tracked targets do not require carrying any sensors. LiFS \cite{wang2016lifs} formulates the relationship between CSIs and a target location based on a signal power fading model. Then it can rely on the change in CSI amplitude to determine the absence or presence of a person in the first Fresnel zone and locate the target. However, this work requires deploying a large number of receivers in advance (11 receiver PCs are used). MaTrack \cite{li2016dynamic} is a AoA-based solution which uses 2D MUSIC algorithm to estimate absolute AoAs and relative ToFs of static and dynamic paths. However, multiple receivers are required to be deployed apart around a surveillance region and the localization accuracy is subject to the precision of random phase shift removal. Rui Zhou et al. \cite{zhou2017device} construct a CSI fingerprint database by modeling the relationship between CSI fingerprints and target locations. When a set of new CSI samples comes, this method performs localization by finding the most similar sample from the fingerprint database. However, constructing such a fingerprint database is a time-consuming task. Widar \cite{qian2017widar} and IndoTrack \cite{li2017indotrack} are DFS-based solutions. They both adopts a reference antenna approach to coarsely separate each antenna's dynamic component from cross-correlation terms. However, the tracking error will accumulate over time. Widar2.0 \cite{qian2018widar2}, mD-Track \cite{xie2019md}, and WiPolar \cite{venkatnarayan2020leveraging} are joint parameter estimation solutions by simultaneously estimating signal attenuation, AoA, ToF, and DFS from CSIs. Widar2.0 could completely mitigate random phase shifts by CSI cross correlation between RX antennas. However, its separated dynamic component is not accurate enough. The localization accuracy of mD-Track and WiPolar is determined by the accuracy of random phase shift measurement. Moreover, the three solutions require huge computations in each estimation and cannot achieve real-time tracking. Our proposed WiDFS scheme only uses a COTS transmitter (with 1 TX antenna) and receiver (with 3 RX antennas). It completely removes the impact of random phase shifts via cross correlation. A novel DFS estimation algorithm is developed to obtain unambiguous DFSs. On this basis, WiDFS separately estimates the AoA and reflection distance to achieve low-cost and real-time passive tracking.

\subsubsection{Deep Learning-based WiFi tracking} 
Recently, many applications have been designed with the help of deep learning technique to achieve fine-grained localization/tracking. FreeTrack \cite{zhou2019freetrack} is a deep neural network (DNN)-based tracking system by feeding into CSI amplitude fingerprints. To reduce the impact of environment change, FreeTrack needs to fine-tune the pre-trained DNN with a few CSI samples from new environments. DLoc \cite{ayyalasomayajula2020deep} designs a novel DL-based framework by treating WiFi localization as an image translation problem. It transforms CSI data into an image and labels training data with a customized robotic mapping platform. FiDo \cite{chen2020fido} could significantly reduce the number of training data in new scenarios. Even at the same position, different people may induce different CSI data. FiDo adopts a domain-adaptive solution and can pinpoint different persons only using few labelled CSI data. Compared to these data-driven systems that are sensitive to the changes in deployment environment and transceiver position, WiDFS is an unsupervised system and can still achieve high-precision passive tracking.

\section{Conclusion}
This paper introduces a device-free WiFi tracking system that can track a moving person in real time at sub-meter position accuracy. The key design is a novel DFS algorithm that enables measuring an unambiguous DFS when CSI cross-correlation is exploited to mitigate the impact of transceiver asynchronization. And novel algorithms of dynamic human component separation and localization parameter estimation are proposed to achieve high-precision and real-time tracking. Our system prototype runs in a practical multipath-rich scenario without requiring any environment-specific training. Some video clips demonstrating real-time tracking results are available from our website\footnote{\begin{scriptsize} {https://sites.google.com/view/andrewzhang/research/wif\mbox{}isensing?authuser=0} \end{scriptsize}}.

One of the important future works is to extend our method to achieve multi-person tracking. To fulfil this task, three TX antennas can be utilized to build a multiple-input-multiple-output (MIMO) system. The AoA resolution of the MIMO WiFi system with 3 TX antennas and 3 RX antennas may be achieved to be equivalent to that of a single-input-multiple-output (SIMO) system (i.e., our WiDFS) with 9 RX antennas. Therefore, the MIMO setup would be a promising way to improve the accuracy in multi-person tracking.

\ifCLASSOPTIONcompsoc
  \section*{Acknowledgments}
\else
  \section*{Acknowledgment}
\fi
This work is partially funded by the NSW Defence Innovation Network and the NSW State Government through Pilot Project grant DINPP-19-20 10.01.

\ifCLASSOPTIONcaptionsoff
  \newpage
\fi



%
\input{WiFi_tracking.bbl}

\bibliographystyle{IEEEtran}
\bibliography{WiFi_tracking}


\vspace{-13em}
\begin{IEEEbiography} [{\includegraphics[width=1in,height=1.25in,clip,keepaspectratio]{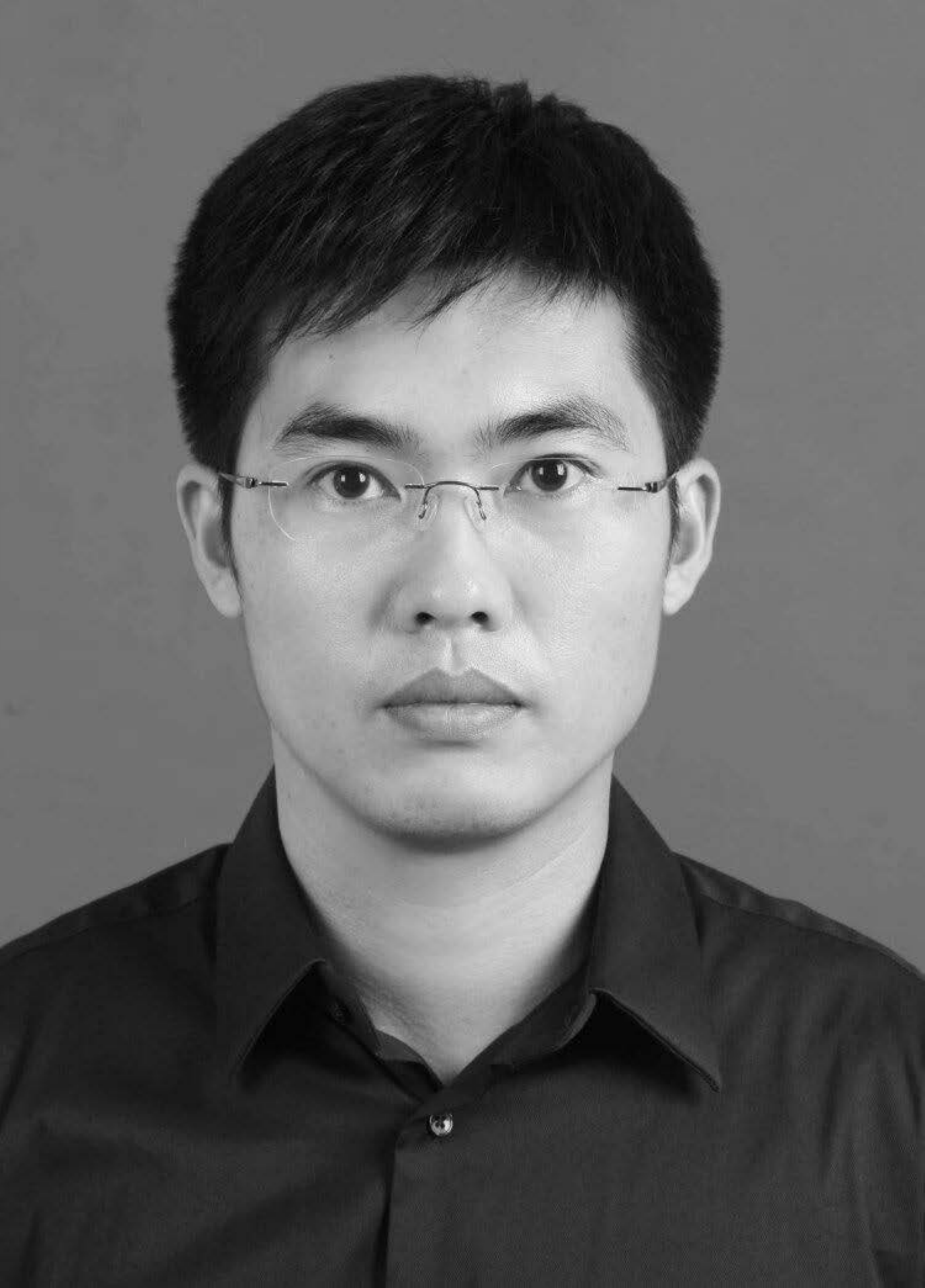}}]{Dr Zhongqin Wang} is currently a research engineer in the School of Electrical and Data Engineering at University of Technology Sydney. He received his M.S. and Ph.D. degrees from Nanjing University of Posts and Telecommunications, Nanjing, China in 2014 and 2020, respectively. He also received a Ph.D. degree from University of Technology Sydney, Australia in 2021. His research interest includes Wireless Sensing based on WiFi, mmWave, and RFID.
\end{IEEEbiography}
\vspace{-15em}
\begin{IEEEbiography}
[{\includegraphics[width=1in,height=1.25in,clip,keepaspectratio]{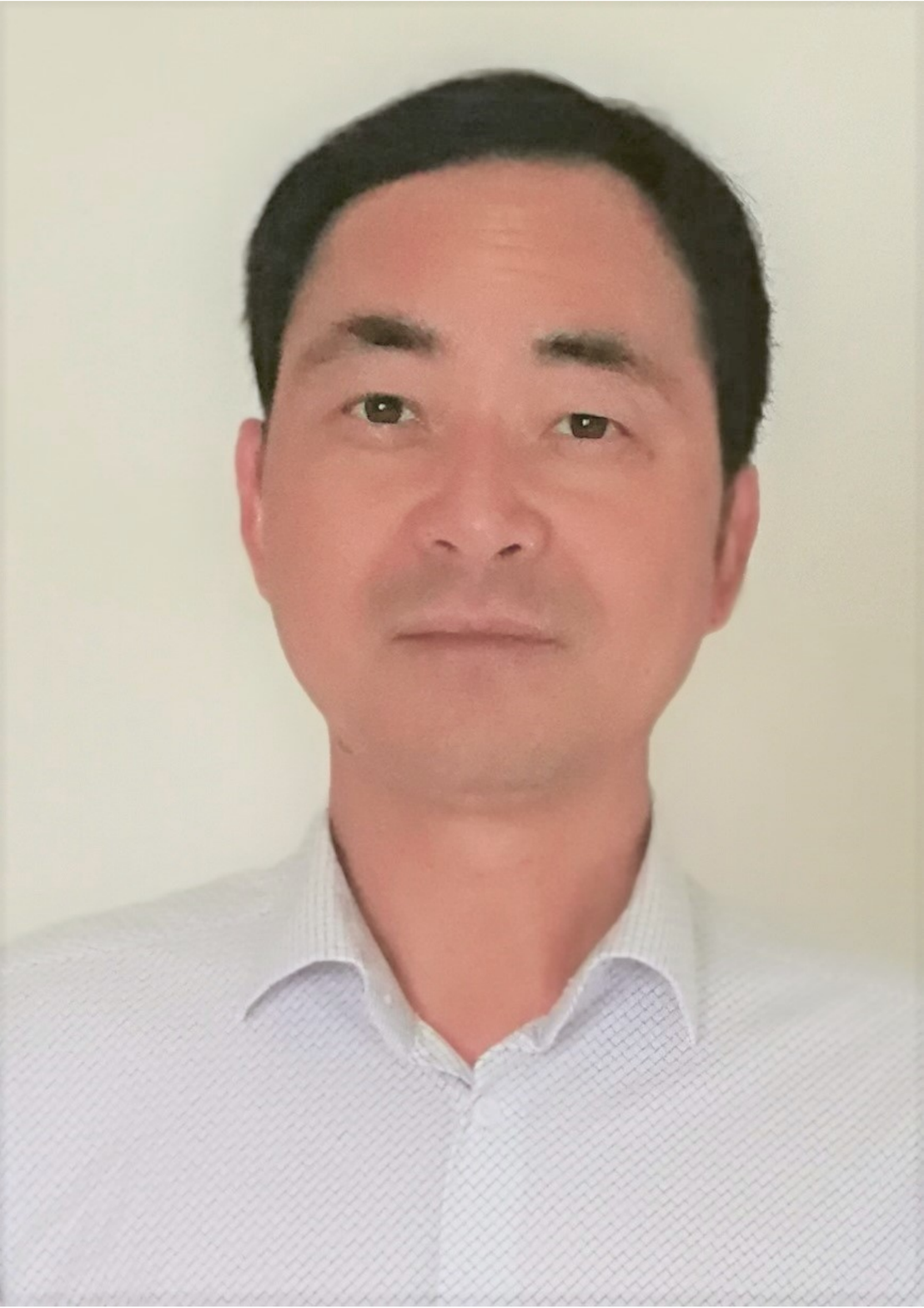}}]{Dr J. Andrew Zhang} (M'04-SM'11) received B.Sc. degree from Xi'an JiaoTong University, China, in 1996, M.Sc. degree from Nanjing University of Posts and Telecommunications, China, in 1999, and Ph.D. degree from the Australian National University, in 2004. Currently, He is an Associate Professor in the School of Electrical and Data Engineering, University of Technology Sydney, Australia. Dr. Zhang's research interests are in the area of signal processing for wireless communications and sensing, and autonomous vehicular networks. He has published 200+ journal and conference papers, and has won 5 best paper awards.
\end{IEEEbiography}
 \newpage
\begin{IEEEbiography} [{\includegraphics[width=1in,height=1.25in,clip,keepaspectratio]{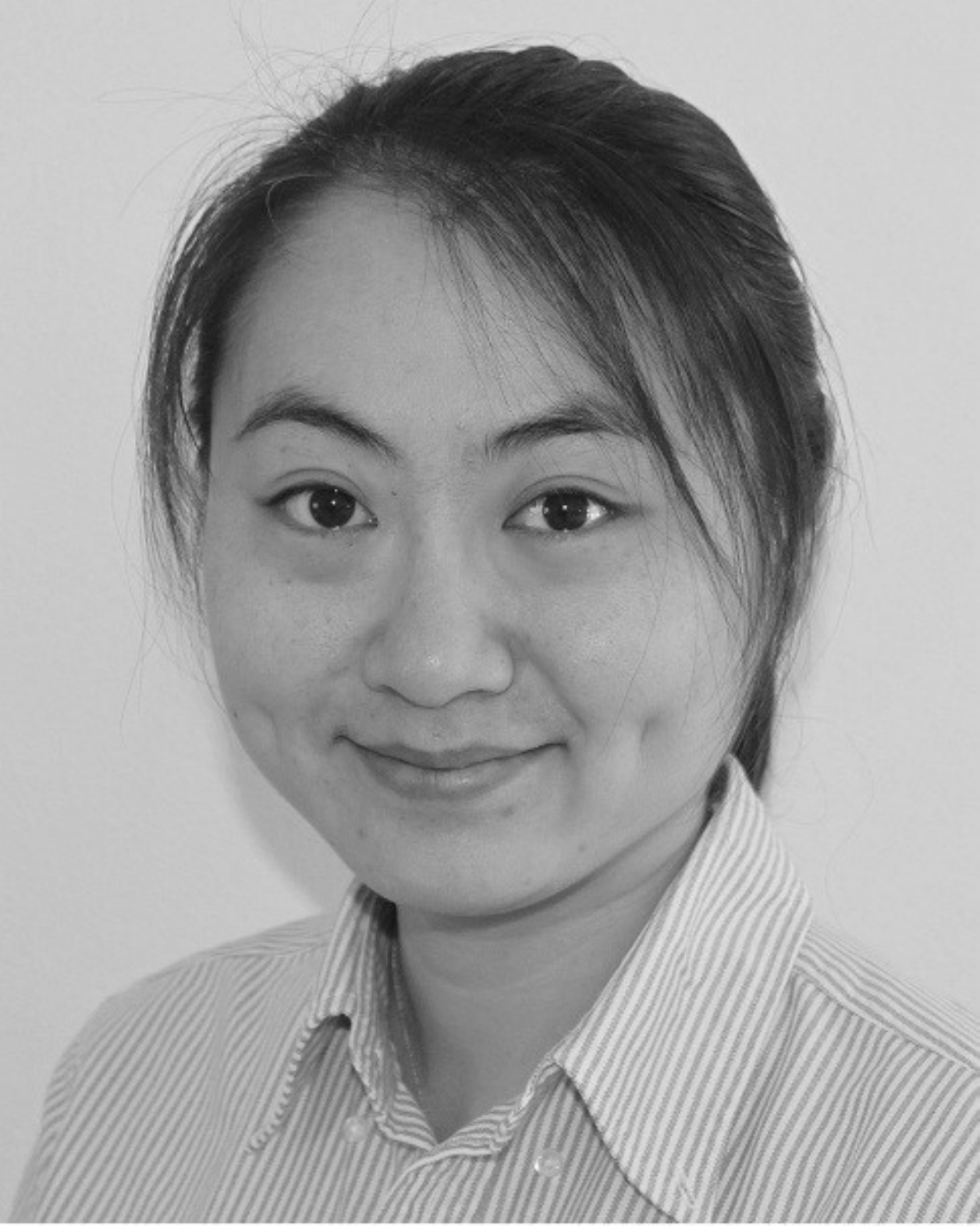}}]{Min Xu} (M'10) is currently an Associate Professor at University of Technology Sydney. She received the B.E. degree from the University of Science and Technology of China, Hefei, China, in 2000, the M.S. degree from National University of Singapore, Singapore, in 2004, and the Ph.D. degree from University of Newcastle, Callaghan NSW, Australia, in 2010. Her research interests include multimedia data analytics, computer vision and machine learning. She has published over 100 research papers in high quality international journals and conferences. She has been invited to be a member of the program committee for many international top conferences, including ACM Multimedia Conference and reviewers for various highly-rated international journals, such as IEEE Transactions on Multimedia, IEEE Transactions on Circuits and Systems for Video Technology and much more. She is an Associate Editor of Journal of Neurocomputing.
\end{IEEEbiography} 
\vspace{-21em}
\begin{IEEEbiography} [{\includegraphics[width=1in,height=1.25in,clip,keepaspectratio]{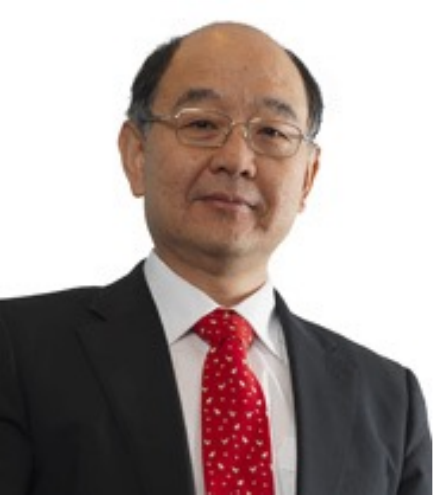}}] {Y. Jay Guo} (Fellow'2014) received a Bachelor Degree and a Master Degree from Xidian University in 1982 and 1984, respectively, and a PhD Degree from Xian Jiaotong University in 1987, all in China. His research interest includes antennas, mm-wave and THz communications and sensing systems as well as big data technologies. He has published four books and over 600 research papers including over 280 IEEE Transactions papers, and he holds 26 patents. He is a Fellow of the Australian Academy of Engineering and Technology, a Fellow of IEEE and a Fellow of IET, and was a member of the College of Experts of Australian Research Council (ARC, 2016-2018). He has won a number of most prestigious Australian Engineering Excellence Awards (2007, 2012) and CSIRO Chairman's Medal (2007, 2012). He was named one of the most influential engineers in Australia in 2014 and 2015, respectively, and one of the top researchers across fields in Australia in 2020. 

He is a Distinguished Professor and the Director of Global Big Data Technologies Centre (GBDTC) at the University of Technology Sydney (UTS), Australia. Prior to this appointment in 2014, he served as a Director in CSIRO for over nine years. Before joining CSIRO, he held various senior technology leadership positions in Fujitsu, Siemens and NEC in the U.K. 

Prof Guo has chaired numerous international conferences and served as guest editors for a number of IEEE publications. He is the Chair of International Steering Committee, International Symposium on Antennas and Propagation (ISAP). He has been the International Advisory Committee Chair of IEEE VTC2017, General Chair of ISAP2022, ISAP2015, iWAT2014 and WPMC'2014, and TPC Chair of 2010 IEEE WCNC, and 2012 and 2007 IEEE ISCIT. He served as Guest Editor of special issues on "Low-Cost Wide-Angle Beam Scanning Antennas","Antennas for Satellite Communications" and "Antennas and Propagation Aspects of 60-90GHz Wireless Communications," all in IEEE Transactions on Antennas and Propagation, Special Issue on "Communications Challenges and Dynamics for Unmanned Autonomous Vehicles," IEEE Journal on Selected Areas in Communications (JSAC), and Special Issue on "5G for Mission Critical Machine Communications", IEEE Network Magazine.
\end{IEEEbiography}




\end{document}